\documentclass[a4paper,11pt]{article}
\usepackage{jcappub}

\usepackage{booktabs}
\usepackage[english]{babel}
\usepackage{amsmath,amssymb,amsbsy,amstext, amsthm, simplewick}
\usepackage{hyperref}
\usepackage{graphicx}
\usepackage{amsfonts}
\usepackage{amssymb}
\usepackage{framed}
\usepackage{enumitem}
\usepackage{soul}
\usepackage{orcidlink}
\usepackage{upgreek}
 \usepackage{exscale,relsize}
 \usepackage[makeroom]{cancel}
\usepackage{soul}
\usepackage{bbold}
\usepackage{lipsum}
\usepackage{mdframed}
\usepackage{mathtools}
\usepackage[dvipsnames]{xcolor}
\usepackage{tikz}
\usepackage{tikz-cd}
\usepackage[export]{adjustbox}
\usepackage{dsfont}
\usepackage[mathscr]{eucal}

 \newcommand{\circled}[2][]{%
  \tikz[baseline=(char.base)]{%
    \node[shape = circle, draw, inner sep = 1pt]
    (char) {\phantom{\ifblank{#1}{#2}{#1}}};%
    \node at (char.center) {\makebox[0pt][c]{#2}};}}
\robustify{\circled}

\newcommand*\diff{\mathop{}\!\mathrm{d}}

\newcommand{\es}[2] {\begin{equation} \label{#1} \begin{split} #2 \end{split} \end{equation}}

\usepackage{colortbl}
\definecolor{lightgreen}{cmyk}{0.2, 0, 0.2, 0.2}
\definecolor{lightgray}{cmyk}{0.1,0.2,0,0.1}
\definecolor{lightgray2}{cmyk}{0.1,0.1,0,0.1}

\makeatletter
\newlength{\apb@width}
\newcommand{\autoparbox}[2][c]{\settowidth{\apb@width}{#2}\parbox[#1]{\apb@width}{#2}}

\newcommand{\Cen}[2]{%
  \ifmeasuring@
    #2%
  \else
    \makebox[\ifcase\expandafter #1\maxcolumn@widths\fi]{$\displaystyle#2$}%
  \fi
}
\makeatother

\newcommand{\beq}{\begin{equation}\begin{aligned}}
\newcommand{\eeq}{\end{aligned}\end{equation}}

\RequirePackage{amsmath}
\RequirePackage{amssymb}
\RequirePackage{epsfig}
\RequirePackage{graphicx}
\RequirePackage[numbers,sort&compress]{natbib}
\RequirePackage[colorlinks=true
  ,urlcolor=blue
  ,anchorcolor=blue
  ,citecolor=blue
  ,filecolor=blue
  ,linkcolor=blue
  ,menucolor=blue
  ,pagecolor=blue
  ,linktocpage=true
  ,pdfproducer=medialab
  ,pdfa=true
]{hyperref}

\def\beq{\begin{equation}}
\def\eeq{\end{equation}}

\def\Beq{\begin{equation}\begin{aligned}}
\def\Eeq{\end{aligned}\end{equation}}

\def\bea{\begin{eqnarray}}
\def\eea{\end{eqnarray}}

\def\beq{\begin{equation}}
\def\eeq{\end{equation}}
\def\bea{\begin{eqnarray}}
\def\eea{\end{eqnarray}}

\def\bp{\boldsymbol{p}}

\def\bk{\boldsymbol{k}}

\DeclareRobustCommand{\SkipTocEntry}[4]{}
\DeclareSymbolFont{extraup}{U}{zavm}{m}{n}
\DeclareMathSymbol{\varheart}{\mathalpha}{extraup}{86}
\DeclareMathSymbol{\vardiamond}{\mathalpha}{extraup}{87}

\title{Gravitational Waves from Isocurvature Perturbations of Spectator Scalar Fields}

\author[a, 1]{Marcos A. G. Garcia \orcidlink{0000-0003-3496-3027}}
\affiliation[a  ]{Departamento de F\'isica Te\'orica, Instituto de F\'isica, \\
Universidad Nacional Aut\'onoma de M\'exico, \\
Ciudad de M\'exico C.P. 04510, Mexico} 

\author[b, 2]{Sarunas Verner \orcidlink{0000-0003-4870-0826}}
\affiliation[b]{Kavli Institute for Cosmological Physics, \\
University of Chicago, 5640 South Ellis Ave., Chicago, IL 60637, USA}

\emailAdd{marcos.garcia@fisica.unam.mx}
\emailAdd{verner@uchicago.edu}

\abstract{We present a mechanism for gravitational wave (GW) production from isocurvature perturbations in spectator scalar fields during inflation. These energetically subdominant fields develop blue-tilted power spectra through inflationary dynamics, generating second-order scalar perturbations that source a stochastic GW background. The mechanism naturally satisfies CMB constraints at large scales while producing enhanced signals at smaller scales across a broad frequency range $10^{-20} - 1$ Hz. We perform comprehensive numerical and analytical calculations of the complete isocurvature spectrum evolution, including gravitational particle production, reheating dynamics, and scalar-induced GW generation. For spectator fields with effective masses $0.5 \lesssim m_{\chi,\mathrm{eff}}/H_I$, the resulting GW energy density reaches $\Omega_{\text{GW}} h^2 \sim 10^{-20}$--$10^{-12}$, accessible to pulsar timing arrays, space-based interferometers, and next-generation CMB experiments. Our analysis reveals that GW-induced constraints exceed current isocurvature bounds. We examine both unstable (curvaton-like) and stable (dark matter) spectator fields, demonstrating strong sensitivity to reheating temperature, inflaton-spectator coupling, and decay dynamics. This framework establishes isocurvature-sourced GWs as a powerful probe of early universe physics, enabling simultaneous constraints on inflationary dynamics, dark matter production, and reheating through coordinated multi-frequency GW observations.}

\begin{document}
\maketitle
\flushbottom
\section{Introduction}
\label{sec:introduction}
Recent gravitational wave (GW) discoveries have highlighted the importance of exploring new sources of GW signals to probe the early universe. This work is motivated by the anticipation of forthcoming results from current and planned CMB polarization experiments, such as LiteBIRD~\cite{Hazumi:2019lys} and the Simons Observatory (SO)~\cite{SimonsObservatory:2018koc}, which aim to measure---and potentially detect---primordial $B$-modes. With additional proposed missions like PICO~\cite{NASAPICO:2019thw} on the horizon, the importance of identifying new probes of inflation that can complement existing observational constraints has grown significantly.

It is well known that single-field slow-roll models of inflation predict a nearly scale-invariant primordial curvature power spectrum across comoving scales of $k \sim 10^{-4} - 1 \, \rm{Mpc}^{-1}$. However, the behavior of the primordial power spectrum on smaller scales $k \geq \, 1 \, \rm{Mpc}^{-1}$ remains largely unconstrained and unprobed by current observations. In this work, we explore contributions to both the curvature power spectrum and gravitational wave signals sourced by spectator scalar fields---fields present during inflation that do not drive the inflationary expansion. These spectator fields can generate substantial isocurvature perturbations that could produce distinctive observable signatures across a wide range of frequencies, potentially detectable by future GW surveys.

The presence of a spectator scalar field during inflation, such as a scalar dark matter candidate~\cite{Chung:1998zb, Chung:1998rq, Peebles:1999fz, Byrnes:2006fr, Markkanen:2018gcw, Kolb:2023ydq, Ling:2021zlj, Garcia:2022vwm, Garcia:2025rut}, the Standard Model (SM) Higgs field \cite{Herranen:2014cua, Espinosa:2007qp}, a curvaton \cite{Linde:1996gt, Enqvist:2001zp, Moroi:2001ct, Lyth:2001nq}, or a Peccei-Quinn field \cite{Peccei:1977ur, Peccei:1977hh,Bao:2022hsg, Chen:2023txq}, leads to the production of isocurvature perturbations~\cite{Ling:2021zlj, Garcia:2023awt}. The \textit{Planck} results constrain the curvature power spectrum to
\begin{equation}
    \label{eq:curvaturespectrum}
    \Delta^2_{\zeta}(k_*) \; \simeq \; 2.1 \times 10^{-9} \, ,
    \end{equation}
evaluated at the {\textit{Planck}} pivot scale $k_* = 0.05 \rm \, {Mpc}^{-1}$. Current constraints on the isocurvature power spectrum from \textit{Planck}~\cite{Planck:2018jri} are expressed as
\begin{equation}
    \label{eq:isocurvaturegen}
    \frac{\Delta^2_{\mathcal{S}}(k_*)}{\Delta^2_{\zeta}(k_*) + \Delta^2_{\mathcal{S}}(k_*)} < 0.038 \, ,
\end{equation}
at the 95\% confidence level, where $\Delta^2_{\mathcal{S}}(k_*)$ represents the isocurvature power spectrum. This constraint implies the following upper limit~\cite{Planck:2018jri}:
\begin{equation}
    \label{eq:isocurvatureupperlimit}
    \Delta^2_{\mathcal{S}}(k_*) < 8.3 \times 10^{-11} \, .
\end{equation}
Although the isocurvature power spectrum must be strongly suppressed at CMB scales of $k_* = 0.05 \, \rm{Mpc}^{-1}$ to satisfy \textit{Planck} constraints, it is possible for isocurvature perturbations to grow significantly at smaller scales. Our work demonstrates how such steep, blue-tilted ultraviolet (UV) spectra can naturally arise during inflation from the dynamics of a spectator scalar field, that we denote by $\chi$. Generally, the isocurvature constraints imply that the spectator scalar field must have an effective mass larger or comparable to the Hubble scale during inflation, with $m_{\chi, \rm eff} \gtrsim 0.5 H_I$~\cite{Chung:2004nh, Redi:2022zkt, Garcia:2023awt}. However, in this paper, we will show that the GWs induced by isocurvature of the spectator fields lead to more stringent bounds, typically requiring $m_{\chi, \rm eff} \gtrsim 0.6-0.7 H_I$ to ensure that the GW contribution is not in conflict with the bounds on the effective number of relativistic species $\Delta N_{\rm eff}$ from \textit{Planck}~\cite{Planck:2018jri}. Additionally, we show that there are complementary constraints on the spectator field mass arising from CMB B-mode polarization measurements by \textit{Planck} \cite{Planck:2018vyg}, as well as projections for future constraints from high-precision CMB polarization measurements by LiteBIRD \cite{Hazumi:2019lys}.

Cosmological modes characterized by high comoving wavenumbers $k \geq 1 \, \rm{Mpc}^{-1}$ probe the smallest physical scales accessible to observational cosmology. These short-wavelength fluctuations encode crucial information about early universe physics, including post-inflationary dynamics and potential insights into reheating mechanisms that remain inaccessible through traditional CMB measurements~\cite{Planck:2018jri}. A moderately amplified primordial power spectrum can lead to an overabundance of dark matter subhalos, opening opportunities for novel astrophysical probes and potentially resolving long-standing structure formation puzzles~\cite{DES:2020fxi, Boddy:2022knd}. 

The landscape of small-scale curvature power spectrum enhancement encompasses several compelling theoretical mechanisms. These include the ultra-slow roll phase in single-field inflation models~\cite{Garcia-Bellido:2017mdw, Ballesteros:2017fsr}, early matter domination scenarios~\cite{Erickcek:2011us, Kumar:2024hsi}, axion models with blue isocurvature power spectra~\cite{Kasuya:2009up, Chung:2015pga, Chung:2021lfg}, and recently proposed stochastic scalar field fluctuations~\cite{Ebadi:2023xhq}. More significant enhancements of the power spectrum with $\Delta_\zeta^2 \geq 10^{-7}$ can induce a stochastic gravitational wave background (SGWB) through second-order scalar perturbations, falling within the sensitivity range of future gravitational wave observatories \cite{Domenech:2021ztg}. In this work, we demonstrate that a spectator scalar field can naturally lead to a curvature power spectrum enhancement up to $\Delta_\zeta^2 \simeq 10^{-7} - 10^{-5}$ across scales $k \simeq 10^{3} - 10^{13} \, \rm{Mpc}^{-1}$ potentially generating GW signatures detectable by next-generation missions. Furthermore, even stronger curvature power spectrum enhancements may be achievable through direct couplings between the inflaton and spectator scalar fields or through the introduction of spectator fields with nonminimal gravitational couplings~\cite{Fairbairn:2018bsw, Garcia:2023qab, Verner:2024agh, Chakraborty:2024rgl}, which we leave for future studies.
 
Recent work~\cite{Ebadi:2023xhq} demonstrated a method for computing GW production in the early Universe from spectator fields with masses below the inflationary energy scale, when these fields obtain blue-tilted power spectra through stochastic effects during inflation. GW production sourced by spectator scalar fields during inflation and reheating was explored in detail in~\cite{Garcia:2025yit}. Our present study extends this framework through comprehensive numerical and analytical analysis of the full isocurvature power spectrum of spectator scalar fields, accounting for perturbation evolution both during and after inflation. We incorporate a detailed treatment of the reheating phase, which significantly impacts the resulting GW signals. Our analysis provides a complete calculation of the secondary GW spectrum induced by isocurvature perturbations at second order, as well as a thorough computation of the energy density associated with gravitationally produced spectator scalar field abundance~\cite{Domenech:2021and, Domenech:2021ztg, Domenech:2023jve}.\footnote{See Refs.~\cite{Gorji:2023ziy,Gorji:2023sil} for scenarios in which spectator fields generate tensor perturbations. Due to their linear coupling to metric tensor modes, such fields can produce a stochastic gravitational-wave background with amplitude $\Omega_{\rm GW}\sim10^{-7}$, comparable to that sourced by curvature perturbations.}

In general, the curvature power spectrum of the spectator field is sensitive to the cosmological history and the nature of the field itself---whether it behaves as a curvaton-like scalar that eventually decays or remains stable as a dark matter candidate~\cite{Ebadi:2023xhq}. During reheating, if the spectator field scales as matter while the inflaton decay products scale as radiation, the energy density ratio between the spectator scalar field (which remains subdominant compared to the inflaton) and the radiation background increases as a function of time. This evolution can potentially lead to GW signals with significantly larger amplitudes. Consequently, we expect the signals to be stronger for higher reheating temperatures, suggesting that future detection of such GWs could provide a novel observational probe of the otherwise elusive reheating epoch.

We compute gravitational particle production to determine the spectator field energy density that sources the GW signals.\footnote{For a recent review on gravitational particle production, see Ref.~\cite{Kolb:2023ydq}.} As the mass of the spectator field increases, the GW signal becomes stronger. However, when the effective spectator field mass exceeds the inflationary Hubble scale, $m_{\chi, \rm{eff}} \geq H_I$, the gravitational particle production becomes exponentially suppressed~\cite{Chung:1998bt, Chung:2018ayg, Racco:2024aac, Verner:2024agh}. This interplay creates an optimal mass range for maximal gravitational wave production, typically within $0.5 H_I \leq m_{\chi, \rm{eff}} \leq 0.7 H_I$, where the isocurvature perturbations are sufficiently large while particle production remains substantial.

This work opens a new avenue for studying various GW signatures from spectator scalar fields. We focus on two distinct scenarios: (1) curvaton-like models with decaying spectator fields where we systematically explore the impact of varying reheating temperatures, and (2) cases where the spectator field is identified as a stable dark matter candidate with reheating temperatures constrained by observed dark matter abundance constraints. For each scenario, we analyze both minimally coupled cases with purely gravitational interactions and models featuring direct coupling between the inflaton and spectator field through an interaction term of the form $\mathcal{L} \supset \frac{1}{2} \sigma \phi^2 \chi^2$. Importantly, we find that isocurvature perturbations can induce a SGWB across a remarkably broad frequency range, spanning from $10^{-20} - 1 \, \rm{Hz}$, potentially detectable by future GW observatories. For models incorporating direct inflaton-spectator coupling through, the amplitude of the GW energy density reaches $\Omega_{\rm GW} h^2 \simeq 10^{-16} - 10^{-12}$ in the nHz - mHz range, making these signals potentially detectable by proposed missions such as THEIA~\cite{Garcia-Bellido:2021zgu} and $\mu$-Ares~\cite{Sesana:2019vho}.

The remainder of this paper is structured as follows. Section~\ref{sec:infandmodel} presents the inflationary dynamics and spectator field properties, along with key aspects of reheating relevant to our study. Section~\ref{sec:gpp} examines gravitational particle production of the spectator field. Isocurvature fluctuations and their power spectra are analyzed in Section~\ref{sec:isocurvaturefluct}, while the resulting curvature power spectrum is detailed in Section~\ref{sec:curvatureperturbations}. Section~\ref{sec:gwspectraldensity} summarizes our computation of the GW spectral density. Secondary GWs induced by scalar perturbations are discussed in Section~\ref{sec:gravwaves1}, and Section~\ref{sec:gwiso} explores GW production from isocurvature fluctuations. Section~\ref{sec:gwsignals} presents the resulting GW signals and detection prospects, and we conclude in Section~\ref{sec:conclusions}.

We adopt natural units, $k_B = \hbar = c = 1$, with metric signature $(-,+,+,+)$.

\section{Inflation and Spectator Scalar Fields}
\label{sec:infandmodel}

\subsection{Inflationary Dynamics}
\label{sec:infdynamics}
We work within the framework of a homogeneous and isotropic Friedmann–Robertson–Walker (FRW) spacetime described by the metric
\begin{equation}
     ds^2 \; = \; a(\eta)^2 \left(-d\eta^2 + d\mathbf{x}^2 \right) \, ,
\end{equation}
where $d\eta \; \equiv \; dt/a$ is the conformal time, $a(\eta)$ is the dimensionless scale factor, and $\mathbf{x}$ represents the comoving spatial coordinates. Conformal time derivatives are denoted by primes, $' \equiv d/d\eta$. The dynamics of the inflaton field are governed by the action  
\begin{equation}
    \label{eq:infaction}
    \mathcal{S}_{\phi} = \int d^4 x \sqrt{-g} \left[-\frac{1}{2}M_P^2 R - \frac{1}{2} (\partial_{\mu} \phi)^2 - V(\phi) \right] \, ,
\end{equation}
where $g = \det g_{\mu \nu}$ is the determinant of the metric, $M_P = 1/\sqrt{8 \pi G_N} \simeq 2.435 \times 10^{18} \, \mathrm{GeV}$ is the reduced Planck mass, $R$ is the Ricci curvature scalar, and $V(\phi)$ is the inflaton potential. The equation of motion for the inflaton field $\phi$, derived from the above action, is  
\begin{equation}
    \label{eq:eom1conf}
    {\phi}'' + 2 \mathcal{H} \phi' + a^2 \frac{dV(\phi)}{d \phi} \; = \; 0 \, ,
\end{equation}
where $\mathcal{H} = a'/a $ is the conformal Hubble parameter. The dynamics of $\mathcal{H}$ are governed by
\begin{equation}
    \label{eq:hubble}
    \mathcal{H}^2 \; = \; (aH)^2 \; = \; \left(\frac{a'}{a} \right)^2 \; = \; \frac{\rho_{\phi}}{3M_P^2} a^2\, ,
\end{equation}
where $H = a'/a^2$ is the Hubble parameter and the inflaton energy density $\rho_{\phi}$ is given by
\begin{equation}
    \label{eq:endeninfl}
    \rho_{\phi} \; = \; \frac{\phi'^2}{2a^2} + V(\phi) \, .
\end{equation}

To characterize the slow-roll regime of inflation, we define the slow-roll parameters:
\begin{equation}
    \label{eq:endenandpress}
    \varepsilon_V \; \equiv \; \frac{1}{2}M_P^2 \left(\frac{V'(\phi)}{V(\phi)}\right)^2, \qquad \eta_V \; \equiv \; M_P^2 \frac{V''(\phi)}{V(\phi)} \, ,
\end{equation}
which quantify the flatness of the potential. During slow-roll inflation ($\varepsilon_V, |\eta_V| \ll 1$),  the number of $e$-folds from a given field value $\phi_{*}$ until the end of inflation at $\phi_{\rm end}$ is approximated by
\begin{equation}
    N_* \; \simeq \; \frac{1}{M_P^2} \int_{\phi_{\rm end}}^{\phi_*} \frac{V(\phi)}{V'(\phi)} d\phi \; \simeq \;  \int_{\phi_{\rm end}}^{\phi_*} \frac{1}{\sqrt{2\varepsilon_V}} \frac{d\phi}{M_P} \, ,
\end{equation}
where $k_* = 0.05~\mathrm{Mpc}^{-1}$ is the pivot scale adopted by the \textit{Planck} collaboration~\cite{Planck:2018jri}, and $\phi_*$ is the inflaton field value at the horizon exit scale $k_*$. Inflation ends at the scale factor $a_{\rm end} \equiv a(\eta_{\rm end})$, which is defined as the point where $d(1/aH)/d\eta = 0$. At this stage, the comoving Hubble scale reaches its minimum value and begins to increase immediately after inflation concludes. Alternatively, the end of inflation can be defined by the condition $V(\phi_{\rm end}) = \phi_{\rm end}'^2/a_{\rm end}^2$, where the kinetic and potential energy densities of the inflaton are equal. At this transition point, the total energy density is given by $\rho_{\rm end} \equiv \rho(a_{\rm end}) = \frac{3}{2}V(\phi_{\rm end})$. The corresponding Hubble parameter at the end of inflation is denoted as $H_{\rm end} \equiv H(a_{\rm end})$.

Observational data impose stringent constraints on inflationary models. The BICEP/Keck collaboration~\cite{BICEP:2021xfz} reports an upper bound on the tensor-to-scalar ratio, $r < 0.036$ at the 95\% confidence level (C.L.). Similarly, combined data from WMAP, \textit{Planck}, and BICEP/\textit{Keck} provide a tight range for the scalar spectral index, $0.958 < n_s < 0.975$ at the 95\% C.L. for $r = 0.004$. These observational boundaries serve as critical benchmarks for viable inflationary scenarios, including the T-model~\cite{Kallosh:2013maa} and the Starobinsky model~\cite{Starobinsky:1980te, Ellis:2013nxa}.

For our numerical analysis, we adopt the T-model of inflation~\cite{Kallosh:2013maa}, characterized by the potential:\footnote{We emphasize that the results presented in this paper are general and applicable to various large-field plateau-like inflationary models.}
\begin{equation}
\label{inf:tmodel}
V(\phi) = \lambda M_P^4 \left[ \sqrt{6} \tanh\left(\frac{\phi}{\sqrt{6}M_P}\right)\right]^2 \, .
\end{equation}
The dimensionless parameter $\lambda$ determines the inflaton mass at the minimum ($\phi = 0$), with $m_\phi = \sqrt{2\lambda} M_P$. This parameter is fixed by the observed amplitude of the primordial curvature power spectrum. The normalization of the potential~(\ref{inf:tmodel}) can be accurately approximated as~\cite{Garcia:2020wiy, Ellis:2021kad}:
\begin{align}
\lambda \; \simeq \; \frac{3\pi^2 \Delta_{\zeta}^2(k_*)}{N_*^2} \; \simeq \; \frac{6.2 \times 10^{-8}}{N_*^2} \, ,
\end{align}
where $N_*$ denotes the number of $e$-folds between horizon crossing of the pivot scale and the end of inflation and $\Delta_{\zeta}^2(k_*) \simeq 2.1 \times 10^{-9}$. For $N_* = 55$ $e$-folds,\footnote{Most of the details discussed in this work are not particularly sensitive to the number of $e$-folds $N_*$.} we obtain the following values for the scalar spectral index and tensor-to-scalar ratio:
\begin{equation}
    n_s \; \simeq \; 0.963 \, , \qquad r \; \simeq \; 0.004 \, .
\end{equation}
These results align well with current CMB observations from \textit{Planck} and BICEP/Keck~\cite{Planck:2018jri, BICEP:2021xfz, Ellis:2021kad}, placing the T-model comfortably within the observationally favored region of parameter space. For $N_* = 55$, the inflaton mass is $m_\phi \simeq \sqrt{2\lambda} M_P \simeq 1.6 \times 10^{13}~\mathrm{GeV}$, with the Hubble scale during inflation given by $H_I \simeq \sqrt{2\lambda} M_P \simeq m_\phi$, and $H_{\rm end} \simeq 6.3 \times 10^{12}~\mathrm{GeV}$ at the end of inflation (with $\phi_{\rm end} = 0.84 M_P$).\footnote{For a recent discussion of constraints on T-model inflation and reheating in light of ACT DR6 and SPT-3G data, see Ref.~\cite{Ellis:2025zrf}.}

\subsection{Reheating}
\label{sec:reheating}
Inflation ends when the accelerated expansion of the universe stops. This moment coincides with the end of the slow-roll regime of $\phi$ and the beginning of its coherent oscillations around the minimum of $V(\phi)$. The dissipation of the energy stored in these oscillations into light degrees of freedom (re)heats the universe. Assuming the inflaton decays into SM particles (directly or indirectly) this reheating populates the universe with a hot thermal plasma in equilibrium necessary to explain primordial nucleosynthesis~\cite{Kawasaki:2000en, deSalas:2015glj, Hasegawa:2019jsa}. 

For this work, we assume that reheating proceeds perturbatively around a quadratic minimum, consistent with our choice of potential (\ref{inf:tmodel}). The inflaton condensate maintains coherence as it decays, and collective effects from parametric resonance of decay product mode functions ({\em preheating}) can be neglected.\footnote{For the T-model (\ref{inf:tmodel}), this assumption holds if we assume two-body decays into fermions with Yukawa coupling $y\lesssim 10^{-5}$~\cite{Garcia:2021iag}, or multi-body decays as in no-scale inflation scenarios~\cite{Ellis:2015kqa,Ellis:2020lnc}.} Under these conditions, the coupled system of Boltzmann equations describing the inflaton energy dissipation into radiation are
\begin{align}
\rho'_{\phi} + 3\mathcal{H}\rho_{\phi} \;&=\; -a\Gamma_{\phi}\rho_{\phi}\,,\\ \label{eq:rhoRreh}
\rho'_R + 4\mathcal{H}\rho_R \;&=\; a\Gamma_{\phi}\rho_{\phi} \,,
\end{align}
where $\Gamma_{\phi}$ denotes the inflaton decay rate, $\rho_{R}$ is the energy density in the primordial radiation, and the energy densities satisfy the Friedmann constraint
\beq
\label{eq:friedreh}
3\mathcal{H}^2M_P^2 \;=\; a^2(\rho_{\phi} + \rho_R)\,.
\eeq
Reheating ends at the crossover time $t_{\rm reh}$ when $\rho_{\phi}(t_{\rm reh})=\rho_R(t_{\rm reh})\equiv \rho_{\rm reh}$. The reheating temperature $T_{\rm reh}$ is defined in terms of the thermodynamic relation 
\beq
\rho_R(t_{\rm reh}) \;=\; \frac{\pi^2}{30}g_{\rm reh}T_{\rm reh}^4\,,
\eeq
where $g_{\rm reh}$ denotes the effective number of relativistic degrees of freedom, with $g_{\rm reh} = 427/4$ in the SM for $T_{\rm reh} > m_t$, our choice for the present paper.

Before the crossover, when $a_{\rm end}<a\ll a_{\rm reh}$, radiation remains subdominant to the inflaton energy density, which redshifts as non-relativistic matter, $\rho_{\phi}\simeq \rho_{\rm end}(a/a_{\rm end})^{-3}$. Substituting this approximation into~(\ref{eq:rhoRreh}) and (\ref{eq:friedreh}) implies that in this regime~\cite{Garcia:2020eof}
\begin{align}
\label{eq:rhoRan}
\rho_{R}(a) \; \simeq\; \frac{2}{5}&\sqrt{3M_P^2\Gamma_{\phi}^2\rho_{\rm end}}\left(\frac{a_{\rm end}}{a}\right)^{4} \times \left[\left(\frac{a}{a_{\rm end}}\right)^{5/2}-1\right]\,.
\end{align}
This expression allows us to identify the maximum temperature of the primordial plasma, given by
\beq \label{eq:Tmax}
T_{\rm max}\;\simeq\; \frac{5}{3}\left(\frac{3}{8}\right)^{8/5}\sqrt{3M_P^2\Gamma_{\phi}^2\rho_{\rm end}} \,, 
\eeq
at $a_{\rm max} \simeq (8/3)^{2/5}a_{\rm end}$, and an approximate expression for the reheating temperature, 
\beq\label{eq:TrehP}
T_{\rm reh} \;\simeq\; \left(\frac{72 \Gamma_{\phi}^2 M_P^2}{5\pi^2 g_{\rm reh}}\right)^{1/4} \,, 
\eeq
at
\beq\label{eq:arehP}
a_{\rm reh} \;\simeq\; \left(\frac{25}{12}\frac{\rho_{\rm end}}{\Gamma_{\phi}^2M_P^2}\right)^{1/3}a_{\rm end}\,.
\eeq
In what follows, we assume that particle production in the visible sector proceeds perturbatively, and we fix the coupling to the inflaton by specifying the reheating temperature. 

\subsection{Spectator Scalar Fields}
\label{sec:spectatorfields}
In this paper, we investigate the dynamics and observational signatures of a spectator scalar field, $\chi$, which remains energetically subdominant throughout inflation and does not drive the inflationary expansion. Such fields can generate significant isocurvature perturbations that produce observable scalar-induced gravitational wave signatures---a central focus of our analysis. We consider a model where the inflaton field, $\phi$, interacts with the spectator field, $\chi$, either gravitationally or through a coupling term of the form~$\frac{1}{2}\sigma \phi^2 \chi^2$.\footnote{For completeness, we note that a nonminimal coupling between the Ricci scalar and the spectator scalar field of the form $\frac{1}{2}\xi R \chi^2$ could substantially alter the dynamics of $\chi$ and potentially enhance the resulting GW signals. However, such scenarios introduce additional complexities that lie beyond the scope of this work and are left for future investigation.}

Our analysis encompasses two main scenarios for the spectator field: (1) cases when $\chi$ is an unstable curvaton-like field that decays early in cosmic history~\cite{Linde:1996gt, Enqvist:2001zp, Moroi:2001ct, Lyth:2001nq}, (2) scenarios where $\chi$ is a stable dark matter candidate~\cite{Chung:1998zb, Chung:1998rq, Peebles:1999fz, Byrnes:2006fr, Markkanen:2018gcw, Kolb:2023ydq, Ling:2021zlj, Garcia:2022vwm, Garcia:2025rut}. While the spectator field could alternatively be the Standard Model Higgs field~\cite{Lu:2019tjj, Litsa:2020mvj, Karam:2021qgn}, a Peccei-Quinn field associated with the strong CP problem \cite{Peccei:1977ur, Peccei:1977hh}, or various scalar fields arising in extensions of the Standard Model such as supersymmetry or supergravity \cite{Martin:1997ns, Freedman:2012zz}, we concentrate our study on the dark matter and curvaton scenarios.

The action describing the spectator scalar field $\chi$ is given by:
\begin{equation}
    \label{eq:genaction}
    \mathcal{S}_{\chi} = \int d^4 x \sqrt{-g} \left[ 
        -\frac{1}{2} (\partial_{\mu} \chi)^2 - \frac{1}{2}m_{\chi}^2 \chi^2  - \frac{1}{2} \sigma \phi^2 \chi^2 \right] \, ,
\end{equation}
where $m_{\chi}$ is the bare mass of the spectator field, and $\sigma$ is the dimensionless coupling constant characterizing the interaction between the inflaton field $\phi$ and the spectator field $\chi$. Varying this action with respect to $\chi$ yields the equation of motion:
\begin{equation}
    \label{eq:eom_general}
    \left( \frac{d^2}{d\eta^2} + 2 \mathcal{H} \frac{d}{d\eta} - \nabla^2 + a^2 m_{\chi}^2 + \sigma a^2 \phi^2 \right)\chi \; = \; 0 \, .
\end{equation}
To simplify our analysis and isolate the effects of the expanding background, we introduce the rescaled field, $X \equiv a \chi$, which transforms the equation of motion to:
\begin{equation}
    \Bigg[ \frac{d^2}{d\eta^2} - \nabla^2 - \frac{a''}{a} - a^2 m_\chi^2 + \sigma a^2 \phi^2  \Bigg] X = 0 \, ,
\end{equation}
or in a more compact covariant form:
\begin{equation}
    \left( \Box - a^2 m_{\chi, \rm eff}^2 \right) X \; = \; 0 \, ,
\end{equation}
where $\Box \equiv g^{\mu\nu} \nabla_\mu \nabla_\nu$, $\nabla_\mu$ denotes the covariant derivative, and the effective mass of the spectator field is given by:
\begin{equation}\label{eq:meffchi}
    m_{\chi, \rm eff}^2 = m_{\chi}^2 + \sigma \phi^2 + \frac{R}{6}  \, .
\end{equation}

The coupling $\sigma$ and the background evolution significantly affect $m_{\chi, \rm eff}^2$. In the regime where $\sigma \phi^2 \ll |R|$ and $m_{\chi} \ll H_I$, the effective mass is dominated by the curvature contribution, $m_{\chi, \rm eff}^2 \simeq m_{\chi}^2 + R/6$. During inflation, the Ricci scalar in a quasi-de Sitter background is given by $R = -12 H_I^2$. This results in a large negative contribution to $m_{\chi, \rm eff}^2$ during inflation, potentially driving tachyonic instabilities ($m_{\chi, \rm eff}^2 < 0$), enhancing gravitational particle production, and generating substantial second-order isocurvature perturbations~\cite{Ling:2021zlj, Chung:2004nh, Chung:2011xd, Garcia:2023awt}. Conversely, when $\sigma \phi^2 \gg |R|$, the effective mass becomes $m_{\chi, \rm eff}^2 \simeq \sigma \phi^2$, suppressing gravitational production and isocurvature perturbations~\cite{Garcia:2022vwm, Garcia:2023awt}. These effects have significant implications for the energy density of the spectator field and the resulting primordial perturbation spectra, which we explore in detail in the following sections.

\section{Gravitational Particle Production}
\label{sec:gpp}
To evaluate the GW signatures arising from the presence of spectator scalar fields, it is essential to track the evolution of the energy density of the spectator field, $\rho_\chi$. These signatures critically depend on the behavior of $\rho_\chi$ during and after inflation. The stability of the spectator field plays a key role: stable fields that persist into the late Universe contribute directly to the matter content, potentially serving as a dark matter candidate, while decaying fields inject energy into the thermal bath, modifying the GW spectrum. Below, we outline the mechanism of gravitational particle production and its role in determining $\rho_\chi(\eta)$.

Gravitational particle production is the primary mechanism generating the spectator field energy density during and after inflation. This phenomenon, fundamentally quantum mechanical in nature, occurs due to the non-adiabatic evolution of the effective mass of the spectator field as spacetime undergoes rapid expansion and subsequent deceleration. The transition from the quasi-de Sitter phase during inflation to radiation domination induces a sudden change in the background geometry, violating the adiabatic condition $|\dot{\omega}_k/\omega_k^2| \ll 1$~\cite{Parker:1969au, Ford:1986sy, Kofman:1997yn, Kolb:2023ydq}.

The efficiency of this production mechanism depends critically on several key parameters, including the bare mass $m_{\chi}$ of the spectator field, the coupling strength $\sigma$ between the inflaton and spectator fields, and the detailed dynamics of the inflationary exit and reheating process. For spectator fields with effective masses near the Hubble scale during inflation $m_{\chi, \rm eff} \sim H_I$, the production is particularly efficient \cite{Chung:1998zb, Chung:1998rq, Ling:2021zlj, Garcia:2022vwm, Garcia:2023awt, Garcia:2023qab}. However, when the effective mass significantly exceeds the Hubble scale $m_{\chi, \rm eff} \gg H_I$, the production becomes exponentially suppressed~\cite{Chung:1998bt, Chung:2018ayg, Racco:2024aac, Verner:2024agh}. The subsequent evolution of $\rho_\chi$ relative to the radiation background $\rho_R$ directly influences the amplitude of scalar-induced GWs, creating a rich phenomenology that spans multiple frequency bands and detection channels. This evolution depends critically on whether the spectator field behaves as stable dark matter or decays as a curvaton-like field---scenarios we systematically investigate in the following sections.

\sloppy To compute the energy density of the spectator scalar field, we begin by introducing the stress-energy tensor for $\chi$, derived from the action through the standard relation $\sqrt{-g} T_{\mu\nu,\chi} = 2\delta (\sqrt{-g} \mathcal{L}_\chi) / \delta g^{\mu\nu}$. This yields~\cite{Birrell:1982ix}:
\begin{equation}
T^\chi_{\mu\nu} \; = \;  \nabla_\mu \chi \nabla_\nu \chi - \frac{1}{2} g_{\mu\nu} g^{\rho\sigma} (\nabla_\rho \chi)(\nabla_\sigma \chi) + \frac{1}{2} g_{\mu\nu} \left( m_\chi^2 + \sigma \phi^2 \right) \chi^2 \, ,
\end{equation}
Here all derivatives are taken with respect to cosmic time. Using this expression, the $T_{00}$ component of the spectator field in the FRW background takes the form:
\begin{equation}
    T_{00}^{(\chi)} (\eta, \mathbf{x}) \; = \; \frac{1}{2a^2} \left[\chi'^2 + (\nabla \chi)^2 + a^2 \left(m_{\chi}^2 + \sigma \phi^2 \right) \chi^2 \right] \, .
\end{equation}
Applying the field redefinition $X = a \chi$ and the relation:
\begin{equation}
    \chi' \; = \; \frac{1}{a} \left(X' - \frac{a'}{a} X \right) \, ,
\end{equation}
we find that the stress-energy tensor in terms of the rescaled field $X$ becomes:
\begin{equation}
    \label{eq:T00gen}
        T_{00}^{(\chi)}(\eta, \mathbf{x}) \; = \;   \frac{1}{2a^4} \Big[ (X')^2 + (\nabla X)^2 
         - 2 \mathcal{H} X X'  + \left(\mathcal{H}^2 - \frac{a^2 R}{6} \right) X^2 + a^2 m_{\chi, \rm eff}^2 X^2 \Big] \, .
\end{equation}

To study the quantum fluctuations of $\chi$, we expand the rescaled field $X$ into its Fourier components:
\begin{equation}
\label{eq:xfourier}
X(\eta, \mathbf{x}) = \int \frac{d^3 k}{(2\pi)^{3/2}} e^{-i \mathbf{k} \cdot \mathbf{x}} \left[ X_k(\eta) \hat{a}_k + X_k^*(\eta) \hat{a}_{-k}^\dagger \right] \, ,
\end{equation}
where $\mathbf{k}$ is the comoving momentum, and $\hat{a}_k^\dagger$ and $\hat{a}_k$ are the creation and annihilation operators, respectively, satisfying the canonical commutation relations:
\begin{equation}
[\hat{a}_k, \hat{a}_{k'}^\dagger] = \delta^{(3)}(\mathbf{k} - \mathbf{k'})\,\ \ \text{and}\ \ [\hat{a}_k, \hat{a}_{k'}] = [\hat{a}_k^\dagger, \hat{a}_{k'}^\dagger] = 0 \, .
\end{equation}
The mode functions $X_k(\eta)$ evolve according to the equation:
\begin{equation}\label{eq:eomX}
X_k''(\eta) + \omega_k^2(\eta) X_k(\eta) = 0 \, ,
\end{equation}
with the time-dependent mode frequency:
\begin{equation}
    \omega_k^2(\eta) = k^2 + a^2(\eta) m^2_{\chi, \rm eff}(\eta) \, .
\end{equation}
For the quantum field theory to be consistent, the mode functions must satisfy the Wronskian normalization condition:
\begin{equation}
  X_k X_k^{\prime *} - X_k^* X_k' \; = \; i \, ,
\end{equation}
which ensures that the canonical commutation relations between the field and its conjugate momentum are preserved throughout the cosmological evolution.

To study the gravitational particle production mechanism, we employ the Bogoliubov transformations, which provide a mathematical framework for tracking the evolution of quantum states in dynamical spacetimes. We expand the field $X_k(\eta)$ in terms of the Bogoliubov coefficients $\alpha_k(\eta)$ and $\beta_k(\eta)$:
\begin{equation}
    X_k(\eta) \; = \; \alpha_k(\eta) f_k(\eta) + \beta_k(\eta) f_k^*(\eta) \, ,
\end{equation}
with the corresponding derivative
\begin{equation}
    X_k'(\eta) \; = \;  -i \omega_k \left[\alpha_k(\eta) f_k(\eta) - \beta_k(\eta) f_k^*(\eta)\right] \, ,
\end{equation}
where the basis functions $f_k$ are defined according to the WKB approximation:
\begin{equation}
    f_k(\eta) \equiv \frac{e^{-i \int^\eta \omega_k d\eta}}{\sqrt{2 \omega_k}} \, .
\end{equation}
The Bogoliubov coefficients satisfy the unitarity condition $|\alpha_k|^2 - |\beta_k|^2 = 1$, which preserves the canonical commutation relations.

We introduce a Bogoliubov rotation that defines time-dependent creation and annihilation operators:
\begin{equation}
\begin{aligned}
    \label{eq:rotbog1}
     \hat{A}_k(\eta)  &=  \alpha_k(\eta) \hat{a}_{\mathbf{k}} + \beta_k^*(\eta) \hat{a}_{-\mathbf{k}}^\dagger \, , \\
     \hat{A}_{-k}^\dagger(\eta) & = \beta_k(\eta) \hat{a}_{\mathbf{k}} + \alpha_k^*(\eta) \hat{a}_{-\mathbf{k}}^\dagger \, .
\end{aligned}
\end{equation}
The set of ladder operators $\hat{A}_k$ is used to define an adiabatic vacuum state and construct a Fock space of multi-particle states. The vacuum state, denoted by $|0\rangle$, is defined as $\hat{A}_k |0\rangle = 0$ for all $\mathbf{k}$, i.e., it is the state annihilated by all $\hat{A}_k$ lowering operators. The vacuum is normalized such that $\langle 0|0 \rangle = 1$. The original operators $\hat{a}_{\mathbf{k}}$ and $\hat{a}_{-\mathbf{k}}^\dagger$ can be expressed in terms of the new set of ladder operators $\hat{A}_k$ and $\hat{A}_{-k}^\dagger$ through the Bogoliubov transformation:
\begin{equation}
    \begin{aligned}
    & \hat{a}_{\mathbf{k}} = \alpha_k^* \hat{A}_k - \beta_k^* \hat{A}_{-k}^\dagger \, , \\
    & \hat{a}_{-\mathbf{k}} = \alpha_k^* \hat{A}_{-k} - \beta_k^* \hat{A}_k^\dagger \, .
    \end{aligned}
\end{equation}
These transformations allow us to express the field $X$ and its derivative $X'$ in terms of the rotated ladder operators,
\begin{equation}
    \begin{aligned}
    X(\eta, \mathbf{x}) & = \int \frac{d^3 k}{(2\pi)^{3/2}} e^{-i \mathbf{k} \cdot \mathbf{x}} \left[f_k(\eta) \hat{A}_k + f_k^*(\eta) \hat{A}_{-k}^{\dagger} \right] \, , \\
     X'(\eta, \mathbf{x}) & = -i \omega_k \int \frac{d^3 k}{(2\pi)^{3/2}} e^{-i \mathbf{k} \cdot \mathbf{x}}  \left[f_k(\eta)  \hat{A}_k - f_k^*(\eta) \hat{A}_{-k}^{\dagger} \right] \, .
    \end{aligned}
\end{equation}
Substitution into (\ref{eq:T00gen}) would then provide us with the energy density of the quantum field. However, the resulting expression contains ultraviolet (UV) divergences and requires renormalization. A detailed discussion of the renormalization process, employing normal-ordered energy operators, is provided in~\cite{Chung:2004nh, Kolb:2023ydq}. In general, renormalization involves normal-ordering with respect to the rotated basis of ladder operators, $\hat{A}_k$ and $\hat{A}_{-k}^\dagger$.

To proceed, it is useful to rewrite the normal-ordered operators in terms of the unrotated operators, $\hat{a}_k$ and $\hat{a}_{-k}^\dagger$, using the Bogoliubov transformations provided in Eqs.~(\ref{eq:rotbog1}). The resulting expression for the normal-ordered energy density of the spectator field is:
\begin{equation}
\begin{aligned}
    \label{eq:endenfullexp}
    \rho_{\chi} (\eta)  = \langle : T_{00}^{(\chi)}(\eta, \mathbf{x}): \rangle \;=\; \frac{1}{2a^2} \int \frac{d^3 k}{(2\pi)^3} \bigg[ & |\omega_k X_k - i X_k'|^2   + \left(\mathcal{H}^2 - \frac{a^2 R}{6} \right) |X_k|^2 \\
    & - \mathcal{H} (X_k X_k'^* + X_k' X_k^* ) - \frac{1}{2\omega_k} \left(\mathcal{H}^2 - \frac{a^2 R}{6} \right)  \bigg] \, ,
\end{aligned}
\end{equation}
where the derivation of this energy density expression, along with details on the normal-ordering procedure, is presented in Appendix~\ref{app:A}. 
We define the comoving number density distribution per logarithmic momentum interval in terms of the number density of the field
\begin{equation}
    n_\chi(\eta) a^3 \; = \; \int_{k_0}^\infty \frac{dk}{k} \mathcal{N}_k\,, \quad \text{with} \quad \mathcal{N}_k = \frac{k^3}{2 \pi^2} n_k \, ,
\end{equation}
which in turn defines the corresponding energy density distribution
\begin{equation}
  \rho_\chi(\eta) a^3 \; = \; \int_{k_0}^\infty \frac{dk}{k} \mathcal{E}_k(\eta), \quad \text{with} \quad \mathcal{E}_k(\eta) = \frac{k^3}{2 \pi^2} \frac{\omega_k(\eta)}{a(\eta)} n_k \, ,
\end{equation}
where $n_k(\eta)$ is the occupation number. In the asymptotic late-time limit, when $\mathcal{H} \to 0$ and $R \to 0$, the occupation number simplifies to the standard form:
\begin{equation}
    n_k = \frac{1}{2 \omega_k} \left| \omega_k X_k - i X_k' \right|^2 \, .
\end{equation}

For our numerical implementation, we initialize the mode functions $X_k$ in the Bunch-Davies vacuum state, which represents the ground state in the asymptotic past when all modes of interest are well within the Hubble horizon:
\begin{equation}
X_k(\eta_0) \; = \; \frac{1}{\sqrt{2\omega_k}} \, , \quad X'_k(\eta_0) = -\frac{i \omega_k}{\sqrt{2\omega_k}} \, ,
\end{equation}
valid under the condition $|\eta_0 \omega_k| \gg 1$, ensuring that the modes begin in the adiabatic regime.
We then numerically evolve the mode functions $X_k$, which follow (\ref{eq:eomX}), and the full energy density expression given in Eq.~(\ref{eq:endenfullexp}) from the inflationary epoch through reheating and into the radiation-dominated era. This treatment captures the full non-perturbative dynamics of the spectator field, essential for accurate predictions of its contribution to the primordial GW spectrum. At late times, when the background becomes approximately Minkowskian, the energy density expression simplifies to
\begin{equation}
    \rho_{\chi} = \frac{1}{2a^4} \int_{k_0}^{\infty} dk \, \frac{k^2}{2\pi^2} \left| \omega_k X_k - i X_k' \right|^2 \, ,
\end{equation}
consistent with the standard energy density of a free scalar field in flat spacetime.

\begin{figure*}[t!]
    \centering
    \includegraphics[width=\linewidth]{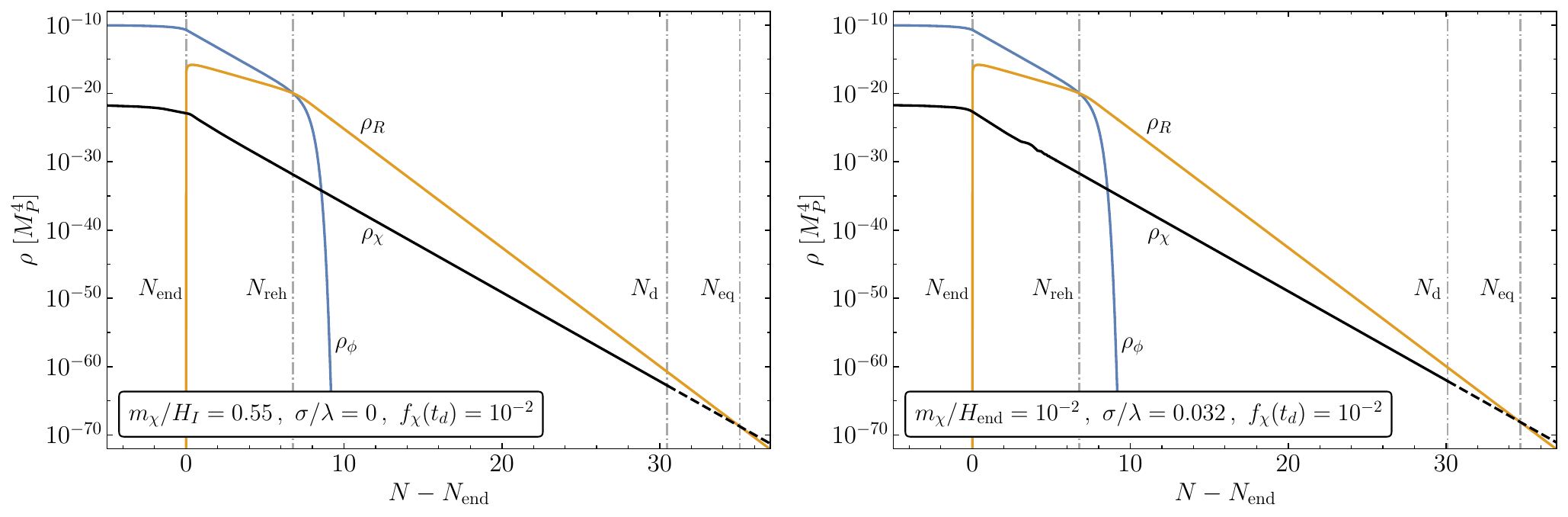}
    \caption{Energy densities of the inflaton, $\rho_{\phi}$, the visible sector radiation, $\rho_R$, and the spectator scalar, $\rho_{\chi}$, as functions of the number of $e$-folds, for a high reheating temperature, $T_{\rm reh}=10^{13}\ {\rm GeV}$. The continuous black curve follows $\rho_{\chi}$ until its decay, which is assumed to occur when $f_{\chi}(t_d)=10^{-2}$. The dashed line follows $\rho_{\chi}$ in the absence of a decay.}
    \label{fig:rhos}
\end{figure*}

Fig.~\ref{fig:rhos} shows the evolution of three energy densities as functions of $e$-folds: the inflaton field $\rho_{\phi}$, the radiation $\rho_{R}$, and the spectator field $\rho_{\chi}$, for both $\sigma=0$ and $\sigma\neq 0$. The inflaton and radiation densities follow the dynamics described in Section~\ref{sec:infandmodel}. During inflation, $\rho_\phi$ remains nearly constant, transitions to matter-like scaling during reheating, with $\rho_{\phi} \propto a^{-3}$, and then decays exponentially after $N_{\rm reh}=\ln a_{\rm reh}$. The radiation energy density grows rapidly at the onset of reheating according to Eq.~(\ref{eq:rhoRan}), then redshifts as $\rho_R \propto a^{-3/2}$ during reheating, and finally it scales as $\rho_R \propto a^{-4}$ after reheating ends. The spectator field energy density exhibits distinct regimes. During inflation, its nearly constant value requires numerical evaluation using the full expression~(\ref{eq:endenfullexp}) to capture quantum effects from gravitational particle production. After inflation ends, $\rho_\chi$ decreases as its oscillating modes become rapidly non-relativistic. The comoving number density freezes out at $n_{\chi}a^3\simeq\text{const.}$ approximately 5 $e$-folds after inflation ends for the parameters shown. Subsequently, particle production becomes negligible and $\chi$ redshifts as pressureless dust.

We distinguish two main scenarios: unstable (curvaton-like) and stable (dark matter) spectator fields. In the unstable case, we assume $\chi$ decays into radiation when its fractional energy density $f_{\chi}\equiv \rho_{\chi}/\rho_{R} \leq 1$. The solid black lines in Fig.~\ref{fig:rhos} show the case where $\chi$ decays at time $t_{\rm d}$ ($e$-fold number $N_{\rm d}$) when $f_{\chi}=10^{-2}$, injecting negligible entropy into the thermal bath. The dashed black line shows the stable scenario where $\chi$ constitutes the dark matter. We track its evolution through matter-radiation equality at $N_{\rm eq}$ and beyond.

We discuss the phenomenology of unstable and stable spectator fields in detail below, after determining the primordial isocurvature spectrum generated by superhorizon mode growth during inflation.
\section{Isocurvature Fluctuations}
\label{sec:isocurvaturefluct}
We assume no initial misalignment for the spectator scalar field, $\chi$, such that $ \langle \chi \rangle = 0 $ and $ \langle \chi^2 \rangle = 0 $ at the beginning of inflation. As inflation proceeds, the condition $ \langle \chi \rangle = 0 $ persists~\cite{Starobinsky:1994bd}, while a non-zero value $ \langle \chi^2 \rangle \neq 0 $ is generated due to efficient excitation of superhorizon modes. These fluctuations are expected to affect the primordial power spectrum of scalar fluctuations.\footnote{See Refs.~\cite{Ema3, Tenkanen:2019aij} for detailed discussions on isocurvature perturbations induced by long-wavelength superhorizon modes.} For a spectator field with vanishing background value, the fluctuations $ \delta \chi $ do not directly source the curvature perturbation. Consequently, these fluctuations can be treated as pure isocurvature perturbations in the uniform density gauge~\cite{Chung:2011xd, Chung:2004nh}. The growth in the energy density of the spectator field is primarily driven by the quadratic fluctuations that source the variance $ \langle \chi^2 \rangle $~\cite{Chung:2004nh, Chung:2015pga, Ling:2021zlj, Redi:2022zkt}. CMB observations impose stringent constraints on isocurvature perturbations, as no significant departures from purely adiabatic initial conditions have been detected at CMB scales. This yields an upper limit on the primordial isocurvature power spectrum, $\Delta_{\mathcal{S}}^2(k_*) \lesssim 8.3 \times 10^{-11}$ at the 95\% confidence level~\cite{Planck:2018jri}, which directly constrains the properties of any viable spectator field model.
    
The amplitude of isocurvature perturbations is directly connected to the effective mass of the spectator field during inflation. For large mass or large coupling strength $\sigma$, the effective mass term $m_{\chi, \rm eff}^2 = m_{\chi}^2 + \sigma \phi^2$ becomes substantial, suppressing fluctuations during inflation and consequently reducing the isocurvature contribution. This suppression mechanism provides a natural means of satisfying CMB constraints, which require isocurvature components to remain subdominant relative to the adiabatic mode~\cite{Planck:2018jri}. Conversely, light spectator fields with weak coupling experience efficient gravitational particle production, generating substantial isocurvature perturbations. While these must remain subdominant at CMB scales, they can produce observable signatures in large-scale structure and induce significant gravitational waves at smaller scales.
    
We compute the isocurvature power spectrum in the uniform density gauge, where the inflaton energy density perturbation vanishes, with $ \delta \rho_{\phi} = 0 $. The spectator field energy density perturbation is given by
\begin{equation}
    \delta \rho_{\chi}(\eta, \mathbf{x}) \; = \; : \rho_{\chi}(\eta, \mathbf{x}) : - \bar{\rho}_{\chi}(\eta) \, ,
\end{equation}
where $: \rho_{\chi}(\eta, \mathbf{x}) :$ denotes the normal-ordered energy density operator and $\bar{\rho}_{\chi}$ is the average energy density of the spectator field. The density contrast of the spectator field is defined as
\begin{equation}
    \delta_{\chi}(\eta, \mathbf{x}) \; = \;  \frac{\delta \rho_{\chi}(\eta, \mathbf{x})}{\bar{\rho}_{\chi}(\eta)} \; = \; \frac{: \rho_{\chi}(\eta, \mathbf{x}) :}{\bar{\rho}_{\chi}(\eta)} - 1 \, .
\end{equation}
The spectator field isocurvature power spectrum is characterized through the two-point correlation function in Fourier space~\cite{Liddle:1999pr,Chung:2004nh,Ling:2021zlj},
\begin{equation}
    \label{eq:isocurvature1}
    \langle \delta_{\chi}(\eta, \mathbf{k}) \delta_{\chi}^*(\eta, \mathbf{k}') \rangle \; \equiv \; \delta^{(3)} (\mathbf{k} - \mathbf{k}') \mathcal{P}_{\mathcal{S}}(\eta, k) \, ,
\end{equation}
where
\begin{equation}
    \mathcal{P}_{\mathcal{S}}  \; = \;  \frac{1}{\rho_{\chi}^2(\eta)} \int d^3 \mathbf{r} \, \langle \delta \rho_{\chi}(\eta, \mathbf{x}) \delta \rho_{\chi}(\eta, \mathbf{x} + \mathbf{r}) \rangle e^{-i \mathbf{k} \cdot \mathbf{r}} \, .
\end{equation}
The dimensionless isocurvature power spectrum is defined as
\begin{equation}
    \label{eq:isocurvaturedimensionless}
    \Delta_{\mathcal{S}}^2(\eta, k) \; = \; \frac{k^3}{2\pi^2} \mathcal{P}_{\mathcal{S}}(\eta, k) \, .
\end{equation}
After expressing the energy density in terms of field mode functions and performing the necessary contractions of creation and annihilation operators, the full isocurvature power spectrum becomes:
\begin{equation}
\begin{aligned}
\label{eq:fullisocurvature}
&\Delta_{\mathcal{S}}^2 (\eta, k) \; = \; \frac{1}{2a^8\bar{\rho}_{\chi}^2(\eta)} \frac{k^3}{2\pi^2} \int \frac{d^3 p}{(2\pi)^3} P_X(p, q) \, ,
\end{aligned}
\end{equation}
where 
\begin{equation}
\begin{aligned}
    P_X(p, q) \; = \;\Bigg| X_{p} X_{q} \left( a^2 m_{\chi, \text{eff}}^2 - p q + \left(\mathcal{H}^2 -\frac{a^2 R}{6} \right) \right) + X_{p}' X_{q}' - \mathcal{H} \left( X_{p}' X_{q} + X_{p} X_{q}' \right)\Bigg|^2 \, ,
\end{aligned}
\end{equation}
with $q \equiv |\mathbf{p - k}|$ denoting the relative momentum magnitude. A detailed derivation is provided in Appendix~\ref{app:B}.
    
In the late-time asymptotic limit where the spacetime approaches Minkowski, $ \mathcal{H} \rightarrow 0 $ and $ R \rightarrow 0 $, and neglecting the momentum $p q$ contribution, the isocurvature power spectrum simplifies to:
\begin{equation}
\begin{aligned}
    \Delta_{\mathcal{S}}^2 (\eta, k) \; = \; \frac{k^3}{ (2\pi)^5 \bar{\rho}_{\chi}^2(\eta) a^8} \int d^3 p \bigg[|X_p'|^2 |X_q'|^2 
    + a^4 m_{\chi, \rm eff}^4 |X_p|^2 |X_q|^2 
    + a^2 m_{\chi, \rm eff}^2 \bigg[ (X_p X_p'^*)(X_q X_q'^*) + \text{h.c.} \bigg] \, ,
\end{aligned}
\end{equation}
which matches previous results~\cite{Garcia:2023awt, Ling:2021zlj}. This approximation applies only to low-momentum modes and serves primarily to estimate $\Delta_{\mathcal{S}}^2(k_*)$ at CMB scales. To capture the full spectral shape, particularly at small scales relevant for GW production, we numerically evolve the complete expression~(\ref{eq:fullisocurvature}) throughout cosmic history.

\begin{figure}[!t]
\centering
\begin{tikzpicture}%
        \fill[SpringGreen] (1.15,1.15) - - (4.8,4.8) - - (4.8,2.5) - - (2.8,0.5) - - (1.8,0.5) - - cycle;
            \draw[->, line width=0.6pt] (0,0) -- (6.3,0);%
            \draw[->, line width=0.6pt] (0,0) -- (0,6.3);%
            \draw[line width=0.6pt] (0,2.3) -- (2.3,0);%
            \draw[line width=0.6pt] (0,2.3) -- (3.7,6);%
            \draw[line width=0.6pt] (2.3,0) -- (6,3.7);%
            \draw[dashed,line width=0.6pt] (0,0) -- (4.85,4.85);%
            \draw[dashed,line width=0.6pt] (4.8,0) -- (4.8,6);%
            \draw[dashed,line width=0.6pt] (0,4.8) -- (6,4.8);%
            \node at (4.8,-0.4) {$k_{\rm UV}$};
            \node at (-0.5,4.8) {$k_{\rm UV}$};v
            \draw[dashed,line width=0.6pt] (0,0.5) -- (6,0.5);%
        \draw[dashed,line width=0.6pt] (0.5,0) -- (0.5,6);%
        \node at (-0.5,0.6) {$k_{\rm IR}$};
            \node at (2.3,-0.3) {$k$};
            \node at (-0.3,2.3) {$k$};
            \node at (6.3,-0.4) {$q$};
            \node at (-0.3,6.3) {$p$};
  \end{tikzpicture}
  \caption{Integration domain of Eq.~(\ref{eq:PSp4}), which corresponds to the case $2k_{\rm IR} < k < k_{\rm UV} - k_{\rm IR}$. This condition ensures that the integration limits cover the regions of interest for both low and high momenta avoiding potential divergences in both IR and UV regimes. The domain is divided into three parts for numerical evaluation: (1) $k_{\rm IR} \leq p \leq k/2$ with $k-p \leq q \leq k+p$, (2) $k/2 \leq p \leq k_{\rm UV}-k$ with $p \leq q \leq k+p$, and (3) $k_{\rm UV}-k \leq p \leq k_{\rm UV}$ with $p \leq q \leq k_{\rm UV}$.} 
  \label{fig:region}
\end{figure}
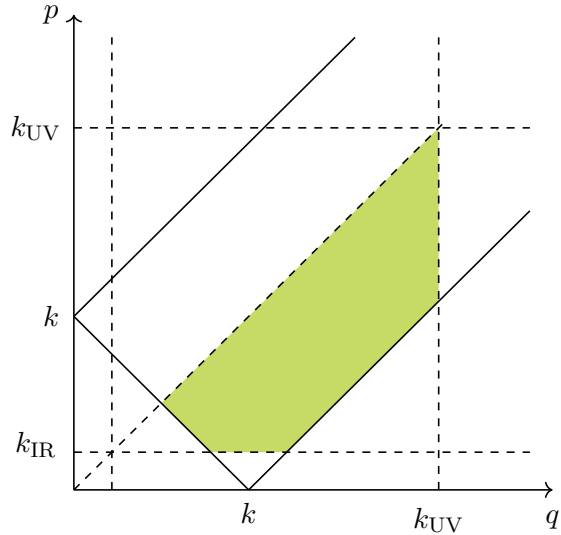

For finite spectator field masses and couplings, numerical evaluation of the isocurvature spectrum requires careful treatment of potential infrared ($p \to 0$) and collinear ($|\bp - \bk| \to 0$) divergences in the momentum integral. The dimensionless isocurvature power spectrum can be expressed as
\begin{equation}
\begin{aligned}
\label{eq:PSp3}
\Delta_{\mathcal{S}}^2 (\eta, k) \; = \; \frac{k^2}{(2\pi)^4 \bar{\rho}_{\chi}^2a^8}\int_0^{\infty} \diff  p\ p \int_{|k-p|}^{k+p}\diff q\ q \ P_X(p,q) \; =\; \frac{2k^2}{(2\pi)^4 \bar{\rho}_{\chi}^2a^8} \, I \big(P_X(p,q)\big) \,, 
\end{aligned}
\end{equation}
where 
\begin{equation}
\begin{aligned}
\label{eq:PSp4}
I \big(f(p,q)\big) & \equiv \bigg( \int_{k_{\rm IR}}^{k/2} \diff p \int_{k-p}^{k+p}\diff  q + \int_{k/2}^{k_{\rm UV}-k} \diff p \int_{p}^{k+p}\diff q + \int_{k_{\rm UV}-k}^{k_{\rm UV}} \diff p \int_{p}^{k_{\rm UV}}\diff q \bigg) p q\,f(p,q)\,.
\end{aligned}
\end{equation}
This decomposition introduces an IR cutoff $k_{\rm IR}$ that regulates potential divergences at small momenta, and a UV $k_{\rm UV}$ that corresponds to the physical decoupling scale, typically the highest frequency mode excited during the reheating process. While the UV cutoff is included for computational completeness, the integral naturally converges in the UV regime due to the structure of the mode functions.\footnote{However, it can lead to a UV divergence in the spectrum $\propto k^3$ if the integral is not renormalized to converge to zero, see Eq.~(\ref{eq:UVk}) and the accompanying discussion.} Expression~(\ref{eq:PSp3}) is valid for wavenumbers satisfying $2k_{\rm IR} < k < k_{\rm UV} - k_{\rm IR}$. The spectrum near boundary values is determined by adapting the integration limits, as illustrated in Fig.~\ref{fig:region}.

\begin{figure*}[t!]
    \centering
    \includegraphics[width=\linewidth]{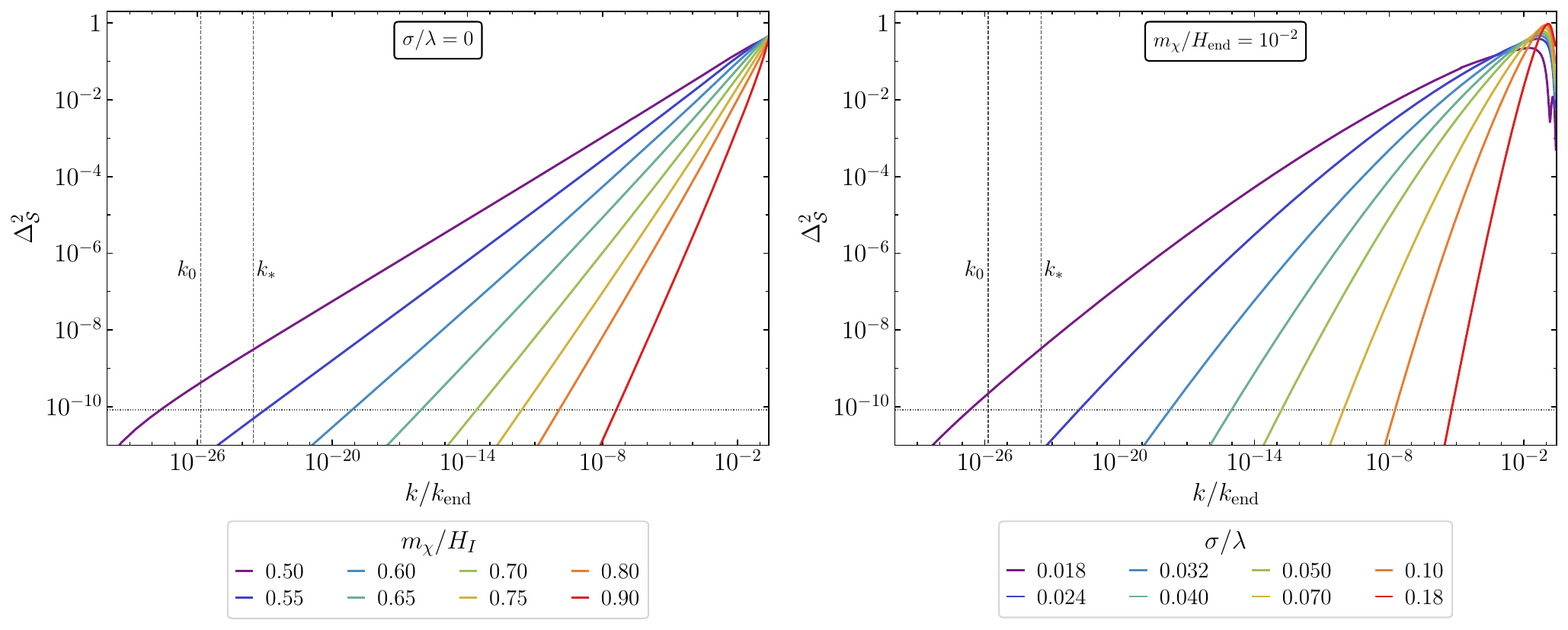}
    \caption{Isocurvature power spectrum for the spectator field $\chi$ for pure minimal gravitational production (left), and for a fixed mass with varying inflaton coupling (right). Each mass parameter (left panel) and coupling strength (right panel) is represented by a different color. Here $N_{\rm tot}=76.5$ $e$-folds of inflation. The vertical dashed lines show the positions of the present horizon scale ($k_0 = 1.4 \times 10^{-26} k_{\rm end}$) and the \textit{Planck} pivot scale ($k_* = 3.2\times 10^{-24}k_{\rm end}$). The horizontal dotted line
    represents the current upper bound on isocurvature perturbations from \textit{Planck}, $\Delta_{\mathcal{S}}^2 \simeq 8.3 \times 10^{-11}$ at the \textit{Planck} pivot scale (\ref{eq:isocurvatureupperlimit}).}
    \label{fig:iso1}
\end{figure*}

Fig.~\ref{fig:iso1} shows the numerically computed late-time isocurvature power spectrum $\Delta_{\mathcal{S}}^2$. The left panel displays results for pure gravitational production ($\sigma=0$) across different spectator field masses. All scales are normalized to the horizon scale at the end of inflation, $k_{\rm end}=a_{\rm end}H_{\rm end}\simeq 1.6\times 10^{22}\ {\rm Mpc}^{-1}$. At the CMB pivot scale $k_{*} = 0.05\ {\rm Mpc}^{-1} \simeq 3.2\times 10^{-24}k_{\rm end}$, the isocurvature amplitude varies dramatically with spectator mass. Only $m_{\chi}/H_I = 0.5$ violates the \textit{Planck} constraint (dotted line), while heavier masses yield strongly suppressed values, reaching $\Delta_{\mathcal{S}}^2(k_*)\simeq 4\times 10^{-31}$ for $m_{\chi}=0.9H_I$. This extreme sensitivity, spanning 22 orders of magnitude for $0.5\leq m_{\chi}/H_I\leq 0.9$, arises from the increasingly blue-tilted spectrum at $k<k_{\rm end}$ for larger masses, even though amplitudes near $k\simeq k_{\rm end}$ remain comparable. The \textit{Planck} isocurvature constraint is satisfied for $m_{\chi}\gtrsim 0.54 H_I$. Our numerical computation evolves the mode equations~(\ref{eq:eomX}) for $N_{\rm tot}=76.5$ e-folds total, ensuring that the present horizon scale $k_0 \simeq 2.3\times 10^{-4}\ {\rm Mpc}^{-1}\simeq 1.4\times 10^{-26}k_{\rm end}$ begins in its vacuum state while maintaining CMB compatibility with $N_*=55$ (see Fig.~\ref{fig:PRspectra} below).

The right panel of Fig.~\ref{fig:iso1} shows isocurvature spectra for fixed mass $m_{\chi}/H_{\rm end}=10^{-2}$ with varying inflaton coupling. We parameterize the coupling strength through $\sigma/\lambda$, where $\lambda$ normalizes the inflationary potential~(\ref{inf:tmodel}). This effective coupling determines the dominant particle production mechanism: tachyonic instability for superhorizon modes ($\sigma/\lambda\ll 1$), perturbative production ($\sigma/\lambda\gtrsim 1$), or parametric resonance of subhorizon modes ($\sigma/\lambda\gg 1$)~\cite{Garcia:2022vwm,Garcia:2023awt}. Remarkably, modest variations in the coupling yield dramatic changes in the isocurvature amplitude at the CMB pivot scale and $\Delta^2_{\mathcal{S}}(k_*)$ spans from $3\times 10^{-9}$ to $6\times 10^{-73}$ for $0.018 \lesssim \sigma/\lambda \lesssim 0.18$. These spectra deviate significantly from power-law behavior because the $\phi$-dependent term in the effective mass~(\ref{eq:meffchi}) becomes less suppressed for modes exiting the horizon near the end of inflation, when $\phi$ approaches its minimum. The \textit{Planck} isocurvature constraint requires $\sigma/\lambda\gtrsim 0.02$.

\begin{figure*}[t!]
    \centering
    \includegraphics[width=\linewidth]{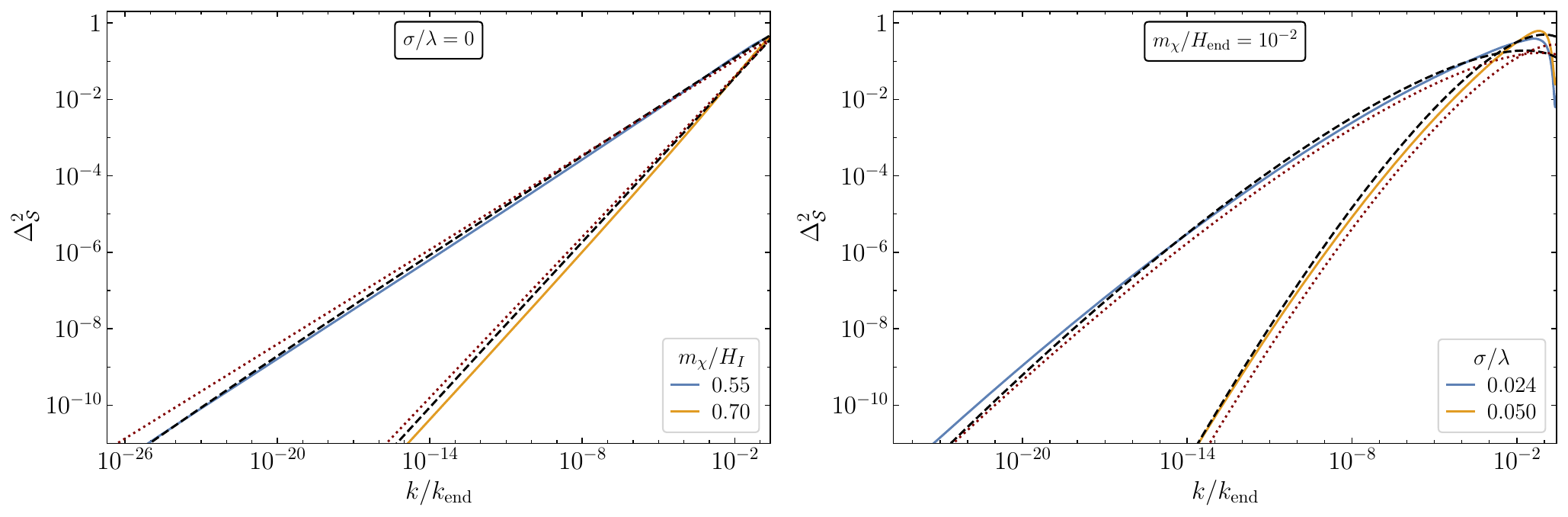}
    \caption{Numerical computation of the exact isocurvature power spectrum (solid lines), compared with two analytical approximations given by Eq.~(\ref{eq:PSan2}) (black dashed lines) and Eq.~(\ref{eq:PSan1}) (red dotted lines), for representative values of spectator field masses and inflaton couplings. For the coupled case shown in the right panel, we implement a phenomenological re-scaling $m_{\chi,{\rm eff}}\rightarrow0.9 m_{\chi,{\rm eff}}$ to improve the accuracy of the analytical approximation.}
    \label{fig:isofits}
\end{figure*}

We exclude modes with $k\gtrsim k_{\rm end}$ from our analysis, as these high-frequency modes are not excited by the tachyonic instability and experience inefficient gravitational particle production inside the horizon. However, numerical evaluation of $\Delta^2_{\mathcal{S}}$ reveals a subtlety. The normal ordering procedure in Eq.~(\ref{eq:fullisocurvature}) yields a finite UV contribution~\cite{Ling:2021zlj}:
\beq
\label{eq:UVk}
\Delta^2_{\mathcal{S}}\;\simeq\; \frac{k^3}{2\pi^2n_{\chi}a^3}\,,\quad (k>k_{\rm end}) \, ,
\eeq
in the Minkowski limit. This asymptotic behavior resembles a perfectly localized white-noise process in configuration space and must be systematically subtracted to obtain the physically relevant renormalized spectrum.

Despite the computational complexity involved in evaluating the time-dependent isocurvature spectrum, analytical approximations to its late-time behavior can be constructed. One such approximation derives from the stochastic formalism for superhorizon mode evolution~\cite{Starobinsky:1986fx, Starobinsky:1994bd}. This analytical expression, based on eigenvalues of the Fokker-Planck equation, takes the form~\cite{Markkanen:2019kpv, Ebadi:2023xhq, Choi:2024bdn, Garcia:2025rut}
\begin{equation}\label{eq:PSan1}
\Delta^2_{\mathcal{S}}(k) \;\simeq\; \frac{4}{\pi} 
\Gamma\left(2 - 2\Lambda \right)  \sin\left(\pi \Lambda \right)
\left( \frac{k}{k_{\text{end}}} \right)^{2 \Lambda} \, ,
\end{equation}
where all background quantities are evaluated at the horizon crossing of the scale $k$, with
\beq\label{eq:nu}
\Lambda \;=\; \frac{2m_{\chi,{\rm eff}}^2}{3H_{\rm I}^2}\,.
\eeq
This approximation assumes an instantaneous transition from inflation to matter domination, neglecting the smooth evolution during reheating. Its accuracy improves for larger $m_{\chi, \rm{eff}}$, when modes rapidly become non-relativistic. Fig.~\ref{fig:isofits} compares the analytical approximation (red dotted curves) with exact numerical results (solid curves), assuming constant $H_I=m_{\phi}$. For $\sigma=0$ and $m_{\chi}\lesssim H_I$ (left panel), Eq.~(\ref{eq:PSan1}) accurately captures the spectral amplitude across most scales. The primary discrepancy is a slightly reduced tilt in the analytical result, producing $\mathcal{O}(1)$ deviations at infrared scales.

The right panel of Fig.~\ref{fig:isofits} shows results for $m_{\chi}\ll H_I$ with $\sigma\neq 0$. To apply Eq.~(\ref{eq:PSan1}), we need the inflaton field value when mode $k$ crosses the horizon, $N_k$ $e$-folds before inflation ends. For T-model inflation, this can be approximated as~\cite{Ellis:2021kad}:
\beq\label{eq:phik}
\frac{\phi_k}{M_P} \;\simeq\; \sqrt{\frac{3}{2}}{\rm cosh}^{-1}\left[ \frac{4}{3}N_k + \cosh\left(\sqrt{\frac{2}{3}}\frac{\phi_{\rm end}}{M_P}\right) \right]\,,
\eeq
with
\beq\label{eq:Nk}
N_k \;\simeq\; N_{\rm tot} - \ln\left(\frac{k}{H_I}\right)\,.
\eeq
The end result is a large deviation (of $\mathcal{O}(10^{3})$ at $k_*$) between the approximate and the exact results, because in this case the isocurvature spectrum does not reach its final value until $\gtrsim 5$ $e$-folds after the end of inflation (see Fig.~\ref{fig:isoT} below). However, rescaling the effective mass by $m_{\chi,{\rm eff}}\rightarrow 0.9 m_{\chi,{\rm eff}}$ largely corrects this discrepancy, as shown in Fig.~\ref{fig:isofits}.

An alternative analytical approximation introduced in \cite{Garcia:2023awt,Pierre:2023jej} is based on the de Sitter solution of the mode function equations (\ref{eq:eomX}). This approach estimates the isocurvature power spectrum as:
\begin{align} \notag
\Delta_{\mathcal{S}}^2(k) \;&\simeq\; \Lambda^2 (N_{\rm tot}-N_k)\, e^{-2\Lambda N_k } 
 \left(1- e^{-\Lambda N_{\rm tot}} \right)^{-2}\\ 
\label{eq:PSan2}
&\simeq\; \Lambda^2 \left(\frac{k}{k_{\rm end}}\right)^{2 \Lambda } \ln \left(\frac{k}{k_{\rm end}}\frac{k_{\rm end}}{H_I}\right) \left(\frac{H_{\rm end}}{H_I}\right)^{2 \Lambda}  \,,
\end{align}
where the second expression follows from Eq.~(\ref{eq:Nk}). The corresponding spectra in Fig.~\ref{fig:isofits} (black dashed curves) demonstrate improved accuracy compared to the stochastic approach. This improvement is evident for both $\sigma=0$ and $\sigma\neq 0$ cases. For the latter case, the accuracy is further enhanced by applying a phenomenological correction to the effective mass, $m_{\chi , \rm eff} \rightarrow 0.9m_{\chi , \rm eff}$.

\begin{figure*}[t!]
    \centering
    \includegraphics[width=\linewidth]{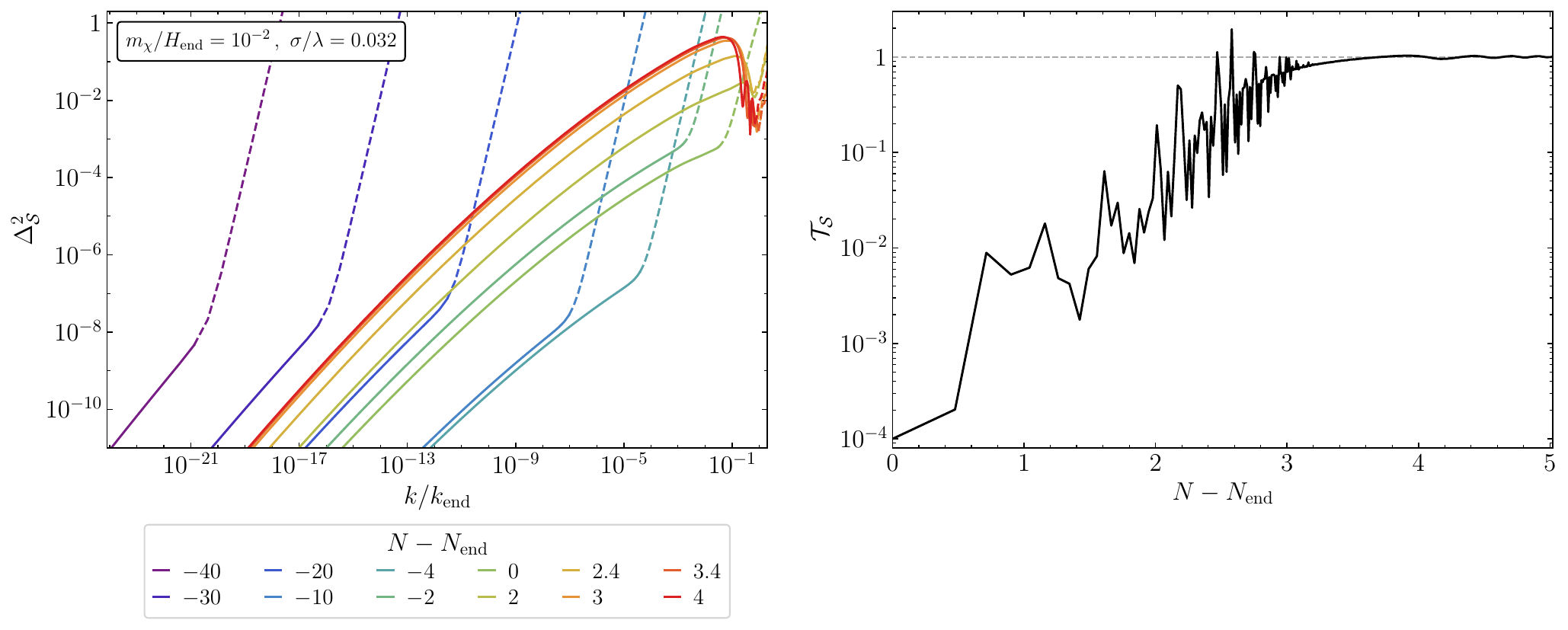}
    \caption{Left: Time ($e$-fold) evolution of the isocurvature power spectrum for the spectator field $\chi$, with $m_{\chi}/H_{\rm end} = 10^{-2}$ and $\sigma/\lambda = 0.032$, shown at different number of $e$-folds during and after inflation. Continuous lines represent super-horizon modes ($k < a H$), which maintain their spectral shape while undergoing amplitude modulation. Dashed lines indicate sub-horizon modes ($k > a H$) that exhibit the characteristic $k^3$ ultraviolet behavior arising from normal-ordering regularization (see the discussion around Eq.~(\ref{eq:UVk})). The spectrum decreases during inflation but rapidly grows during the early reheating phase as the field becomes effectively non-relativistic. Right: Time evolution of the function $\mathcal{T}_{\mathcal{S}}$ during reheating (see Eq. (\ref{eq:TS})).}
    \label{fig:isoT}
\end{figure*}

While our previous discussion focused on the late-time isocurvature spectrum, computing the corresponding GW signals requires the complete time evolution of the isocurvature spectrum, not just its late-time asymptotic value. Our numerical analysis reveals that the spectrum~(\ref{eq:fullisocurvature}) evolves with a characteristic profile that can be factorized for super-horizon modes: 
\beq
\label{eq:TS}
\Delta_{\mathcal{S}}^2(\eta,k) \;=\; \mathcal{T}_{\mathcal{S}}(\eta)\Delta_{\mathcal{S}}^2(k)\,,\quad (k<aH) \, ,
\eeq
where $\mathcal{T}_{\mathcal{S}}(\eta)$ encodes the time dependence of the spectrum.

For the range of parameters we have considered, with $m_{\chi,{\rm eff}}\gtrsim 10^{-1}H_{\rm end}$, the late-time value is reached within a few $e$-folds after the end of inflation, $N_{\rm NR}-N_{\rm end}\lesssim 6$, where NR denotes the time at which $\mathcal{T}_{\mathcal{S}} \simeq 1$, and the spectator field becomes effectively non-relativistic, with $m_{\chi}\gg H$, where $H$ is the instantaneous Hubble parameter during reheating.

The left panel of Fig.~\ref{fig:isoT} shows the evolution of the isocurvature spectrum as a function of the $e$-fold variable $N$ for $m_{\chi}/H_{\rm end}=10^{-2}$, $\sigma/\lambda=0.032$, both during and after inflation. For all curves shown, the super-horizon isocurvature modes (solid lines) evolve smoothly, while sub-horizon modes (dashed lines) are disregarded in our computations due to their unphysical $k^3$ ultraviolet behavior that arises from the normal-ordering regularization procedure. For a fixed large-scale mode, the spectral shape is established at horizon crossing, with only the amplitude varying---decreasing during inflation and rapidly increasing during reheating. This behavior justifies our factorization approach in Eq.~(\ref{eq:TS}). The right panel of Fig.~\ref{fig:isoT} provides the evolution of the function $\mathcal{T}_{\mathcal{S}}$ during reheating. In all cases studied, the spectrum reaches its asymptotic value within $N_{\rm NR}-N_{\rm end}\simeq 5-6$.

\section{Curvature Perturbations}
\label{sec:curvatureperturbations}
After inflation ends, the universe transitions to reheating as the inflaton decays into SM particles.
The spectator field, decoupled from the radiation bath, evolves as pressureless matter. In curvaton-like scenarios, the spectator field eventually decays into radiation, potentially leaving distinctive imprints on primordial perturbations \cite{Lyth:2001nq, Enqvist:2001zp}. Alternatively, a stable spectator field can serve as a dark matter candidate produced through gravitational particle production during inflation, with its abundance set by the freeze-in mechanism~\cite{Hall:2009bx, Bernal:2017kxu}.

We analyze perturbation evolution in the multi-component system using the Newtonian gauge, where the perturbed line element is:
\begin{equation}
    ds^2 = a(\eta)^2 \left[-(1+2\Psi) d \eta^2 + (1-2\Phi) \delta_{ij} dx^i dx^j \right] \, ,
\end{equation}
with scalar metric perturbations $\Psi$ and $\Phi$. To understand curvature perturbations at scales $k \geq 1 \, \rm{Mpc}^{-1}$, we track the gauge-invariant curvature perturbation on uniform-density hypersurfaces, defined as 
\cite{Wands:2000dp, Malik:2008im}
\begin{equation}
\zeta \; = \; -\Psi - \mathcal{H} \frac{\delta \rho}{\bar{\rho}'} \, ,
\end{equation}
where $\delta \rho$ represents the total density perturbation and $\bar{\rho}$ is the background energy density.

In this section, our analysis focuses on the scenario of rapid reheating, which maximizes the resulting curvature power spectrum amplitude, providing an upper bound on potential observational signatures. That is, we assume reheating is completed before the relevant modes for the enhancement of the curvature spectrum re-enter the horizon.\footnote{It is important to clarify our methodological approach: throughout this section, we employ the rapid reheating approximation primarily for analytical treatment in deriving the curvature power spectrum. However, when computing GW signals in subsequent sections, we implement a more physically realistic model with continuous reheating dynamics (down to $T_{\rm reh}\sim {\rm TeV}$) to capture the full time-dependent evolution of metric perturbations.} Following reheating, the radiation and spectator field components evolve independently, governed by separate continuity equations:
\begin{equation}
\rho_R' \; = \;  -4 \mathcal{H} \rho_R \, , \quad \rho_\chi' \; = \;  -3 \mathcal{H} \rho_\chi \, ,
\end{equation}
showing the radiation and matter-like scaling behaviors, respectively.

We express the total curvature perturbation in terms of the individual gauge-invariant fluctuations of each component:
\begin{equation}
    \zeta_R \; = \; -\Psi - \mathcal{H} \frac{\delta \rho_R}{\bar{\rho}_R'} \, , \quad \zeta_\chi \; = \; -\Psi - \mathcal{H} \frac{\delta \rho_\chi}{\bar{\rho}_\chi'} \, ,
\end{equation}
where $ \zeta_R$ and $\zeta_\chi$ represent the curvature perturbations on uniform-density hypersurfaces of radiation and the spectator field, respectively. The total curvature perturbation can then be expressed as a weighted average of these individual components,
\begin{equation}
\begin{aligned}
\zeta = \frac{\bar{\rho}'_R}{\bar{\rho}'_R + \bar{\rho}'_\chi} \, \zeta_R 
+ \frac{\bar{\rho}'_\chi}{\bar{\rho}'_R + \bar{\rho}'_\chi} \, \zeta_\chi \; = \; \frac{4 \bar{\rho}_R  \zeta_R + 3 \bar{\rho}_\chi \zeta_\chi}{4 \bar{\rho}_R + 3 \bar{\rho}_\chi} 
\, .
\end{aligned}
\end{equation}

To highlight the role of isocurvature perturbations, we introduce the gauge-invariant entropy perturbation~\cite{Wands:2000dp, Gordon:2000hv}:
\begin{equation}
\mathcal{S} \; \equiv \; 3(\zeta_\chi - \zeta_R) \, ,
\end{equation}
which measures the relative perturbation between the two components. This formulation allows us to express the total curvature perturbation as~\cite{Ebadi:2023xhq}:
\begin{equation}
\zeta \; = \; \zeta_R + \frac{\bar{\rho}_\chi}{4 \bar{\rho}_R + 3 \bar{\rho}_\chi} \mathcal{S}\; = \; \zeta_R + \frac{f_\chi}{4 + 3 f_\chi} \, \mathcal{S} \, ,
\end{equation}
where we have made use of the energy density ratio $f_{\chi} = \rho_{\chi}/\rho_R$.

In the absence of energy transfer between radiation and the spectator field, both $\zeta_R$ and $\zeta_\chi$ remain conserved on super-horizon scales \cite{Malik:2008im, Lyth:2003im}. Consequently, the evolution of the total curvature perturbation $\zeta$ is governed entirely by the time-dependent relative energy density ratio $f_{\chi}$, which increases with time due to the different scaling behaviors of radiation and matter.

Since the fluctuations in the spectator field $\chi$ are statistically uncorrelated with those in the inflaton $\phi$ (and hence radiation bath), the total curvature power spectrum is simply the sum of the two uncorrelated contributions:
\begin{equation}
\mathcal{P}_\zeta(k) \; = \; \mathcal{P}_{\zeta_R}(k) 
+ \left( \frac{f_\chi}{4 + 3 f_\chi} \right)^2 \mathcal{P}_{S}(k) \, , 
\end{equation}
where
\begin{equation}
\langle \zeta(\mathbf{k}) \zeta(\mathbf{k}') \rangle \equiv \delta^{(3)}(\mathbf{k} + \mathbf{k}') \mathcal{P}_\zeta(k) \, .
\end{equation}
The dimensionless power spectrum becomes
\begin{equation}
\label{eq:deltazetafull}
\Delta^2_\zeta \; = \; \Delta^2_{\zeta_R} 
+ \left( \frac{\bar{\rho}_\chi}{4 \bar{\rho}_R + 3 \bar{\rho}_\chi} \right)^2 \Delta^2_{\mathcal{S}} \, , 
\end{equation}
where $\Delta^2_\zeta(k) = \frac{k^3}{2\pi^2} \mathcal{P}_\zeta(k)$, with $\Delta^2_{\zeta_R}(k)$ and $\Delta^2_{\mathcal{S}}(k)$ defined analogously.

We determine the inflationary contribution to the curvature power spectrum $\Delta_{\zeta_{\phi}}^2 = \Delta_{\zeta_{R}}^2$ using the Mukhanov-Sasaki formalism~\cite{Mukhanov:1981xt, Sasaki:1986hm} and solving the equations for the gauge-invariant curvature perturbations generated during inflation. In conformal time, the Mukhanov-Sasaki equation for the mode function $v_k(\eta)$ of the canonical variable $v = z \zeta$, where $z \equiv a \phi'/\mathcal{H}$, is given by
\begin{equation}
v_k'' + \left( k^2 - \frac{z''}{z} \right) v_k \; = \; 0 \, ,
\label{eq:MukhanovSasaki}
\end{equation}
where primes denote derivatives with respect to conformal time $\eta$. To solve this equation, we impose Bunch–Davies initial conditions deep inside the horizon, where the mode function behaves as a free plane wave:
\begin{equation}
v_k(\eta) \rightarrow \frac{1}{\sqrt{2k}} e^{-ik\eta} \, ,~~\text{as }~~k\eta \to -\infty \, .
\end{equation}
This corresponds to the quantum vacuum of the perturbations in the asymptotic past and ensures the proper normalization of the power spectrum on subhorizon scales.

The power spectrum of the curvature perturbation $\zeta_{\phi} = \zeta_r = v_k / z$ is then computed as
\begin{equation}
\Delta_{\zeta_r}^2(k) = \frac{k^3}{2\pi^2} \left| \frac{v_k}{z} \right|^2 \Big|_{k \ll aH} \, ,
\end{equation}
evaluated in the superhorizon limit $k \ll aH$ after horizon crossing. For the T-model with $N_* = 55$ $e$-folds described in Section~\ref{sec:infdynamics}, this yields the expected nearly scale-invariant spectrum at observable CMB scales $k \lesssim 1 \, \rm{Mpc}^{-1}$, as shown in Fig.~\ref{fig:PRspectra} below.

The isocurvature contribution in Eq.~(\ref{eq:deltazetafull}) is determined using the formalism developed in Section~\ref{sec:isocurvaturefluct}, which captures the evolution of perturbations in the spectator field. Fig.~\ref{fig:rhos} illustrates the post-inflationary dynamics of the background energy densities, assuming a high reheating temperature that allows us to track the evolution of the energy density ratio $f_{\chi}$ as a function of the number of $e$-folds. The spectator field energy density (black curve) is assumed to decay instantaneously into radiation at time $t_d$. In the figure, this decay occurs when $f_{\chi}\simeq(\rho_{\chi,{\rm end}}/\rho_{\rm end})(a_d/a_{\rm end}) = 10^{-2}$, an approximation valid for $m_{\chi,{\rm eff}}\gtrsim 10^{-1}H_{\rm end}$ or larger values. The dashed black line shows the alternative scenario where $\chi$ remains stable as dark matter through matter-radiation equality (at time $t_{\rm eq}$) and beyond.

To connect conformal time evolution and observable scales, we express the fractional energy density $f_{\chi}$ as a function of comoving wavenumber $k$. Different perturbation modes re-enter the horizon at different cosmic epochs, with a mode of wavenumber $k$ crossing the horizon when $k = a_k H_k$, where $a_k$ and $H_k$ represent the scale factor and Hubble parameter at horizon re-entry. During radiation domination, this simplifies to $k / k_{\rm end} = a_{\rm end} / a_k$, where $k_{\rm end}$ denotes the mode re-entering immediately after inflation ends.

\begin{figure*}[t!]
    \centering
    \includegraphics[width=\linewidth]{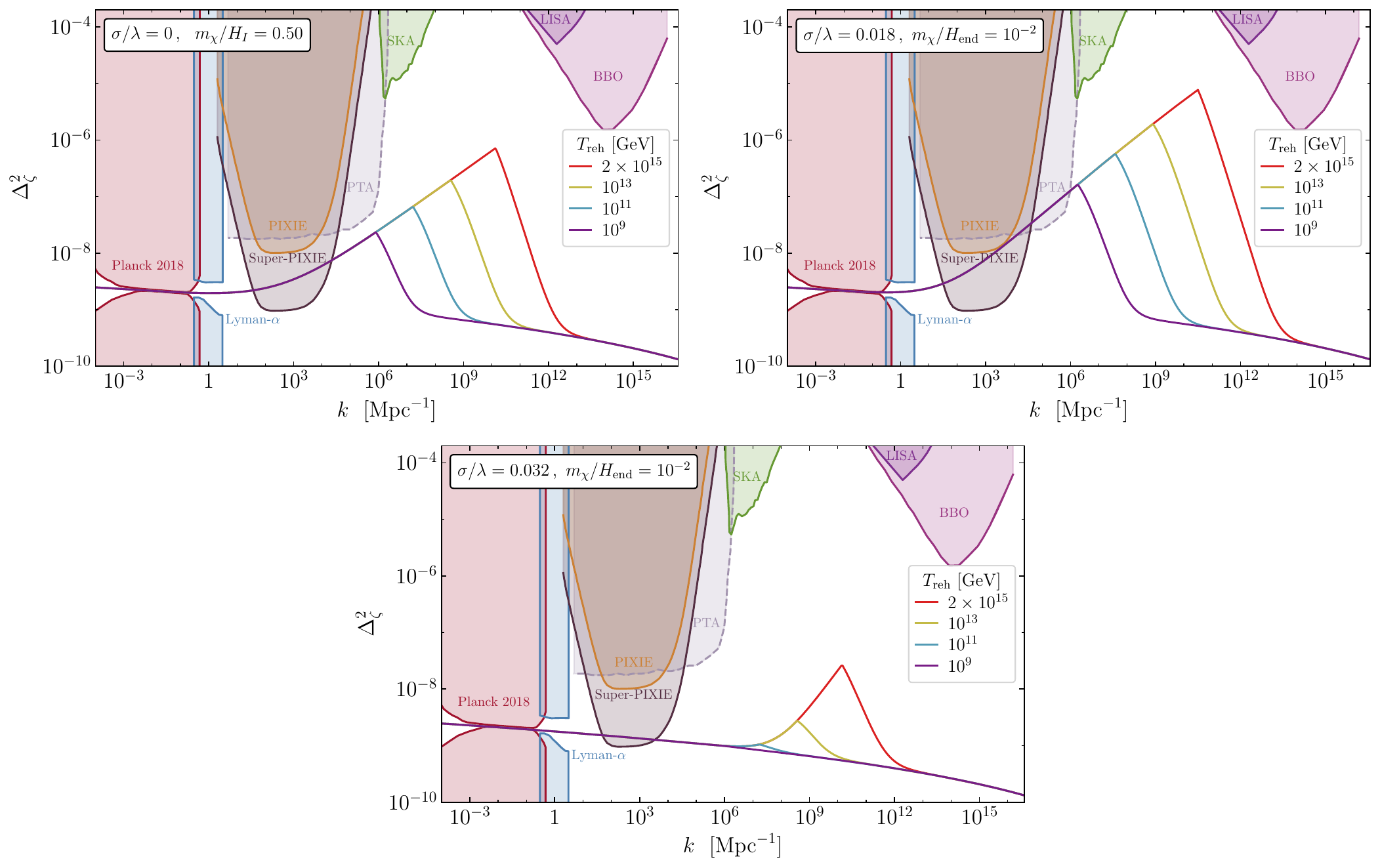}
    \caption{Curvature power spectrum for selected spectator field masses and couplings. 
    Each panel displays four different reheating temperatures, with the largest corresponding to instantaneous reheating, $a_{\rm reh}=a_{\rm end}$. Background curves show existing and projected sensitivity for various cosmological probes. The spectrum exhibits a characteristic peak at scale $k_d$ defined in (\ref{eq:kd}), with position dependent on reheating temperature through (\ref{eq:TrehP})-(\ref{eq:arehP}). For scales smaller than $k_d$ the spectrum obtains a red tilt as 
    $f_{\chi}\ll 1$, when these modes re-enter the horizon, following $\Delta^2_{\zeta}\simeq (k_d/k)^2\Delta^2_{\mathcal{S}}/16$. Shown for reference are sensitivity curves for CMB spectral distortion experiments (PIXIE and Super-PIXIE~\cite{Chluba:2019kpb, Kogut:2019vqh}), PTA constraints on enhanced DM substructure~\cite{Lee:2020wfn}, and projected sensitivities of gravitational wave observatories including SKA~\cite{Carilli:2004nx, Janssen:2014dka, Weltman:2018zrl}, LISA~\cite{LISA:2017pwj, Baker:2019nia}, and BBO~\cite{Crowder:2005nr, Corbin:2005ny}.}
    \label{fig:PRspectra}
\end{figure*}

Since the spectator field scales as matter while radiation dominates, the density ratio grows as $f_\chi \sim a$. This allows us to relate its value at any time $t < t_d$ through
\begin{equation}
\label{eq:fchit}
f_\chi(t) \; = \; f_\chi(t_d) \left( \frac{k_d}{k} \right) \, ,
\end{equation}
where $k$ is the mode re-entering at time $t$, and 
\begin{align}\label{eq:kd}
k_d \; \equiv \; a_dH_d \; \simeq \; f^{-1}_{\chi}(t_d)\left(\frac{\rho_{\chi,{\rm end}}}{\rho_{\rm end}}\right)\left(\frac{a_{\rm end}}{a_{\rm reh}}\right)^{1/2}k_{\rm end}\,,
\end{align}
represents the wavenumber of the mode re-entering the horizon at the decay time. The relation (\ref{eq:fchit}) captures the scale-dependent enhancement of the isocurvature contribution. Incorporating it into Eq.~(\ref{eq:deltazetafull}) for the curvature power spectrum yields
\es{eq:Delta_zeta_1}{
\Delta_\zeta^2(k) = 
\begin{cases}
    \Delta_{\zeta_R}^2(k) + \left(f_\chi(t_d) \over {4+3 f_\chi(t_d)}\right)^2 \Delta_{\mathcal S}^2(k),~k < k_d,\\
     \Delta_{\zeta_R}^2(k) + \left(f_\chi(t_d)(k_d/k) \over {4+3f_\chi(t_d) (k_d/k)}\right)^2 \Delta_{\mathcal S}^2(k),~k > k_d.
\end{cases}
}
Modes re-entering before spectator decay ($k < k_d$) maintain constant $f_\chi$, while those re-entering afterward ($k > k_d$) experience suppressed isocurvature contributions due to the decreasing density ratio.

Fig.~\ref{fig:PRspectra} shows three different benchmark scenarios for the full curvature power spectrum, assuming an unstable spectator field with $f_{\chi}(t_d)=1$ that decays just before domination.
We consider four reheating temperatures, with the highest corresponding to instantaneous reheating. The top left panel displays pure gravitational production ($\sigma=0$) with a bare mass of $\chi$ near the isocurvature limit. The blue-tilted isocurvature spectrum remains CMB-compatible at large scales due to strong suppression of $\Delta^2_{\mathcal{S}}$. For $k\gtrsim 1\ {\rm Mpc}^{-1}$, modes re-entering after $t_d$ yield $\Delta^2_{\zeta}\simeq \Delta^2_{\mathcal{S}}/49$. In this range, the spectrum intersects with sensitivity contours for future CMB spectral distortion experiments (PIXIE and Super-PIXIE~\cite{Chluba:2019kpb, Kogut:2019vqh}), while narrowly avoiding constraints from Pulsar Timing Arrays (PTA) probes for enhanced dark matter substructure~\cite{Lee:2020wfn}. The spectrum peaks at scale $k_d$, defined in Eq. (\ref{eq:kd}), which depends on the reheating temperature through Eqs.~(\ref{eq:TrehP})-(\ref{eq:arehP}). At smaller scales ($k > k_d$), the spectrum becomes red-tilted as $\Delta^2_{\zeta}\simeq (k_d/k)^2\Delta^2_{\mathcal{S}}/16$ when $f_{\chi}\ll 1$. Eventually, the isocurvature contribution becomes subdominant and the spectrum approaches the pure T-model value $\Delta_{\zeta_R}^2$. We also show sensitivity curves for gravitational wave detectors: SKA~\cite{Carilli:2004nx, Janssen:2014dka, Weltman:2018zrl}, LISA~\cite{LISA:2017pwj, Baker:2019nia}, and BBO~\cite{Crowder:2005nr, Corbin:2005ny}.

The top right and bottom panels of Fig.~\ref{fig:PRspectra} show scenarios with a light spectator field ($m_{\chi}=10^{-2}H_{\rm end}$) and varying inflaton coupling. For $\sigma/\lambda=0.018$ (top right), the curvature spectrum reaches $\Delta_{\zeta}^2\sim 10^{-5}$ while remaining CMB-compatible. Increasing the coupling to $\sigma/\lambda=0.032$ (bottom) suppresses the spectrum by two orders of magnitude, illustrating the strong sensitivity to coupling strength.

\section{Gravitational Wave Spectral Density}
\label{sec:gwspectraldensity}
In this section, we outline the procedure for calculating the GW spectral density, following the formalism developed in Refs.~\cite{Maggiore:2007ulw, Domenech:2021ztg}. Our primary focus is on GW signals generated by the dynamics of spectator scalar fields during and after inflation, which provide a unique observational window into early universe physics.

Using the Newtonian gauge for scalar perturbations and the Transverse-Traceless (TT) gauge for tensor perturbations, the perturbed line element is expressed as:
\begin{equation}
\begin{aligned}
    \label{eq:lineelement}
    ds^2 \; = \; a(\eta)^2 \bigg[&-(1+2\Phi)d\eta^2 + \left[(1 - 2\Phi)\delta_{ij} +h_{ij} \right]dx^i dx^j \bigg] \, ,
\end{aligned}
\end{equation}
where $\Phi$ denotes the scalar metric fluctuations, and $h_{ij}$ represents the tensor perturbations
satisfying the transverse $\partial^i h_{ij}$ and traceless $h^i_{~i} = 0$ conditions. We assume vanishing anisotropic stress, which implies $\Psi = \Phi$ in the metric. This section focuses on tensor perturbations and their role in GW generation, while subsequent sections will address secondary gravitational waves induced by scalar metric fluctuations.

In the weak-field approximation, gravitational waves propagate in an effective Ricci-flat spacetime without direct matter sources. The background metric $\bar{g}_{\mu\nu}$ is perturbed by tensor fluctuations $h_{\mu\nu}$, yielding the total metric
\begin{equation}
g_{\mu\nu}(x) \; = \; \bar{g}_{\mu\nu}(x) + h_{\mu\nu}(x), \quad |h_{\mu\nu}| \ll 1 \, ,
\end{equation}
where the perturbations are assumed to be small in amplitude ($|h_{\mu\nu}| \ll 1$). Expanding the Einstein tensor to second order in $h_{\mu\nu}$ yields quadratic terms that act as an effective energy-momentum source. After integrating out high-frequency modes, the effective Einstein equation becomes:
\begin{equation}
    \bar{G}_{\mu\nu}[\bar{g}_{\mu\nu}] \; = \; M_{P}^{-2} t_{\mu\nu}^{\rm GW}[h_{\mu\nu}] \, , 
    \label{eq:effectiveEinstein}
\end{equation}
where $\bar{G}_{\mu\nu}$ represents the Einstein tensor of the background metric, and $t_{\mu\nu}^{\rm GW}$ denotes the (pseudo)-energy-momentum tensor of gravitational waves. Averaging over GW wavelengths, the (pseudo)-energy-momentum tensor for GWs is given by~\cite{Isaacson:1968hbi, Isaacson:1968zza}:
\begin{equation}
    \label{eq:GWtensor}
    t_{\mu\nu}^{\rm GW} \; = \; \frac{M_{P}^2}{4} \left\langle 
    \partial_\mu h^{\alpha\beta} \partial_\nu h_{\alpha\beta} 
    - \frac{1}{2} \bar{g}_{\mu\nu} \partial_\sigma h^{\alpha\beta} \partial^\sigma h_{\alpha\beta} 
    \right\rangle \, ,
\end{equation}
where index contractions are performed using the background metric $\bar{g}_{\mu\nu}$, and angular brackets denote spatial averaging over several wavelengths. From Eq.~(\ref{eq:GWtensor}), the energy density of GWs can be expressed as:
\begin{equation}
\begin{aligned}
\label{eq:endenGW}
& \rho_{\text{GW}} \; = \; t_{00}^{\rm GW} \; = \; \frac{M_P^2}{4} \bigg( \langle \dot{h}_{ij}(\eta, \mathbf{x}) \dot{h}_{ij}(\eta, \mathbf{y}) \rangle - \frac{1}{2} \langle \dot{h}_{ij}(\eta, \mathbf{x}) \dot{h}_{ij}(\eta, \mathbf{y}) \rangle + \frac{1}{2} \langle \partial_l h_{ij}(\eta, \mathbf{x}) \partial^l h_{ij}(\eta, \mathbf{y}) \rangle \bigg) \, .
\end{aligned}
\end{equation}

The tensor perturbations $h_{ij}$ can be decomposed into Fourier modes as
\begin{equation}
\label{eq:hijfour}
h_{ij}(\eta, \mathbf{x}) \; = \; \sum_\lambda \int \frac{d^3k}{(2\pi)^{3/2}} e^{-i \mathbf{k} \cdot \mathbf{x}} \varepsilon_{ij}^\lambda(\mathbf{k}) h_\lambda(\eta, \mathbf{k}) \, ,
\end{equation}
where $\lambda = (+, \times)$ the two independent polarization states, and $\varepsilon_{ij}^\lambda(\mathbf{k})$ represents the corresponding polarization tensors. To ensure that $h_{ij}$ remains a real-valued field, the following reality conditions must be satisfied,
\begin{equation}
\varepsilon_{ij}^\lambda(\mathbf{k}) \; = \; \varepsilon_{ij}^\lambda(-\mathbf{k}) \quad \text{and} \quad h_\lambda^*(\eta, \mathbf{k}) = h_\lambda(\eta, -\mathbf{k}) \, .
\end{equation}
The polarization tensors are defined as:
\begin{equation}
\begin{aligned}
& \varepsilon_{ij}^+(\mathbf{k}) \; = \; \frac{1}{\sqrt{2}} \left[ e_i(\mathbf{k}) e_j(\mathbf{k}) - \bar{e}_i(\mathbf{k}) \bar{e}_j(\mathbf{k}) \right] \,, \\
& \varepsilon_{ij}^\times(\mathbf{k}) \; = \; \frac{1}{\sqrt{2}} \left[ e_i(\mathbf{k}) \bar{e}_j(\mathbf{k}) + \bar{e}_i(\mathbf{k}) e_j(\mathbf{k}) \right] \,.
\end{aligned}
\end{equation}
Using the relations $e_i(\mathbf{k}) k^i = \bar{e}_i(\mathbf{k}) k^i = 0$ and $e_i(\mathbf{k}) \bar{e}^i(\mathbf{k}) = 1$, the properties of the polarization tensors are:
\begin{equation}
\begin{aligned}
&\varepsilon_{ij}^{\lambda}(\mathbf{k}) \varepsilon^{\lambda' ij}(\mathbf{k}) \; = \; \delta^{\lambda \lambda'} \,, \\
&\delta_{ij} \varepsilon^{+ij}(\mathbf{k}) \; = \; \delta_{ij} \varepsilon^{\times ij}(\mathbf{k}) = 0 \,, \\
&k^i \varepsilon_{ij}^+(\mathbf{k}) \; = \; k^i \varepsilon_{ij}^\times(\mathbf{k}) = 0 \, .
\end{aligned}
\end{equation}

The two-point correlation function for the GW modes is given by
\begin{equation}
\begin{aligned}
\label{eq:gwpowerspectrum1}
\langle h_{ij}(\eta, \mathbf{x}) h_{ij}(\eta, \mathbf{y}) \rangle \; = \; \sum_{\lambda, \lambda'} \int \frac{d^3k d^3k'}{(2\pi)^3} e^{-i\mathbf{k} \cdot \mathbf{x}} e^{-i\mathbf{k}' \cdot \mathbf{y}} \delta_{\lambda \lambda'} \langle h_\lambda(\eta, \mathbf{k}) h_{\lambda'}(\eta, \mathbf{k}') \rangle \, .
\end{aligned}
\end{equation}
The corresponding GW power spectrum is defined as
\begin{equation}
\label{eq:gwpowerspectrum2}
\langle h_\lambda(\eta, \mathbf{k}) h_{\lambda'}(\eta, \mathbf{k}') \rangle \; = \; \delta_{\lambda \lambda'} \delta^{(3)}(\mathbf{k} + \mathbf{k}') \mathcal{P}_\lambda(\eta, k) \,, 
\end{equation}
while the dimensionless power spectrum is given by
\begin{equation}
\label{eq:gwdimensionlessspectrum}
\Delta^2_\lambda(\eta, k) \; = \; \frac{k^3}{2\pi^2} \mathcal{P}_\lambda(\eta, k) \, .
\end{equation}
Substituting Eq.~(\ref{eq:gwpowerspectrum2}) into Eq.~(\ref{eq:gwpowerspectrum1}), we find that
\begin{equation}
\langle h_{ij}(\eta, \mathbf{x}) h_{ij}(\eta, \mathbf{y}) \rangle \; = \; \sum_\lambda \int \frac{d^3k}{(2\pi)^3} \mathcal{P}_\lambda(\eta, k) \, .
\end{equation}
Similarly, for the gradient terms, we can write
\begin{equation}
\langle \partial_l h_{ij}(\eta, \mathbf{x}) \partial_l h_{ij}(\eta, \mathbf{y}) \rangle \; = \; \frac{k^2}{a^2} \sum_\lambda \int \frac{d^3k}{(2\pi)^3} \mathcal{P}_\lambda(\eta, k) \, .
\end{equation}
Moreover, assuming a freely propagating wave approximation
\begin{equation}
\dot{h}_\lambda(\eta, \mathbf{k}) \; \simeq \; \frac{k}{a} h_\lambda(\eta, \mathbf{k}) \,,
\end{equation}
the energy density of GWs~(\ref{eq:endenGW}) simplifies to
\begin{equation}
\rho_{\text{GW}} \; = \; M_P^2 \int d\ln k \, \frac{k^3}{8\pi^2} \frac{k^2}{a^2} \sum_\lambda \mathcal{P}_\lambda(\eta, k) \, .
\end{equation}
The GW signal strength is defined as the energy density per unit logarithmic interval of wavenumber normalized to the total energy density,
\begin{equation}
\Omega_{\text{GW}}(k) \; \equiv\;  \frac{1}{3M_P^2H^2} \frac{d \rho_{\text{GW}}}{d\ln k} \, ,
\end{equation}
which, using the dimensionless power spectrum~(\ref{eq:gwdimensionlessspectrum}), becomes
\begin{equation}
\label{eq:gwgeneral}
\Omega_{\text{GW}}(k) \; = \; \frac{1}{12} \left( \frac{k}{a(\eta)H(\eta)} \right)^2 \sum_\lambda \Delta_\lambda^2 \,.
\end{equation}
With the critical energy density defined as $\rho_c = 3H_0^2M_P^2$ and the present-day Hubble parameter $H_0 = 100 \, h \, \rm{km \, s^{-1} \, Mpc^{-1}} $, the present-day GW spectrum can finally be expressed as
\begin{equation}
\label{eq:gwpresentday}
\Omega_{\text{GW},0}(k) h^2 \; = \; \frac{1}{12} \left( \frac{k}{a_0 H_0} \right)^2  \Delta_h^2 \,, 
\end{equation}
where $a_0$ is the present-day scale factor and $\sum_\lambda \Delta_\lambda^2 = \Delta_h^2$ represents the total gravitational wave power across both polarization states.

The gravitational wave spectrum $\Omega_{\text{GW},0}(k) h^2$ provides a cosmological probe that reveals the energy distribution of gravitational radiation across frequency scales. For spectator scalar fields, this spectrum encodes the field dynamics during inflation and reheating. Distinctive features in $\Omega_{\text{GW}}(k)$ could signal specific events in cosmic history, such as phase transitions or enhanced particle production, that occurred as particular scales entered the horizon. The precise relationship between observed frequencies and cosmological epochs provides a complementary window into early universe physics beyond CMB observations, and we discuss it in detail below.
\section{Secondary Gravitational Waves from Scalar Perturbations}
\label{sec:gravwaves1}

\subsection{Scalar Field Fluctuations from Spectator Scalar Fields}
\label{sec:scalarflucspectfield}
We analyze the perturbations of the energy-momentum tensor in a two-field system with the inflaton $\phi$ and the spectator scalar field $\chi$. The total energy-momentum tensor is $T_{\mu \nu} = T_{\mu \nu}^{(\phi)} + T_{\mu \nu}^{(\chi)}$. The matter perturbations are given by\footnote{The gauge freedom is used to set $B = 0$ in the line element term $2\partial_i B dx^i d\eta$, simplifying the analysis by cleanly separating the scalar perturbations. For a detailed discussion on gauge choices, see Ref.~\cite{Baumann:2022mni}.}
\begin{equation}
    \label{eq:generalmatterperturbations}
    T^0_0 \; \equiv \; -\left(\bar{\rho} + \delta \rho \right) \, , \qquad T^0_i \; \equiv \; \left(\bar{\rho} + \bar{P} \right)v_i \, ,
\end{equation}    
where $\bar{\rho}$ represents the background energy density, $\delta \rho$ is the perturbed energy density, $\bar{P}$ is the background pressure, and $v_i$ denotes the fluid velocity. To compute $\delta T_{0}^0$ and $\delta T_{i}^0$ for both the inflaton and the spectator scalar field, we expand the background as $\phi(\eta, \mathbf{x}) = \bar{\phi}(\eta) + \delta \phi(\eta, \mathbf{x})$, while the spectator field $\chi$ has no background value, with $\bar{\chi}(\eta) = 0$ and $\delta \chi(\eta, \mathbf{x})= \chi (\eta, \mathbf{x})$.

The energy-momentum tensor of the inflaton field is
\begin{equation}
    \label{eq:genstresstensor}
    T_{\mu\nu}^{(\phi)} \; = \; \nabla_\mu \phi \nabla_\nu \phi - g_{\mu\nu} \left( \frac{1}{2} g^{\sigma \rho} \nabla_\sigma \phi \nabla_\rho \phi + V(\phi) \right) \, ,
\end{equation}
with an analogous form for the spectator field $\chi$, where its mass and coupling terms are incorporated in $V(\chi)$. In the uniform density gauge, where $\delta \rho_\phi = 0$, the perturbed $\delta T_i^0$ component of the inflaton energy-momentum tensor is
\begin{equation}
\delta T_{i}^{0(\phi)}\; = \; (\bar{\rho}_\phi + \bar{P}_\phi)v_i^{(\phi)} \; \simeq \; -\frac{1}{a^2} \bar{\phi}' \partial_i \delta \phi \, .
\end{equation}
Using the continuity equation for background energy densities
\begin{equation}
\label{eq:continuitygen}
    \bar{\rho}_{\phi}' + 3 \mathcal{H} \left(\bar{\rho}_{\phi} + \bar{P}_{\phi} \right) \; = \; 0 \, ,~~\bar{\rho}_{\chi}' + 3 \mathcal{H} \left(\bar{\rho}_{\chi} + \bar{P}_{\chi} \right) \; = \; 0 \, ,
\end{equation}
we find that
\begin{equation}
    \label{eq:rhobarv}
    \bar{\rho}_{\phi}' v_i^{(\phi)} \; = \; -3 \mathcal{H} \left( \bar{\rho}_{\phi} + \bar{P}_{\phi} \right) v_i^{(\phi)} \; = \; \frac{3 \mathcal{H}}{a^2} \bar{\phi}' \partial_i \delta \phi \, .
\end{equation}
The gauge-invariant Poisson equation relates the scalar potential $\Phi$ to energy-momentum tensor perturbations,
\begin{equation}
    \nabla^2 \Phi \; = \; 4 \pi G a^2 \left(\delta \rho + \bar{\rho}' v \right) \, .
\end{equation}
In the uniform density gauge with $\delta \rho_\phi = 0$, spectator field contributions dominate:
\begin{equation}
\begin{aligned}
    \label{eq:gaugeinvpoisson}
    \nabla^2 \Phi \; \simeq \; 4 \pi G a^2 \left[\delta \rho_{\chi} + \bar{\rho}_{\chi}' v_{\chi} \right] \; = \; &4 \pi G a^2 \left[\delta \rho_{\chi} + \frac{3 \mathcal{H}}{a^2} \chi' \chi \right] \, .
\end{aligned}
\end{equation}
As shown in Eq.~(\ref{eq:rhobarv}), the inflaton contribution $\bar{\rho}_{\phi}' v_i^{(\phi)} \propto \delta \phi$ is subdominant to spectator field perturbations and can be neglected. Similarly, the term $\frac{3 \mathcal{H}}{a^2} \chi' \chi$ vanishes as $\mathcal{H} \to 0$ during late inflation and reheating. The Poisson equation reduces to
\begin{equation}
     \nabla^2 \Phi \; \simeq \; \frac{a^2}{2M_P^2} \delta \rho_{\chi} \, .
\end{equation}
Importantly, this gauge-independent expression ensures consistency between the uniform-density gauge used for the isocurvature spectrum in Section~\ref{sec:isocurvaturefluct} and the Newtonian gauge employed in subsequent GW calculations.
\subsection{General Framework}
\label{sec:genframe}
Scalar perturbations source secondary GWs at second order in perturbation theory~\cite{Ananda:2006af, Baumann:2007zm, Domenech:2021ztg}. We review the general formalism and examine how the evolving isocurvature power spectrum influences scalar perturbations, $\Phi$, which in turn source tensor perturbations, $h_{ij}$.

We work in the Newtonian gauge for scalar perturbations, with the line element defined in Eq.~(\ref{eq:lineelement}). The equation of motion for secondary GWs sourced by scalar perturbations is given by~\cite{Domenech:2021ztg}\footnote{In general, this expression also contains a term $8 \partial_a (\Phi - \Psi) \partial_b \Phi$ within the curly brackets. However, since the difference between the two scalar perturbations $\Phi - \Psi$ is non-zero only at second order, the term $\partial_a (\Phi - \Psi) \partial_b \Phi$ represents a third-order contribution and is subdominant compared to the remaining terms.}
\begin{equation}
    \begin{aligned}
    \label{eq:hijincosmictime}
    \ddot{h}_{ij} + 3H\dot{h}_{ij} - \frac{\nabla^2 h_{ij}}{a^2} \; = \; a^{-2} P^{ab}_{~~ij} \bigg\{4 \partial_a \Phi \partial_b \Phi + \frac{2}{M_P^2} \partial_a \delta \chi \partial_b \delta \chi \bigg\} \, ,
    \end{aligned}
\end{equation}
where $P^{ab}_{~~ij}$ is the projection operator, and $\delta \chi$ represents the dark matter fluctuations. Here, we neglected the $\delta \phi$ contribution since it is negligible compared to $\chi$ at the background level. Expressing the equation in terms of conformal time, we obtain:
\begin{equation}
    {h}_{ij}'' + 2\mathcal{H} h'_{ij} - \nabla^2 h_{ij} \; = \; P^{ab}_{~~ij} \left\{4 \partial_a \Phi \partial_b \Phi + \frac{2}{M_P^2} \partial_a \chi \partial_b \chi \right\} \, .
\end{equation}
Two key contributions to the secondary gravitational waves arise from scalar perturbations: one from the Newtonian scalar perturbation, $\Phi$, and the other from the spectator scalar field fluctuations,
$\delta \chi$. In general, the $\delta \chi$ contribution leads to high frequency SGWB that could only be probed by resonant cavity experiments. We leave this analysis for future work and focus exclusively on GWs induced by isocurvature perturbations.

To compute the tensor perturbations $h_{ij}$ and the GW energy spectra, we transform the equations to Fourier space. Transforming the left-hand side using Eq.~(\ref{eq:gaugeinvpoisson}), along with the Fourier transform of the energy density fluctuations
\begin{equation}
    \delta \rho_{\chi}(\eta, \mathbf{x}) \; = \; \int \frac{d^3 k}{(2\pi)^{3/2}} e^{-i \mathbf{k} \cdot \mathbf{x}} \delta \rho_{\chi}(\eta, \mathbf{k}) \, ,
\end{equation}
we obtain the equation of motion for each polarization mode. Multiplying both sides by $\varepsilon^{\lambda}_{ij}$ and using the projection property $\tilde{P}_{~~ij}^{ab} \varepsilon^{\lambda}_{ij} = \varepsilon^{\lambda}_{ab}$, where $\tilde{P}_{~~ij}^{ab}$, we derive:
\begin{equation}
    \label{eq:hijgensol}
    h_{\mathbf{k}, \lambda}''(\eta) + 2\mathcal{H}  h_{\mathbf{k}, \lambda}'(\eta) + k^2 h_{\mathbf{k}, \lambda}(\eta) \; = \; S_{\lambda}(\eta, \mathbf{k}) \, .
\end{equation}
where the source term $S_\lambda(\eta, \mathbf{k})$ is given by:
\begin{equation}
\begin{aligned}
\label{eq:sourceterm}
    S_{\lambda}(\eta, \mathbf{k}) = -\int \frac{d^3 p}{(2\pi)^3} \frac{a^4}{M_P^4} \frac{\varepsilon_\lambda^{ab} p_a q_b}{p^2 q^2} \delta \rho_\chi(\eta, \mathbf{p}) \delta \rho_\chi(\eta, \mathbf{q}) \, .
\end{aligned}
\end{equation}

To solve the equation of motion with the source term in Eq.~(\ref{eq:hijgensol}), we use the Green's function method. Given two independent homogeneous solutions, $h_1$ and $h_2$, the Green's function is constructed as
\begin{equation}
\mathcal{G}_{\mathbf k}(\eta, \tilde{\eta}) \; = \; \frac{1}{W(h_1, h_2, \tilde{\eta})} 
\left[ h_1(\eta) h_2(\tilde{\eta}) - h_1(\tilde{\eta}) h_2(\eta) \right] \, ,
\end{equation}
where the Green's function satisfies
\begin{equation}
\mathcal{G}_{\mathbf{k}}''(\eta, \tilde{\eta}) + 2 \mathcal{H} \mathcal{G}_{\mathbf{k}}'(\eta, \tilde{\eta}) + k^2 \mathcal{G}_{\mathbf{k}}(\eta, \tilde{\eta}) \; = \; \delta(\eta - \tilde{\eta}) \, .
\end{equation}
The Wronskian determinant ensures the proper normalization:
\begin{equation}
W(h_1, h_2, \tilde{\eta}) \; = \;  h_1'(\tilde{\eta}) h_2(\tilde{\eta}) - h_1(\tilde{\eta}) h_2'(\tilde{\eta}) \, .
\end{equation}
With appropriate initial conditions $h_\lambda(\eta_i, \mathbf{k}) = h_\lambda'(\eta_i, \mathbf{k}) = 0$, the particular solution for the tensor perturbation is obtained through the convolution integral:
\begin{equation}
h_\lambda(\eta, \mathbf{k}) \; = \;  \int_{\eta_i}^\eta d\tilde{\eta} \, \mathcal{G}_{\mathbf k}(\eta, \tilde{\eta}) S_\lambda(\tilde{\eta}, \mathbf{k}) \, .
\end{equation}
This general form is used for our numerical computation $h_\lambda(\eta, \mathbf{k})$.

\subsection{Computation of the 4-point Correlation Functions}
\label{sec:correlationfunctions}

In this section, we primarily follow Refs.~\cite{Garcia-Saenz:2022tzu, Adshead:2021hnm, Unal:2018yaa, Atal:2021jyo,Domenech:2021and} that discuss the non-Gaussianities in the primordial scalar sector. We assume statistical homogeneity and isotropy of the scalar field perturbation, $\delta \rho_{\chi}$, which appears in the source term of the gravitational wave equation~(\ref{eq:sourceterm}). Assuming vanishing mean $\langle \delta \rho_{\chi} \rangle = 0$, the two-point function of induced gravitational waves can be written as
\begin{equation}
\begin{aligned}
\label{eq:4pointcorr1}
\langle h_\lambda(\eta, \mathbf{k}_1) h_\lambda(\eta, \mathbf{k}_2) \rangle \; \sim \; \int \frac{d^3 q_1}{(2\pi)^{3/2}} \frac{d^3 q_2}{(2\pi)^{3/2}} \langle \delta \rho_{\chi}(\mathbf{q}_1) \delta \rho_{\chi}(\mathbf{k}_1 - \mathbf{q}_1) \delta \rho_{\chi}(\mathbf{q}_2) \delta \rho_{\chi}(\mathbf{k}_2 - \mathbf{q}_2) \rangle \, .
\end{aligned}
\end{equation}

The four-point function can be decomposed into a disconnected part (present even for Gaussian fluctuations) and a connected part (which captures the non-Gaussianity):
\begin{equation}
\begin{aligned}
    \langle  \delta \rho_{\chi}(\mathbf{k}_1)  \delta \rho_{\chi}(\mathbf{k}_2)  \delta \rho_{\chi}(\mathbf{k}_3)  \delta \rho_{\chi}(\mathbf{k}_4) \rangle &\; = \; \langle  \delta \rho_{\chi}(\mathbf{k}_1)  \delta \rho_{\chi}(\mathbf{k}_2)  \delta \rho_{\chi}(\mathbf{k}_3)  \delta \rho_{\chi}(\mathbf{k}_4) \rangle_d \\
    &\; + \; \langle  \delta \rho_{\chi}(\mathbf{k}_1)  \delta \rho_{\chi}(\mathbf{k}_2)  \delta \rho_{\chi}(\mathbf{k}_3)  \delta \rho_{\chi}(\mathbf{k}_4) \rangle_c \, ,
\end{aligned}   
\end{equation}
where the connected piece vanishes for purely Gaussian initial conditions. These components are given by:
\begin{equation}
    \begin{aligned}
    \langle \delta \rho_{\chi}(\mathbf{k}_1) \delta \rho_{\chi}(\mathbf{k}_2) \delta \rho_{\chi}(\mathbf{k}_3) \delta \rho_{\chi}(\mathbf{k}_4) \rangle_c \; = \; \delta^{(3)}(\mathbf{k}_1 + \mathbf{k}_2 + \mathbf{k}_3 + \mathbf{k}_4) 
    \mathcal{T}(\mathbf{k}_1, \mathbf{k}_2, \mathbf{k}_3, \mathbf{k}_4) \,,
    \end{aligned}
\end{equation}
and
\begin{align} \notag
        &\langle \delta \rho_{\chi}(\mathbf{k}_1) \delta \rho_{\chi}(\mathbf{k}_2) \delta \rho_{\chi}(\mathbf{k}_3) \delta \rho_{\chi}(\mathbf{k}_4) \rangle_d \; = \;  \langle \delta \rho_{\chi}(\mathbf{k}_1) \delta \rho_{\chi}(\mathbf{k}_2) \rangle \langle \delta \rho_{\chi}(\mathbf{k}_3) \delta \rho_{\chi}(\mathbf{k}_4) \rangle \\ 
        +& \langle \delta \rho_{\chi}(\mathbf{k}_2) \delta \rho_{\chi}(\mathbf{k}_3) \rangle \langle \delta \rho_{\chi}(\mathbf{k}_4) \delta \rho_{\chi}(\mathbf{k}_1) \rangle + \langle \delta \rho_{\chi}(\mathbf{k}_1) \delta \rho_{\chi}(\mathbf{k}_3) \rangle \langle \delta \rho_{\chi}(\mathbf{k}_2) \delta \rho_{\chi}(\mathbf{k}_4) \rangle \, .
    \end{align}
Here, $\mathcal{T}(\mathbf{k}_1, \mathbf{k}_2, \mathbf{k}_3, \mathbf{k}_4)$ denotes the connected trispectrum, and the two-point function is defined as
\begin{equation}
\langle \delta \rho_{\chi}(\mathbf{k}_1) \delta \rho_{\chi}(\mathbf{k}_2) \rangle \; = \;  \delta^{(3)}(\mathbf{k}_1 + \mathbf{k}_2) \bar{\rho}_{\chi}^2 \mathcal{P}_{\mathcal{S}}(\mathbf{k}_1) \, ,
\end{equation} 
in agreement with Eq.~(\ref{eq:isocurvature1}).

We consider small primordial non-Gaussianities as perturbative corrections to the Gaussian statistics, such that the connected component provides a nonzero but typically subleading contribution to the GW signal. Accordingly, we split the gravitational wave power spectrum into disconnected and connected contributions:
\begin{equation}
    \mathcal{P}_\lambda(k) \; = \; \mathcal{P}_\lambda(k)|_d + \mathcal{P}_\lambda(k)|_c \, .
\end{equation}
The disconnected piece arises from products of the scalar power spectrum, while the connected part encodes contributions from the primordial trispectrum. Quantifying the contribution to the GW spectrum from the trispectrum requires the characterization of the non-Gaussian probability distribution of $\delta\rho_{\chi}$. Moreover, the factorization property of the isocurvature spectrum (\ref{eq:TS}), crucial for our analysis of the disconnected contribution, might not be present for the trispectrum. We therefore quantify only the pure isocurvature contribution in our analysis below, leaving for a separate study the computation of the connected spectrum (which we expect to be subdominant, see~\cite{Garcia-Saenz:2022tzu}). This implies that our resulting GW spectra must be taken as conservative estimates.
%

Therefore, in practice, we want to compute the contribution
\begin{align} \notag
        & \langle \delta \rho_{\chi}(\mathbf{q}_1) \delta \rho_{\chi}(\mathbf{k}_1 - \mathbf{q}_1) \delta \rho_{\chi}(\mathbf{q}_2) \delta \rho_{\chi}(\mathbf{k}_2 - \mathbf{q}_2) \rangle_d \; = \; \langle \delta \rho_{\chi}(\mathbf{q}_1) \delta \rho_{\chi}(\mathbf{k}_1 - \mathbf{q}_1)  \rangle \langle \delta \rho_{\chi}(\mathbf{q}_2) \delta \rho_{\chi}(\mathbf{k}_2 - \mathbf{q}_2)  \rangle \\ 
         + &\langle \delta \rho_{\chi}(\mathbf{k}_1 - \mathbf{q}_1)  \delta \rho_{\chi}(\mathbf{q}_2) \rangle \langle \delta \rho_{\chi}(\mathbf{k}_2 - \mathbf{q}_2) \delta \rho_{\chi}(\mathbf{q}_1) \rangle + \langle \delta \rho_{\chi}(\mathbf{q}_1) \delta \rho_{\chi}(\mathbf{q}_2) \rangle \langle \delta \rho_{\chi}(\mathbf{k}_1 - \mathbf{q}_1)  \delta \rho_{\chi}(\mathbf{k}_2 - \mathbf{q}_2) \rangle \, .
    \end{align}
We now disregard contributions proportional to $\delta^{(3)}(\mathbf{k}_1)$ and $\delta^{(3)}(\mathbf{k}_2)$, which have no support for finite values of the external momentum and correspond to unphysical disconnected zero-momentum modes. After dropping these terms, the disconnected part of the four-point function simplifies to
\begin{align}
\label{eq:4pointsplit}
        \notag
        &\langle \delta \rho_{\chi}(\mathbf{q}_1) \delta \rho_{\chi}(\mathbf{k}_1 - \mathbf{q}_1) \delta \rho_{\chi}(\mathbf{q}_2) \delta \rho_{\chi}(\mathbf{k}_2 - \mathbf{q}_2) \rangle_d \; = \; \\
        &\delta^{(3)}(\mathbf{k}_1 + \mathbf{k}_2) 
        \big[ \delta^{(3)}(\mathbf{q}_1 + \mathbf{q}_2) + \delta^{(3)}(\mathbf{k}_1 + \mathbf{q}_2 - \mathbf{q}_1) \big] 
      \bar{\rho}_{\chi}^4 \mathcal{P}_{S}(\mathbf{q}_1) \mathcal{P}_{S}(|\mathbf{k}_1 - \mathbf{q}_1|) \, .
    \end{align}

\subsection{Gravitational Wave Spectrum}
\label{sec:gwspectrum}
We now proceed to explicitly compute the $4$-point correlation functions and evaluate the GW spectrum. In general, the procedure is rather involved but for clarity, we include all the relevant details of the computation here. In general, we need to compute the GW spectrum
\begin{equation}
\begin{aligned}
    \label{eq:gwspectrgenexp}
   \langle h_\lambda(\eta, \mathbf{k}_1) h_\lambda(\eta, \mathbf{k}_2) \rangle \; = \; \int_0^{\eta} d\eta_1 \int_0^{\eta} d\eta_2 \  \mathcal{G}_{\mathbf k}(\eta, \eta_1) \mathcal{G}_{\mathbf k}(\eta, \eta_2) \langle S_{ \lambda}(\eta_1, \mathbf{k}_1) S_{\lambda}(\eta_2, \mathbf{k}_2) \rangle \, ,
\end{aligned}
\end{equation}
where 
\begin{equation}
\begin{aligned}
    \label{eq:s2point}
    \langle S_{ \lambda}(\eta_1, \mathbf{k}_1) S_{\lambda}(\eta_2, \mathbf{k}_2) \rangle &\; = \;  \frac{a(\eta_1)^4}{M_P^4} \frac{a(\eta_2)^4}{M_P^4} \int \frac{d^3q_1}{(2\pi)^{3/2}} \frac{d^3q_2}{(2\pi)^{3/2}} \frac{Q_{\lambda}(\mathbf{k}_1, \mathbf{q}_1) Q_{\lambda}(\mathbf{k}_2, \mathbf{q}_2)}{q_1^2 |\mathbf{k}_1 - \mathbf{q}_1|^2 q_2^2 |\mathbf{k}_2-\mathbf{q}_2|^2} \\
     &\qquad \; \times \;  \mathcal{T}_{\mathcal{S}}(\eta_1) \mathcal{T}_{\mathcal{S}}(\eta_2) \langle \delta \rho_{\chi}(\mathbf{q}_1) \delta \rho_{\chi}(|\mathbf{k}_1-\mathbf{q}_1|) \delta \rho_{\chi}(\mathbf{q}_2) \delta \rho_{\chi}(|\mathbf{k}_2-\mathbf{q}_2|) \rangle \, .
\end{aligned}    
\end{equation}
Here, we introduced the projection factor
\begin{equation}
    Q_{\lambda}(\mathbf{k}, \mathbf{q}) \; \equiv \; \varepsilon_{ij}^{\lambda}(\mathbf{k}) q_i q_j \; = \; -\varepsilon_{\lambda}^{ij}(\mathbf{k})(\mathbf{k}-\mathbf{q})_i q_j \, ,
\end{equation}
where the second equality follows from \(\varepsilon_\lambda^{ij}(\mathbf{k}) k_i = 0\). If we explicitly set \(\hat{k} = \hat{z}\), we have 
$\mathbf{q} = q (\sin \theta \cos \phi, \sin \theta \sin \phi, \cos \theta)$, 
where $\theta$ and $\phi$ are polar and azimuthal angles. This leads to the expressions:
\begin{equation}
\begin{aligned}
\label{eq:qexpressions}
Q_+(\mathbf{k}, \mathbf{q}) \; = \; \frac{q^2}{\sqrt{2}} \sin^2 \theta \cos(2\phi) \, , \qquad Q_\times(\mathbf{k}, \mathbf{q}) \; = \; \frac{q^2}{\sqrt{2}} \sin^2 \theta \sin(2\phi) \, .
\end{aligned}
\end{equation}
Since $\varepsilon_\lambda(\mathbf{k})$ is orthogonal to $\mathbf{k}$, we have
\begin{equation}
Q_\lambda(\mathbf{k}, \mathbf{q}) = Q_\lambda(\mathbf{k}, \mathbf{q} + c\mathbf{k}) \, ,
\end{equation}
for any constant \(c\). \(Q_\lambda(\mathbf{k}, \mathbf{q})\) is also symmetric under 
\(\mathbf{k} \to -\mathbf{k}\) and \(\mathbf{q} \to -\mathbf{q}\),
\begin{equation}
Q_\lambda(\mathbf{k}, \mathbf{q}) = Q_\lambda(-\mathbf{k}, \mathbf{q}) = Q_\lambda(\mathbf{k}, -\mathbf{q}) = Q_\lambda(-\mathbf{k}, -\mathbf{q}) \, .
\end{equation}
We factor out the time and momentum dependence using the function $\mathcal{T}_{\mathcal{S}}$ defined in Eq.~(\ref{eq:TS}).

One can show that the full dimensionless GW power spectrum can be expressed as 
\begin{equation}\label{eq:deltah2}
\begin{aligned}
&\Delta_h^2(k, N) \;=\; 2 \left[\frac{1}{8}\int^NdN'\, \mathcal{G}_{\mathbf k}(N,N')\, \mathcal{T}_{\mathcal{S}}(N')\left(\frac{a(N')H(N')}{k}\right)^2\left(\frac{\bar{\rho}_{\chi}(N')}{H^2(N')M_P^2}\right)^2\right]^2g(k)
\, ,
\end{aligned}
\end{equation}
where 
\beq
\label{eq:fgen}
g(k) \;=\; k^2\int_0^{\infty} dp\, p\int_{|k-p|}^{k+p}dq\, q\,\frac{(k^4-2k^2(p^2+q^2)+(p^2-q^2)^2)^2}{p^7q^7}\,\Delta_{\mathcal{S}}^2(p)\Delta_{\mathcal{S}}^2(q)\,.
\eeq
We can now express the GW energy density~(\ref{eq:gwgeneral}) as    
\begin{equation}
    \label{eq:gwiso}
    \Omega_{{\rm GW}, \, \mathcal{S}} \; = \; \frac{1}{12} \left(\frac{k}{a H} \right)^2 \Delta_h^2(k, N) \; \equiv \; \mathcal{I}(k, N) g(k) \, , 
\end{equation}
where we have factored out the contribution $g(k)$. Note that this calculation is performed in terms of the number of $e$-folds, and $\mathcal{G}_{\mathbf k}(N,N')$ is the Green's function that uses the number of $e$-folds as the time variable. The full details of this derivation are provided in Appendix~\ref{app:C}. We note that here the factors of $k$ in $g(k)$ are chosen to make it dimensionless. We now proceed and show how to compute this contribution and what signals we expect arising from the isocurvature power spectrum of spectator scalar fields.

\section{Gravitational Wave Production from Isocurvature Fluctuations}
\label{sec:gwiso}
We now focus on evaluating the time-dependent contribution to $\Omega_{{\rm GW}, \, \mathcal{S}}$, namely $\mathcal{I}(k, N)$. We assume the spectator field can either decay when its energy density constitutes a fraction of the background radiation density, $f_{\chi}<1$, or remain stable as a DM component.

\begin{figure*}[t!]
    \centering
    \includegraphics[width=\linewidth]{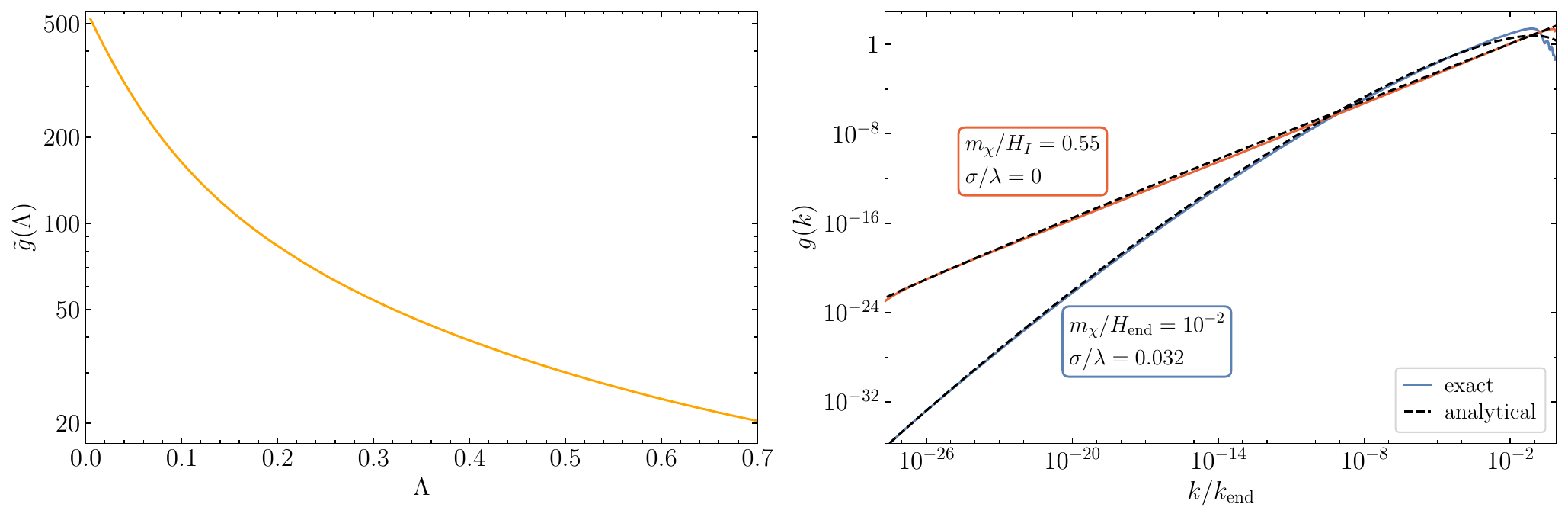}
    \caption{Left: The kernel function $\tilde{g}(\Lambda)$, defined in (\ref{eq:gnu}), which characterizes the momentum-space convolution integral for induced gravitational waves. Right: Comparison between the exact numerical computation (solid curves) and analytical approximation (black dashed curves) for the isocurvature contribution to the gravitational wave energy spectrum, shown for two representative parameter sets. For $\sigma\neq0$ case, we implement the phenomenological rescaling $m_{\chi,{\rm eff}}\rightarrow0.9 m_{\chi,{\rm eff}}$ to improve the accuracy of the analytical approximation (see Fig.~\ref{fig:isofits}).}
    \label{fig:gapp}
\end{figure*}

As demonstrated earlier, the rigid time evolution of $\Delta_{\mathcal{S}}^2$, encoded in Eq.~(\ref{eq:TS}), 
allows us to decompose the gravitational wave power spectrum into time-independent and time-dependent parts. The time-independent part, $g(k)$, can be further simplified by introducing the dimensionless variables  $u=p/k$ and $v=q/k$:
\begin{align}
\label{eq:guv}
g(k) \;=\; \int_0^{\infty} du \int_{|1-u|}^{1+u}dv\ \frac{(1-2(u^2+v^2)+(u^2-v^2)^2)^2}{u^6v^6} \Delta_{\mathcal{S}}^2(uk)\Delta_{\mathcal{S}}^2(vk) \; \simeq \; \tilde{g}(\Lambda) \left( \Delta_{\mathcal{S}}^2(k) \right)^2\,,
\end{align}
where $\Lambda$ was defined in (\ref{eq:nu}). To derive the second equality, we apply the approximation from Eq.~(\ref{eq:PSan2}), neglecting the mild scale-dependence of the logarithmic terms inside the integral~(\ref{eq:guv}). The function $\tilde{g}(\Lambda)$ is defined as
\beq
\label{eq:gnu}
\tilde{g}(\Lambda) \;=\; \int_0^{\infty} du \int_{|1-u|}^{1+u}dv \frac{(1-2(u^2+v^2)+(u^2-v^2)^2)^2}{u^{6-2\Lambda}\, v^{6-2\Lambda}}\,,
\eeq
which generally depends on $k$ through the relation~(\ref{eq:nu}). This integral is evaluated numerically using similar techniques as for Eq.~(\ref{eq:PSp3}), with the notable difference that the integrand vanishes at the boundary, so no regularization is required. The left panel of Fig.~\ref{fig:gapp} illustrates the functional dependence of $\tilde{g}(\Lambda)$ on $\Lambda$ across a range sufficiently broad to encompass all scenarios considered in this work.

The right panel of Fig.~\ref{fig:gapp} displays the time-independent isocurvature contribution to the GW spectrum, $g(k)$, for representative parameter choices of the spectator field. The red and blue curves show the full numerical computation of the integral in Eq.~(\ref{eq:fgen}), using the numerical power spectra shown in Fig.~\ref{fig:iso1}.  The dashed black curves represent the analytical approximation~Eq.~(\ref{eq:guv}), computed using the analytical spectrum~(\ref{eq:PSan2}) with parameter $\Lambda$ from Eq.~(\ref{eq:nu}). As before, Eq.~(\ref{eq:phik}) is used in the case with $\sigma\neq0$ to evaluate $\phi_k$ in $m_{{\rm eff},\chi}$. Similar to Fig.~\ref{fig:isofits}, we implement the phenomenological rescaling $m_{\chi,{\rm eff}}\rightarrow0.9 m_{\chi,{\rm eff}}$ for $\sigma\neq0$, which effectively accounts for the non-instantaneous transition characterized by $N_{\rm NR}-N_{\rm end}$ (see Fig.~\ref{fig:isoT}). Despite the computational complexity involved in the numerical evaluation of $g(k)$, in both cases shown in the figure the analytical result closely approximates the exact value for the isocurvature contribution to $\Omega_{{\rm GW},\mathcal{S}}$.

Having determined the time-independent factor in (\ref{eq:gwiso}), we now proceed to compute $\mathcal{I}(k,N)$. To simplify the integration, we partition the cosmic evolution into three distinct phases guided by Fig.~\ref{fig:rhos}. First, we consider the early time interval from the beginning of inflation to the freeze-out of the isocurvature spectrum, $N\in(0,N_{\rm NR})$.  In this regime, the integral requires full numerical treatment, accounting for the time-dependence of $\mathcal{T}_{\mathcal{S}}$, the smooth transition from inflation to reheating in the Green's function $\mathcal{G}_{\bk}$, and the background quantities appearing inside the square brackets in Eq.~(\ref{eq:deltah2}). Second, we evaluate the contribution during reheating, $N\in(N_{\rm NR},N_{\rm reh})$, when the expansion of the universe is dominated by coherent oscillations of the inflaton field, scaling as non-relativistic matter. During this phase, we can reliably approximate $\mathcal{T}_{\mathcal{S}}\simeq 1$ and treat the energy density $\rho_{\chi}$ as scaling like non-relativistic matter, significantly simplifying the computation. Finally, we incorporate the late-time contribution for $N>N_{\rm reh}$, during which similar simplifications apply, including the assumption of radiation domination until either the decay of $\chi$ or $\chi$-radiation equality in the DM scenario. Explicitly,
\begin{align} \notag
\mathcal{I}(k,N) \simeq &\frac{1}{384} \left(\frac{k}{a(N)H(N)}\right)^2\Bigg[\int_0^{N_{\rm NR}}dN'\,\mathcal{G}_{\bk}(N,N')\,\mathcal{T}_{\mathcal{S}}(N')\left(\frac{a(N')H(N')}{k}\right)^2\left(\frac{\rho_{\chi}(N')}{H^2(N')M_P^2}\right)^2\\ 
& + \int_{N_{\rm NR}}^{N_{\rm reh}}dN'\,\mathcal{G}_{\bk}(N,N') \left(\frac{a(N')H(N')}{k}\right)^2\left(\frac{\rho_{\chi}(N')}{H^2(N')M_P^2}\right)^2 \notag \\ \displaybreak[0]
&\ + \int_{N_{\rm reh}}^{N}dN'\,\mathcal{G}_{\bk}(N,N') \left(\frac{a(N')H(N')}{k}\right)^2\left(\frac{\rho_{\chi}(N')}{H^2(N')M_P^2}\right)^2\Bigg]^2\\ \label{eq:I1}
&\equiv\; \bigg[ \mathcal{J}_{\rm NR}(k,N) + \mathcal{J}_{\rm reh}(k,N) + \mathcal{J}_{\rm rad}(k,N)\bigg]^2 \, .
    \end{align}
We now proceed to evaluate each of the $\mathcal{J}$ defined above separately.

\subsection{Early Time Contribution ($\mathcal{J}_{\rm NR}(k,N)$)}
We begin with the early contribution $\mathcal{J}_{\rm NR}(k,N)$, which is determined by integrating from the beginning of cosmic dynamics up to the freeze-out of the isocurvature power spectrum. It is important to clarify that this upper time limit $N_{\rm NR}$ applies only to the integration variable $N'$ in (\ref{eq:I1}), while our objective is to evaluate the GW spectrum at the present epoch, corresponding to a much later $e$-fold number $N$. We can express this component more explicitly by rewriting it as follows:
\begin{align}\notag \displaybreak[0]
&\mathcal{J}_{\rm NR}(k,N) \;=\;  \frac{1}{8\sqrt{6}} \left(\frac{k}{a(N)H(N)}\right) \int_0^{N_{\rm NR}}dN'\,\mathcal{G}_{\bk}(N,N')\,\mathcal{T}_{\mathcal{S}}(N')\left(\frac{a(N')H(N')}{k}\right)^2\left(\frac{\rho_{\chi}(N')}{H^2(N')M_P^2}\right)^2\\ \notag \displaybreak[0]
&=\;  \frac{1}{8\sqrt{6}} \left(\frac{k_{\rm end}}{k}\right)\left(\frac{a_{\rm end}H_{\rm end}}{a(N)H(N)}\right) \int_0^{N_{\rm NR}}dN'\,\mathcal{G}_{\bk}(N,N')\,\mathcal{T}_{\mathcal{S}}(N')\left(\frac{a(N')H(N')}{a_{\rm end}H_{\rm end}}\right)^2\left(\frac{\rho_{\chi}(N')}{H^2(N')M_P^2}\right)^2\\ \notag \displaybreak[0]
&=\;  \frac{1}{8\sqrt{6}} \left(\frac{k_{\rm end}}{k}\right) \left(\frac{a_{\rm reh}H_{\rm reh}}{a(N)H(N)}\right) \left(\frac{a_{\rm end}H_{\rm end}}{a_{\rm reh}H_{\rm reh}}\right) \mathcal{T}_{\rm R}(N;N_{\rm reh}) \mathcal{T}_{\rm M}(N_{\rm reh};N_{\rm NR})\\ \label{eq:JNR}
&\quad \times \int_0^{N_{\rm NR}}dN'\,\mathcal{G}_{\bk}(N_{\rm NR},N')\,\mathcal{T}_{\mathcal{S}}(N')\left(\frac{a(N')H(N')}{a_{\rm end}H_{\rm end}}\right)^2\left(\frac{\rho_{\chi}(N')}{H^2(N')M_P^2}\right)^2 \, .
\end{align}
Here, $\mathcal{T}_{\rm R,M}$ represent the tensor transfer functions for radiation and matter domination epochs, respectively. These functions, along with the Green's function $\mathcal{G}_{\bk}(N,N')$, are obtained by solving the homogeneous equation of motion for tensor modes, that is, Eq.~(\ref{eq:hijincosmictime}) without the source term on the right-hand side. In terms of the $e$-fold variable and the Fourier modes $h_\lambda(N, \mathbf{k})$, this requires solving
\beq
\label{eq:hkode}
\frac{d^2 h_{\lambda}}{dN^2} + (3-\varepsilon_H)\frac{dh_{\lambda}}{dN} + \left(\frac{k}{aH}\right)^2 h_{\lambda} \;=\; 0\,,
\eeq
with $\varepsilon_H$ the first Hubble flow function,
\beq
\varepsilon_H \; = \; -\frac{\dot{H}}{H^2} \; = \; -\frac{1}{H}\frac{dH}{dN} \, .
\eeq

During radiation domination, Eq.~(\ref{eq:hkode}) reduces to
\beq
\frac{d^2 h_{\lambda}}{dN^2} + \frac{dh_{\lambda}}{dN} + \left(\frac{k}{k_{\rm reh}}\right)^2 e^{2(N-N_{\rm reh})} h_{\lambda} \;=\;0\,,
\eeq
where $k_{\rm reh}=a_{\rm reh}H_{\rm reh}$. Integrating for a solution of the form $h_{\lambda}(N,\bk)=\mathcal{T}_R(N;N_{\rm reh})h_{\lambda}(N_{\rm reh},\bk)$ leads to
\begin{align} 
\mathcal{T}_{\rm R}(N;N_{\rm reh}) \;=\; \left(\frac{aH}{a_{\rm reh}H_{\rm reh}}\right) \left[ \cos\left(\frac{k}{k_{\rm reh}}\left(1 - \frac{a_{\rm reh}H_{\rm reh}}{aH}\right)\right) \right. \left. - \left(\frac{k_{\rm reh}}{k}\right)
\sin\left(\frac{k}{k_{\rm reh}}\left(1 - \frac{a_{\rm reh}H_{\rm reh}}{aH}\right)\right)
\right] \, .
\label{eq:TRcompact}
\end{align}
In Eq.~(\ref{eq:JNR}) this transfer function must be evaluated at the present epoch.\footnote{We follow the standard approach of approximating no expansion between matter-radiation equality and the present day~\cite{Saikawa:2018rcs, Domenech:2021ztg}.} Since all observationally relevant modes are well within our current horizon, we can simplify the calculation by considering only sub-horizon modes, where $k\gg a_0H_0$. Under this approximation, Eq.~(\ref{eq:TRcompact}) reduces to
\begin{align} \mathcal{T}_{\rm R}(N;N_{\rm reh}) \;\simeq\; \left(\frac{aH}{a_{\rm reh}H_{\rm reh}}\right) \cos \left(\frac{k}{k_{\rm reh}} \left(1-\frac{a_{\rm reh}H_{\rm reh}}{aH}\right)\right)\,. 
\end{align}
Averaging over the rapid oscillations, we obtain 
\beq\label{eq:TRav}
\overline{\mathcal{T}_{\rm R}(N;N_{\rm reh})} \;\simeq\; \frac{1}{\sqrt{2}}\left(\frac{aH}{a_{\rm reh}H_{\rm reh}}\right)\,.
\eeq

To determine the matter domination (reheating) transfer function, we solve the equation
\begin{align} \label{eq:heqM}
\frac{d^2 h_{\lambda}}{dN^2} + \frac{3}{2}\frac{dh_{\lambda}}{dN} +\left(\frac{k}{a_{\rm NR}H_{\rm NR}}\right)^2 e^{(N-N_{\rm NR})} h_{\lambda} \;=\;0\,,
\end{align}
obtained from Eq.~(\ref{eq:hkode}). In this case, we need the expression for sub- and super-horizon modes. Direct integration yields the following analytical solution:
\begin{align} \notag
&\mathcal{T}_{\rm M}(N;N_{\rm NR})\;=\; \left[ \frac{3}{8} \left(\frac{aH}{k}\right)^3 -\frac{1}{2}\left(\frac{aH}{ k}\right)\left(\frac{aH}{a_{\rm NR}H_{\rm NR}}\right)^2+\frac{3}{2}\left(\frac{aH}{ k}\right) \left(\frac{aH}{a_{\rm NR}H_{\rm NR}}\right) \right] \sin \left(2 \left(\frac{k}{aH}-\frac{k}{a_{\rm NR}H_{\rm NR}}\right)\right)\\  \displaybreak[0]
& +\left[ \left(\frac{aH}{a_{\rm NR}H_{\rm NR}}\right)^2 -\frac{3}{4} \left(\frac{aH}{k}\right)^2 + \frac{3}{4} \left(\frac{aH}{k}\right)^2 \left(\frac{aH}{a_{\rm NR}H_{\rm NR}}\right) \right] \cos \left(2 \left(\frac{k}{aH}-\frac{k}{a_{\rm NR}H_{\rm NR}}\right)\right)\\  \label{eq:TMapp}
&\simeq\; \begin{cases}
1\,, & \dfrac{k}{aH}\ll 1\,,\\[7pt]
 \left(\dfrac{aH}{a_{\rm NR}H_{\rm NR}}\right)^2\cos \left(2 \left(\dfrac{k}{aH}-\dfrac{k}{a_{\rm NR}H_{\rm NR}}\right)\right)\,, & \dfrac{k}{aH}\gg 1\,.
\end{cases}
\end{align}

\begin{figure}[!t]
\centering
    \includegraphics[width=0.7\textwidth]{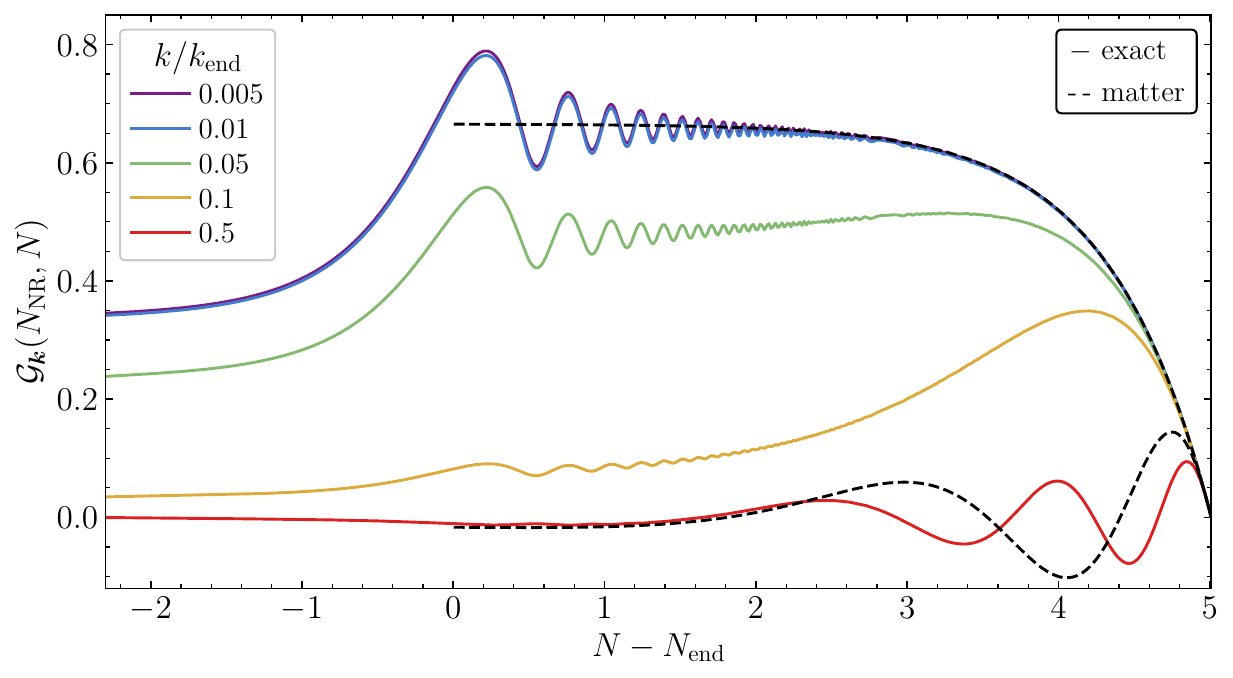}
    \caption{Tensor Green's function evaluated at $N_{\rm NR}=\ln(150)$ across different wavelength scales during the transition from inflation to reheating. The solid lines correspond to the numerical solution for (\ref{eq:wronsk}), while the dashed lines show the analytical solution obtained from (\ref{eq:Gmatter}), which assumes pure matter domination immediately following inflation. The analytical results are displayed only for the smallest and largest scales, demonstrating how the approximation accuracy varies with $k$. }
    \label{fig:green}
\end{figure}

The remaining ingredient in (\ref{eq:JNR}), the Green's function $\mathcal{G}_{\bk}(N,N')$, requires numerical evaluation from the inflationary epoch through isocurvature freeze-out. Following standard formalism, this function is constructed from two linearly independent solutions to the homogeneous equation (\ref{eq:hkode}):
\begin{align}
\label{eq:wronsk}
\mathcal{G}_{\bk}(N,N') \;=\;  \frac{h_{\lambda}^{(1)}(N',{\bk})h_{\lambda}^{(2)}(N,{\bk}) - h_{\lambda}^{(1)}(N,{\bk})h_{\lambda}^{(2)}(N',{\bk})}{W_{12}(N',{\bk})}\,,
\end{align}
where $W_{12}(N',\bk)$ denotes of the Wronskian of $h^{(1,2)}_{\lambda}$. Fig.~\ref{fig:green} shows the form of the numerically integrated Green's function toward the end of inflation up to $N_{\rm NR}$ for a selection of scales. We note that the transition from inflation to the matter-dominated reheating induces an oscillation in the Green's function. This effect is missed if one approximates the post-inflationary evolution simply as a matter-dominated era starting at the end of inflation. This approximation is discussed in detail in the next section (see Eqs.~(\ref{eq:Gmatter})), and is illustrated by dashed black curves for the largest and smallest $k$ values. Despite initial discrepancies, the analytical approximation converges to good agreement with the numerical solution by $N_{\rm NR}$, validating its application for calculating the reheating contribution to the GW spectrum.

For completeness, we now summarize our result for the early-time contribution to the present-day GW spectrum, which can be written in the following compact form:
\begin{equation}
    \begin{aligned}
        \mathcal{J}_{\rm NR}(k,N) & \;\simeq \; \frac{1}{16\sqrt{3}} \left(\frac{k_{\rm end}}{k}\right)\left(\frac{a_{\rm end}H_{\rm end}}{a_{\rm reh}H_{\rm reh}}\right)\mathcal{T}_{\rm M}(N_{\rm reh};N_{\rm NR}) \\
        &\qquad\;\times\;\int_0^{N_{\rm NR}}dN'\,\mathcal{G}_{\bk}(N_{\rm NR},N')\,\mathcal{T}_{\mathcal{S}}(N') \left(\frac{a(N')H(N')}{a_{\rm end}H_{\rm end}}\right)^2\left(\frac{\rho_{\chi}(N')}{H^2(N')M_P^2}\right)^2\,.
    \end{aligned}
\end{equation}

\subsection{Reheating contribution ($\mathcal{J}_{\rm reh}(k,N)$)}
Let us now proceed to evaluate the second integral in Eq.~(\ref{eq:I1}), which represents the reheating contribution to the gravitational wave spectrum. Following an approach analogous to that of the previous section, we can express this term as:
\begin{equation}
    \begin{aligned}
    \label{eq:Jrehs}
        &\mathcal{J}_{\rm reh}(k,N) \;=\;  \frac{1}{8\sqrt{6}}  \left(\frac{k_{\rm end}}{k}\right) \left(\frac{a_{\rm end}H_{\rm end}}{a(N)H(N)}\right)  \int_{N_{\rm NR}}^{N_{\rm reh}} dN'\,\mathcal{G}_{\bk}(N,N')\left(\frac{a(N')H(N')}{a_{\rm end}H_{\rm end}}\right)^2\left(\frac{\rho_{\chi}(N')}{H^2(N')M_P^2}\right)^2\\ 
    &\simeq\;  \frac{1}{16\sqrt{3}} \left(\frac{k_{\rm end}}{k}\right) \left(\frac{a_{\rm end}H_{\rm end}}{a_{\rm reh}H_{\rm reh}}\right) \int_{N_{\rm NR}}^{N_{\rm reh}} dN'\,\mathcal{G}_{\bk}(N_{\rm reh},N')\left( \frac{a(N')H(N')}{a_{\rm end}H_{\rm end}} \right)^2\left(\frac{\rho_{\chi}(N')}{H^2(N')M_P^2}\right)^2\\ 
    &\simeq\;  \frac{1}{16\sqrt{3}} \left(\frac{k_{\rm end}}{k}\right) \left(\frac{a_{\rm end}H_{\rm end}}{a_{\rm reh}H_{\rm reh}}\right)\left(\frac{\rho_{\chi,{\rm NR}}}{H^2_{\rm NR} M_P^2}\right)^2 \int_{N_{\rm NR}}^{N_{\rm reh}} dN'\,\mathcal{G}_{\bk}(N_{\rm reh},N') e^{-(N'-N_{\rm end})} \,.
    \end{aligned}
\end{equation}
In going from the first to the second lines above, we have applied the radiation domination transfer function given in (\ref{eq:TRav}). From second to third line we use the fact that for the T-model potential described in Eq.~(\ref{inf:tmodel}), the coherent oscillations of the inflaton field effectively mimic a matter-dominated era. Furthermore, for $N>N_{\rm NR}$, the spectator field energy density $\rho_\chi$ redshifts as non-relativistic matter. The remaining element needed for computing the reheating contribution is the matter domination tensor Green's function, which can be determined analytically in this case. Solving Eq.~(\ref{eq:hkode}) yields two linearly independent functions:
\begin{equation}\label{eq:Gmatter}
    \begin{aligned}
        &h_{\lambda}^{(1)}(N,\bk) \;=\; e^{-\frac{3}{2}(N-N_{\rm end})}\left[ \cos\left(\frac{2k}{a(N)H(N)}\right) \right.  \left.+\; \frac{2k}{a(N)H(N)}\, \sin\left( \frac{2k}{a(N)H(N)} \right) \right]\,,\\ 
    &h_{\lambda}^{(2)}(N,\bk) \;=\; e^{-\frac{3}{2}(N-N_{\rm end})}\left[ \sin\left(\frac{2k}{a(N)H(N)}\right) \right. 
    \left.-\; \frac{2k}{a(N)H(N)}\, \cos\left( \frac{2k}{a(N)H(N)} \right) \right] \,,
    \end{aligned}
\end{equation}
where $a(N)H(N)=e^{-\frac{1}{2}(N-N_{\rm end})}a_{\rm end}H_{\rm end}$ and $W_{12}(N,\bk)=4e^{-\frac{3}{2}(N-N_{\rm end})}(k/k_{\rm end})^3$. Substituting these into Eq.~(\ref{eq:wronsk}) provides the Green's function for arbitrary $k$. For super-horizon modes during reheating, this expression simplifies to
\beq
\mathcal{G}_{\bk}(N,N') \;\simeq\; \frac{2}{3}\left(1-e^{\frac{3}{2}(N'-N)}\right)\,.
\eeq
Consequently, for modes that remain super-horizon through the end of reheating, the integral in Eq.~(\ref{eq:Jrehs}) can be further simplified to
\begin{align} \notag
     \mathcal{J}_{\rm reh}(k,N)\bigg|_{k\ll k_{\rm reh}} &\; \simeq\; \frac{1}{24\sqrt{3}} \left(\frac{k_{\rm end}}{k}\right) \left(\frac{a_{\rm end}H_{\rm end}}{a_{\rm reh}H_{\rm reh}}\right)\left(\frac{\rho_{\chi,{\rm NR}}}{H^2_{\rm NR} M_P^2}\right)^2  e^{-(N_{\rm NR}-N_{\rm end})} \\  \label{eq:Jrehf}
     &\; \simeq\; \frac{1}{24\sqrt{6}}\left(\frac{k_{\rm end}}{k}\right) \left(\frac{\rho_{\rm end}}{\rho_{\rm reh}}\right)^{1/6}\left(\frac{\rho_{\chi,{\rm NR}}}{H^2_{\rm NR} M_P^2}\right)^2  e^{-(N_{\rm NR}-N_{\rm end})}\,,
 \end{align}
indicating that this contribution scales as $\sim T_{\rm reh}^{-2/3}$.

\subsection{Radiation contribution ($\mathcal{J}_{\rm rad}(k,N)$)}
\label{sec:Jrad}
We now turn to simplifying the third integral contribution in Eq.~(\ref{eq:I1}), which represents the post-reheating phase. Since at this stage the universe is dominated by the relativistic decay products of the inflaton, we can express this contribution as:
\begin{align}\notag
\mathcal{J}_{\rm rad}(k,N) \;&=\;  \frac{1}{8\sqrt{6}} \left(\frac{k_{\rm end}}{k}\right) \left(\frac{a_{\rm end}H_{\rm end}}{a(N)H(N)}\right)  \int_{N_{\rm reh}}^N dN'\,\mathcal{G}_{\bk}(N,N')\left(\frac{a(N')H(N')}{a_{\rm end}H_{\rm end}}\right)^2\left(\frac{\rho_{\chi}(N')}{H^2(N')M_P^2}\right)^2\\ \notag
&=\;  \frac{1}{8\sqrt{6}} \left(\frac{k_{\rm end}}{k}\right) \left(\frac{a_{\rm end}H_{\rm end}}{a(N)H(N)}\right) \left(\frac{a_{\rm reh}H_{\rm reh}}{a_{\rm end}H_{\rm end}}\right)^2 \left(\frac{\rho_{\chi,{\rm NR}}}{H^2_{\rm NR} M_P^2}\right)^2  \int_{N_{\rm reh}}^N dN'\,\mathcal{G}_{\bk}(N,N') \\ \label{eq:Jrads}
&\equiv\;  \frac{1}{8\sqrt{6}} \left(\frac{k_{\rm end}}{k}\right) \left(\frac{a_{\rm reh}H_{\rm reh}}{a(N)H(N)}\right) \left(\frac{a_{\rm reh}H_{\rm reh}}{a_{\rm end}H_{\rm end}}\right) \left(\frac{\rho_{\chi,{\rm NR}}}{H^2_{\rm NR} M_P^2}\right)^2  \gamma(N;k) \,.
\end{align}
The radiation domination Green's function for the tensor modes is evaluated in the same way as in the matter case. It is given by
\begin{align} \notag 
\mathcal{G}_{\bk}(N,N') &\;=\; \left(\frac{k_{\rm reh}}{k}\right)e^{-(N-N_{\rm reh})} \sin\left[\left(e^{N-N_{\rm reh}}-e^{N'-N_{\rm reh}}\right)\frac{k}{k_{\rm reh}}\right]\label{eq:greenrad}\\
&\;=\; \left(\frac{a(N)H(N)}{k}\right)\sin\left[\frac{k}{a(N)H(N)}-\frac{k}{a(N')H(N')}\right]\,.
\end{align}
Its integral defines in turn the function $\gamma(N;k)$, which can be written in terms of the trigonometric integral functions, 
\begin{align}\notag
\gamma(N;k) &\;=\; \left(\frac{a_{\rm reh}H_{\rm reh}}{k}\right) e^{(N-N_{\rm reh})}\Bigg\{\left[{\rm Ci}\left(e^{N-N_{\rm reh}}\left(\tfrac{k}{k_{\rm reh}}\right)\right) - {\rm Ci}\left(\tfrac{k}{k_{\rm reh}}\right) \right]\sin\left(e^{N-N_{\rm reh}}\left(\tfrac{k}{k_{\rm reh}}\right)\right)\\
&\qquad \qquad \qquad\qquad \qquad - \left[{\rm Si}\left(e^{N-N_{\rm reh}}\left(\tfrac{k}{k_{\rm reh}}\right)\right) - {\rm Si}\left(\tfrac{k}{k_{\rm reh}}\right) \right]\cos\left(e^{N-N_{\rm reh}}\left(\tfrac{k}{k_{\rm reh}}\right)\right) \Bigg\}\\ \label{eq:gammaapp}
&\;\simeq\; \begin{cases}
N-N_{\rm reh}\,, & k/aH\ll 1\,,\\
-\left(\frac{k_{\rm reh}}{k}\right)e^{-(N-N_{\rm reh})}\cos\left((1-e^{N-N_{\rm reh}})\left(\tfrac{k}{k_{\rm reh}}\right)\right)\,, & k/aH\gg1 \,.
\end{cases}
\end{align}
As demonstrated explicitly below, the radiation contribution becomes dominant in scenarios with a late-time decaying spectator field $\chi$. When this decay occurs at an energy density ratio $f_{\chi}$, we can significantly simplify the background functions in Eq.~(\ref{eq:Jrads}) by using the standard radiation-dominated universe relations:
\beq
\left(\frac{a_{\rm reh}H_{\rm reh}}{a(N)H(N)}\right)  \left(\frac{\rho_{\chi,{\rm NR}}}{H^2_{\rm NR} M_P^2}\right) \;=\; \frac{3}{2}f_{\chi}\,,
\eeq
where we used $H\propto \rho_R^{1/2}\propto a^{-2}$ and $\rho_{\chi}\propto a^{-3}$. Substituting this result into Eq.~(\ref{eq:Jrads}) leads to
\begin{align} 
\mathcal{J}_{\rm rad}(k,N) \;&=\; \frac{3f_{\chi}}{16\sqrt{6}} \left(\frac{k_{\rm end}}{k}\right)  \left(\frac{a_{\rm reh}H_{\rm reh}}{a_{\rm end}H_{\rm end}}\right)  \left(\frac{\rho_{\chi,{\rm NR}}}{H^2_{\rm NR} M_P^2}\right)  \gamma(N;k) \\ 
&=\;  \frac{\sqrt{3} f_{\chi}}{16} \left(\frac{k_{\rm end}}{k}\right)   \left(\frac{\rho_{\rm reh}}{\rho_{\rm end}}\right)^{1/6}  \label{eq:Jradf}
\left(\frac{\rho_{\chi,{\rm NR}}}{H^2_{\rm NR} M_P^2}\right)  \gamma(N;k) \,.
\end{align}
For super-horizon modes at the time of decay, this contribution then scales as $\sim T_{\rm reh}^{2/3}$.

\begin{figure}[!t]
\centering
    \includegraphics[width=0.8 \textwidth]{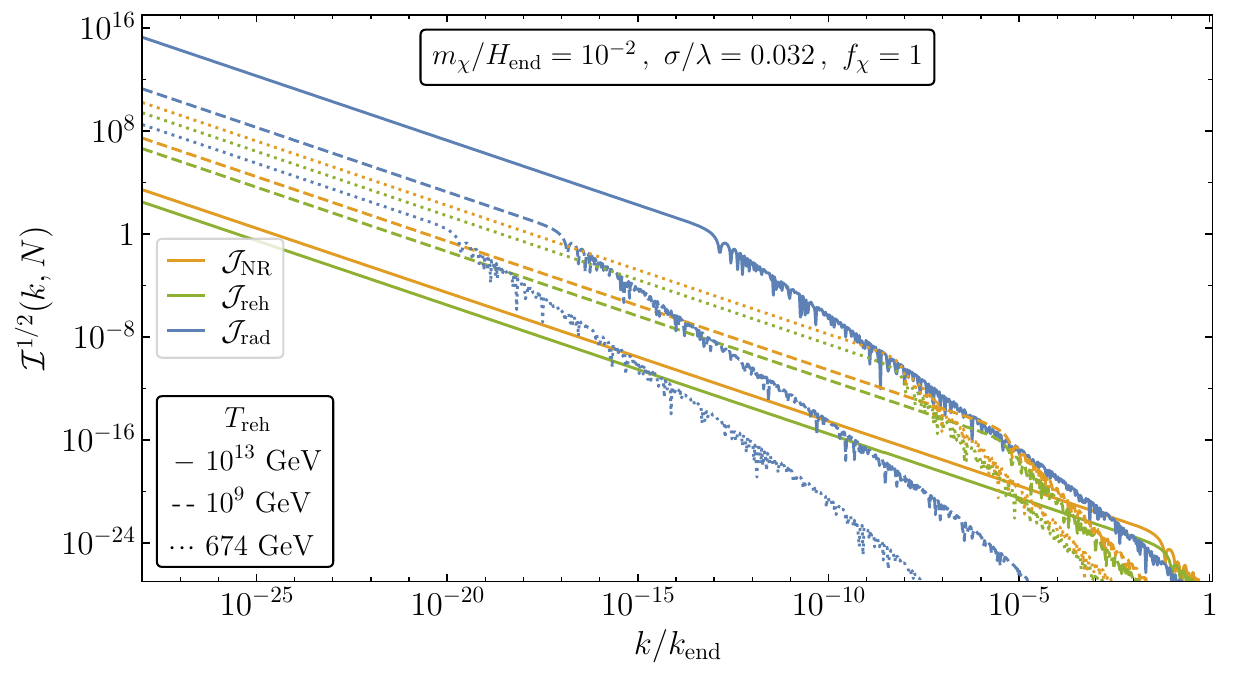}
    \caption{Contributions to the time-dependent factor $\mathcal{I}(k,N)$ of the GW energy spectrum,  evaluated at the present epoch for three representative reheating temperatures and $f_{\chi}(t_{\rm d})=1$. The plot shows the relative magnitudes of the early-time contribution ($\mathcal{J}_{\rm NR}$, occurring before isocurvature freeze-out), the reheating contribution ($\mathcal{J}_{\rm reh}$), and the radiation contribution ($\mathcal{J}_{\rm rad}$, from the subsequent radiation-dominated era). For the scenario with the lowest reheating temperature $T_{\rm reh}$ shown, a stable spectator field would produce precisely the observed DM relic abundance.}
    \label{fig:I1}
\end{figure}

Fig.~\ref{fig:I1} illustrates the scale dependence of the three contributions to $\mathcal{I}(k,N)$ in Eq.~(\ref{eq:I1}), namely $\mathcal{J}_{\rm NR}$, $\mathcal{J}_{\rm reh}$, $\mathcal{J}_{\rm rad}$. To demonstrate their relative significance, we fix the spectator field parameters at $m_{\chi}/H_{\rm end}=10^{-2}$ and $\sigma/\lambda=0.032$ and set the spectator field to decay at its latest possible time with $f_{\chi}=1$. We examine three representative reheating temperatures, $T_{\rm reh}=10^{13}\,{\rm GeV}$ (high reheating temperature where $N_{\rm reh}\gtrsim N_{\rm NR}$), $T_{\rm reh}=10^{9}\,{\rm GeV}$ (intermediate value with $N_{\rm reh}\gg N_{\rm NR}$), and $T_{\rm reh}=674\,{\rm GeV}$ (the specific temperature at which a stable 
$\chi$ would precisely account for the observed DM relic abundance). Across all scenarios, we observe that $\mathcal{J}_{\rm NR}\gtrsim\mathcal{J}_{\rm reh}$, with the early-time contribution exceeding the reheating contribution by a factor of $6-9$. This factor exhibits a weak (increasing) dependence on $T_{\rm reh}$ and emerges from our numerical evaluation of $\mathcal{J}_{\rm NR}$ through the complex inflation-reheating-NR transition. The consistency of this proportionality factor indicates that our analytical insights regarding $\mathcal{J}_{\rm reh}$ effectively extend to the early-time contribution as well.

The $k$-dependence of each contribution is clearly visible in Fig.~\ref{fig:I1}. For $\mathcal{J}_{\rm reh}$, we observe two distinct scaling regimes separated by $k_{\rm reh}$, the horizon scale at reheating. Modes that remain super-horizon at the end of reheating exhibit the $k^{-1}$ scaling derived in Eq.~(\ref{eq:Jrehf}), with amplitude proportional to $T_{\rm reh}^{-2/3}$. Modes that enter the horizon during reheating experience stronger suppression, scaling as $k^{-3}$, as verified through simplification of the matter-dominated Green's function. Importantly, provided $\chi$ does not decay before $N_{\rm reh}$, the GW spectrum generated from $\mathcal{J}_{\rm NR}+\mathcal{J}_{\rm reh}$ remains independent of the post-reheating expansion history, making these results equally applicable to scenarios with a stable spectator field.

The late-time radiation contribution $\mathcal{J}_{\rm rad}$ is shown in Fig.~\ref{fig:I1} (blue curves). It also exhibits two scaling regimes, now separated by the horizon scale at $\chi$ decay, $k_{\rm d}=a(N_{\rm d})H(N_{\rm d})$ (see Fig.~\ref{fig:rhos}). Modes that remain super-horizon at decay time scale as $k^{-1}$, consistent with Eqs.~(\ref{eq:gammaapp}) and (\ref{eq:Jradf}). Sub-horizon modes follow a steeper $k^{-2}$ in addition to an oscillatory features. Importantly, the amplitude of this contribution to $\Omega_{{\rm GW},\mathcal{S}}$ increases with reheating temperature as $T_{\rm reh}^{2/3}$, as discussed above, making $\mathcal{J}_{\rm rad}$ dominant for high reheating temperatures (e.g.~$T_{\rm reh}=10^{13}$ and $10^9$ GeV in the figure) but subdominant to $\mathcal{J}_{\rm NR}$ and $\mathcal{J}_{\rm rad}$ at lower temperatures. Consequently, for the dark matter scenario depicted in Fig.~\ref{fig:I1} (and all other cases examined in the following section), the GW spectrum depends exclusively on the evolution of the spectator field and background dynamics up to the end of reheating.

\section{Gravitational Wave Signals}
\label{sec:gwsignals}

\subsection{Observational Prospects and Constraints}
\label{sec:gwmissions}
The stochastic GW background generated by spectator fields provides a promising target for current and future GW observatories across a wide frequency spectrum. We summarize here the key observational windows and constraints relevant to our predicted signals.

The integrated GW energy density is bounded by constraints on additional relativistic degrees of freedom, parameterized by 
$\Delta N_{\text{eff}}$~\cite{Luo:2020fdt, Maggiore:1999vm}:
\begin{equation}
    \int_{f_{\text{min}}}^{\infty} \frac{df}{f} \Omega_{\rm GW}(f) h^2 \leq 5.6     \times 10^{-6} \Delta N_{\text{eff}} \, .
\end{equation}
Current bounds from BBN ($\Delta N_{\mathrm{eff}}^{\mathrm{BBN}} \simeq 0.4$)~\cite{Cyburt:2015mya} and \textit{Planck}+BAO ($\Delta N_{\mathrm{eff}}^{\mathrm{Planck+BAO}} \simeq 0.28$)~\cite{Planck:2018vyg} constrain the integrated GW background. Future CMB experiments will significantly improve these limits, with CMB-HD targeting  $\Delta N_{\mathrm{eff}}^{\mathrm{Proj.}} = 0.014$~\cite{CMB-HD:2022bsz}, and other missions such as PICO~\cite{Alvarez:2019rhd} and LiteBIRD~\cite{Hazumi:2019lys} providing complementary constraints. CMB spectral distortion experiments including PIXIE and VOYAGE 2050~\cite{Kogut:2019vqh, Sesana:2019vho}  will probe energy injections in the early universe at frequencies $10^{-15} - 10^{-9} \, \mathrm{Hz}$, offering sensitivity to the spectator-induced GW background in regimes inaccessible to interferometers.

In the nHz regime ($10^{-9} - 10^{-7} \, \mathrm{Hz}$), 
Pulsar Timing Arrays (PTAs) including NANOGrav~\cite{NANOGrav:2023hvm}, EPTA~\cite{EPTA:2015qep, EPTA:2015gke}, and future SKA observations~\cite{Weltman:2018zrl} provide sensitivity to the gravitational wave backgrounds potentially generated by our mechanism. These observations are complemented by astrometric missions like Gaia and THEIA, which effectively transform the Milky Way into a galactic-scale detector~\cite{Garcia-Bellido:2021zgu}.

At lower frequencies ($10^{-7} - 10^{-4} \, \mathrm{Hz}$), 
proposed space-based observatories such as $\mu$-Ares~\cite{Sesana:2019vho} would explore the decihertz window with unprecedented sensitivity, potentially accessing gravitational wave signals from intermediate-mass black hole mergers and early universe physics, including our spectator field mechanism. Mid-frequency detectors ($10^{-4} - 1 \, \mathrm{Hz}$) such as AION~\cite{Badurina:2021rgt} and AEDGE~\cite{AEDGE:2019nxb} 
will bridge the sensitivity gap between PTAs and space-based interferometers. At millihertz frequencies, space-based interferometers including LISA~\cite{LISA:2017pwj}, BBO~\cite{Crowder:2005nr}, DECIGO and Ultimate-DECIGO~\cite{Yagi:2011wg} will target regions where many of our predicted signals peak.

At higher frequencies ($1 - 10^4 \, \mathrm{Hz}$), 
ground-based detectors including Advanced LIGO/Virgo~\cite{LIGOScientific:2019lzm} 
and future facilities such as Einstein Telescope (ET)~\cite{Punturo:2010zz} 
and Cosmic Explorer (CE)~\cite{Reitze:2019iox} extend coverage into the kilohertz regime. For ultra-high frequencies ($> 10^5 \, \mathrm{Hz}$), resonant cavity experiments provide unique sensitivity windows for GWs produced by the highest-energy processes~\cite{Herman:2022fau, Aggarwal:2020olq, Ringwald:2022xif}.

\subsection{Curvaton-like scenarios with decaying spectator fields}

Eq.~(\ref{eq:gwiso}) factorizes the GW spectrum into the momentum integral $g(k)$ and the time-evolution factor $\mathcal{I}(k,N)$. While both functions require numerical evaluation in general, Section~\ref{sec:gwiso} provides accurate analytical approximations for practical calculations.

First, consider an unstable spectator field $\chi$ with high reheating temperature. As established in Section~\ref{sec:Jrad}, the radiation contribution $\mathcal{J}_{\rm rad}$ dominates the GW signal, yielding:
\begin{equation}
\label{eq:GWrad}
\Omega_{{\rm GW},\mathcal{S}} \;\simeq\; g(k) \mathcal{J}_{\rm rad}^{2}(k,N) \; \simeq\; \frac{3 f_{\chi}^2}{256} \tilde{g}(\Lambda) \gamma^2(N;k)  \left(\frac{\rho_{\rm reh}}{\rho_{\rm end}}\right)^{1/3} \left(\frac{\rho_{\chi,{\rm NR}}}{H^2_{\rm NR} M_P^2}\right)^2   \left(\frac{k_{\rm end}}{k}\right)^2 \left(\Delta_{\mathcal{S}}^2(k)\right)^2\,.
\end{equation}
Using approximations~(\ref{eq:PSan2}) and (\ref{eq:gammaapp}), this result implies that
\beq
\Omega_{{\rm GW},\mathcal{S}} \;\propto\; \begin{cases}
k^{4\Lambda-2}\ln^2(k/H_I)\,, & k/a_{\rm d} H_{\rm d} \ll 1 \, ,\\
k^{4\Lambda-4}\ln^2(k/H_I)\,, & k/a_{\rm d} H_{\rm d} \gg 1 \,,
\end{cases}
\eeq
with a prefactor that scales with the reheating temperature as $T_{\rm reh}^{4/3}$.

\subsubsection{$\sigma = 0$}
\begin{figure*}[!t]
\centering
    \includegraphics[width=0.7 \textwidth]{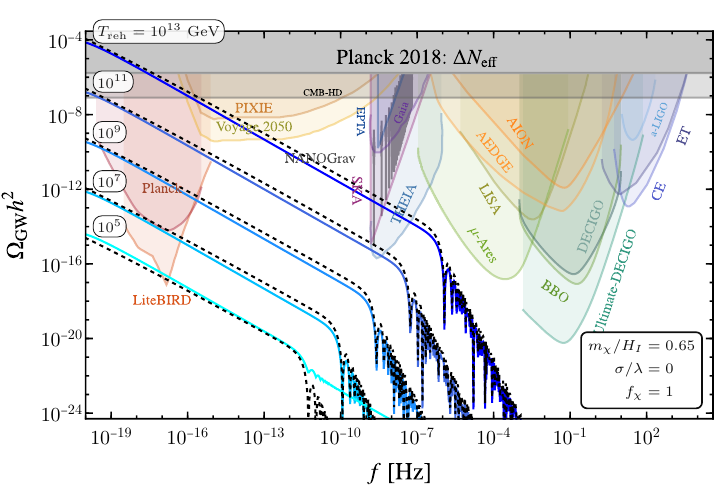}
    \caption{Gravitational wave spectrum sourced by the isocurvature fluctuations of the unstable spectator scalar field $\chi$, which decays right before matter-radiation equality, $f_{\chi}(t_{\rm d})=1$. Here we fix the spectator mass in an instance of pure gravitational production, and vary the reheating temperature in the range where the radiation contribution $\mathcal{J}_{\rm rad}$ is dominant. The continuous curves are the exact numerical evaluation of $\Omega_{{\rm GW},\mathcal{S}}$. The dashed curves correspond to the analytical approximation (\ref{eq:GWrad}).}
    \label{fig:GW_Tm_rad}
\end{figure*}

Fig.~\ref{fig:GW_Tm_rad} presents the GW spectrum for a heavy spectator field with $m_{\chi}/H_I=0.65$ and purely gravitational production ($\sigma=0$). The solid curves represent our full numerical calculations, focusing on scenarios where the radiation contribution $\mathcal{J}_{\rm rad}$ dominates and $f_{\chi}=1$. To compare with current and planned GW detectors, we have translated the scale dependence into frequency using the relation
\begin{align}
    f \; = \; \frac{1}{2\pi}\left(\frac{k_{\rm end}}{\rm Hz}\right)\left(\frac{k}{k_{\rm end}}\right)\,{\rm Hz} \; \simeq\; 2.43\times 10^7\, \left(\frac{k}{k_{\rm end}}\right)\,{\rm Hz}\;\simeq\; 2.43\times 10^7\,\left(\frac{k}{1.6\times 10^{22}\ {\rm Mpc}^{-1}}\right)\, {\rm Hz}\,.
\end{align}
For these parameter choices, the radiation-dominated regime requires $T_{\rm reh}\gtrsim (493\,{\rm GeV})(\mathcal{J}_{\rm NR}/\mathcal{J}_{\rm reh})^{3/4}$. The black dashed curves show the analytical approximation from Eq.~(\ref{eq:GWrad}), which accurately reproduces the numerical results with most discrepancies stemming from the approximation in Eq.~(\ref{eq:guv}) for the isocurvature-dependent factor $g(k)$ in $\Omega_{{\rm GW},\mathcal{S}}$.

We examine five representative reheating temperatures: $T_{\mathrm{reh}} = \{10^{13}, 10^{11}, 10^9, 10^7, 10^5\} \, \mathrm{GeV}$, for which super-horizon modes at decay are dominated by the radiation contribution, shown in Fig.~\ref{fig:GW_Tm_rad}. The results clearly follow the predicted $\sim T_{\mathrm{reh}}^{4/3}$ scaling and closely match the analytical expressions for super-horizon scales. For this particular case, $\Lambda = 0.29$, corresponding to a red-tilted spectrum $\Omega_{\mathrm{GW,S}} \propto f^{4\Lambda-2}\simeq f^{-0.83}$ with an infrared cutoff at the scale that exits the horizon at the beginning of inflation. This red tilt makes the model most stringently constrained at large scales by CMB observations. The highest reheating temperature ($T_{\mathrm{reh}} = 10^{13} \, \mathrm{GeV}$) produces a spectrum that exceeds current bounds on additional relativistic degrees of freedom from \textit{Planck}+BAO measurements~\cite{Planck:2018vyg}. For $T_{\mathrm{reh}} = 10^{11} \, \mathrm{GeV}$, the model remains viable under current constraints but could be tested by forthcoming measurements of $\Delta N_{\mathrm{eff}}$ from experiments such as CMB-HD~\cite{CMB-HD:2022bsz}. Interestingly, at $T_{\mathrm{reh}} = 10^9 \, \mathrm{GeV}$, the secondary tensor spectrum would still be in tension with \textit{Planck} non-detection  of primordial tensor modes~\cite{Planck:2018jri}. Only for $T_{\mathrm{reh}} \lesssim 10^7 \, \mathrm{GeV}$ does the model satisfy all current observational constraints, while remaining potentially detectable by next-generation CMB observatories such as LiteBIRD~\cite{Hazumi:2019lys} for $T_{\mathrm{reh}} \gtrsim 10^5 \, \mathrm{GeV}$.

Examining the high-frequency behavior in Fig.~\ref{fig:GW_Tm_rad}, we observe that the spectral tilt steepens significantly for modes that were already sub-horizon at decay time, particularly for the higher reheating temperatures. This behavior is consistent with our analytical approximation, which predicts $\Omega_{\mathrm{GW,\mathcal{S}}} \propto f^{-2.82}$ in this regime. Consequently, these spectra fall below the sensitivity thresholds of all proposed detectors for $f \gtrsim 10^{-6} \, \mathrm{Hz}$. This steep suppression in Fig.~\ref{fig:GW_Tm_rad} characterizes all cases except the lowest temperature scenario ($T_{\mathrm{reh}} = 10^5 \, \mathrm{GeV}$). For this case, while the infrared spectrum remains dominated by the radiation contribution $\mathcal{J}_{\mathrm{rad}}$, the spectrum at $k > k_d$ is instead dominated by the early-time contribution $\mathcal{J}_{\mathrm{NR}}$, which maintains the same spectral tilt as the infrared modes. This transition in the dominant contribution explains why the analytical expression in Eq.~(\ref{eq:GWrad}) fails to accurately describe the spectrum in this regime. It is worth noting that the gravitational wave spectra for scenarios with earlier spectator field decay can be straightforwardly obtained by scaling these results by a factor of $f_\chi^2 < 1$, reflecting the reduced energy density available for gravitational wave production when the spectator field decays before achieving its maximum possible energy density fraction.

\begin{figure*}[!t]
\centering
    \includegraphics[width=0.7 \textwidth]{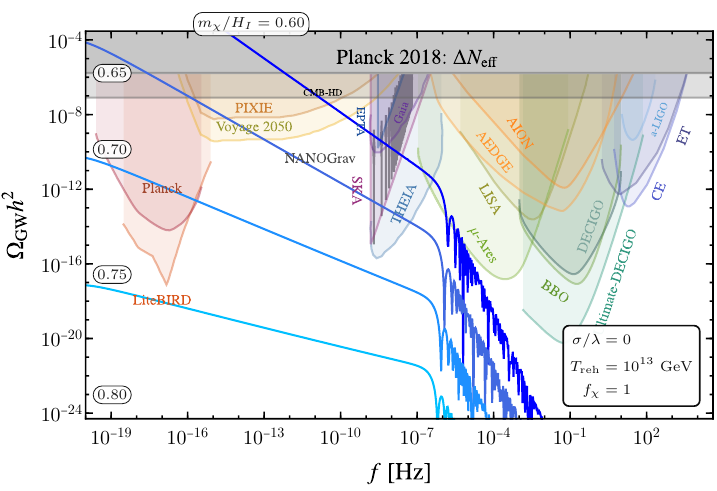}
    \caption{Gravitational wave spectrum sourced by the isocurvature fluctuations of the unstable spectator scalar field $\chi$, which decays right before matter-radiation equality, $f_{\chi}(t_{\rm d})=1$. We fix here the reheating temperature and vary the spectator mass, assuming pure gravitational production.}
    \label{fig:GW_mT_rad}
\end{figure*}

Figure~\ref{fig:GW_mT_rad} shows how the GW spectrum depends on spectator field mass at fixed reheating temperature $T_{\mathrm{reh}} = 10^{13} \, \mathrm{GeV}$. Larger spectator masses suppress the isocurvature power spectrum at large scales, reducing the gravitational wave signal. As $m_\chi$ increases, the spectral tilt of $\Delta_{\mathcal{S}}^2$ increases, making $\Omega_{\mathrm{GW},\mathcal{S}}$ less red-tilted. The GW spectrum approaches scale invariance for $m_\chi / H_I \simeq 1.28$, though with amplitudes far below experimental sensitivities. Notably, for purely gravitational production ($\sigma = 0$), any signals potentially detectable by space-based interferometers or PTAs are already excluded by constraints on $\Delta N_{\mathrm{eff}}$ and CMB observations.

\begin{figure*}[!t]
\centering
    \includegraphics[width=0.8 \textwidth]{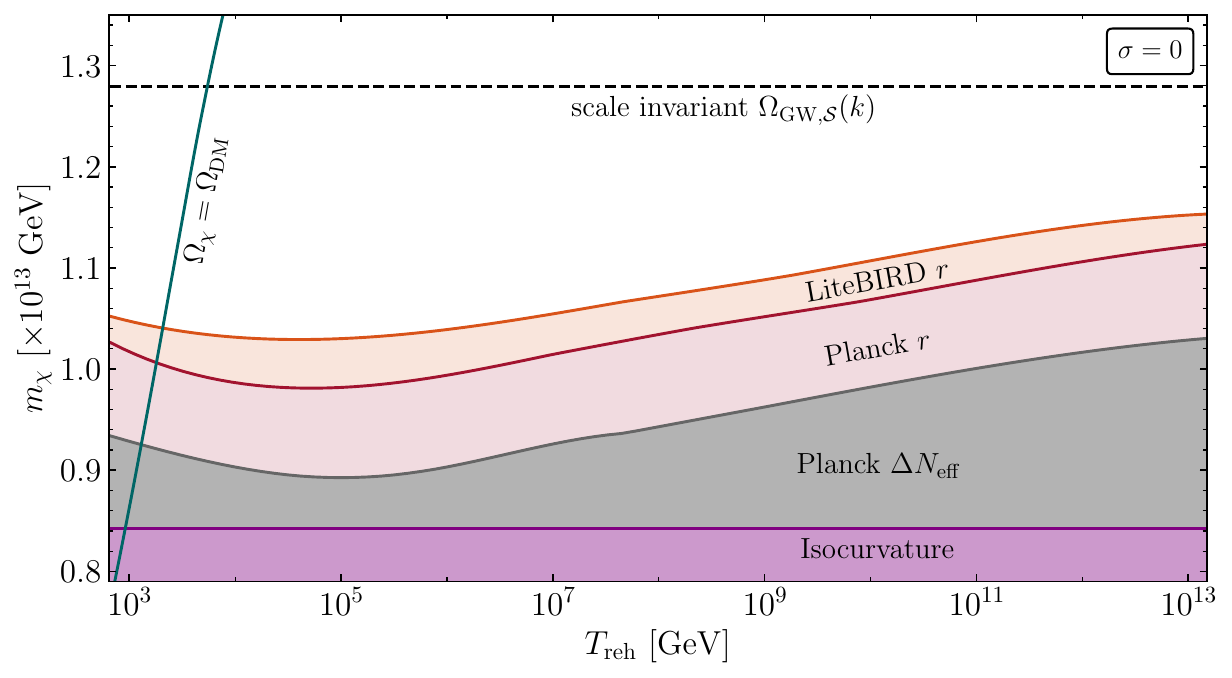}
    \caption{Parameter space constraints on purely gravitational production of heavy spectator fields, assuming $f_{\chi}(t_{\rm d})=1$ and considering only the spectrum inside our present horizon with $N_{\rm tot}=76.5$. The purple region indicates exclusion from CMB isocurvature constraints, which require $m_{\chi} \gtrsim 0.54 H_I \simeq 8.4 \times 10^{12} \, \rm{GeV}$. The gray region shows exclusion from $\Delta N_{\rm eff}$ bounds, ruling out combinations of small masses and high reheating temperatures that produce excessive gravitational wave backgrounds. The red region represents constraints from Planck/BICEP2/Keck non-observation of primordial B-mode polarization, excluding most masses below $10^{13} \, \rm{GeV}$ except at very low reheating temperatures. The orange shaded area indicates the projected sensitivity of LiteBIRD. The teal curve denotes the dark matter relic abundance boundary, where spectator fields with parameters to the right (left) produce $\Omega_{\chi}h^2>\Omega_{\rm DM}h^2=0.12$ ($<\Omega_{\rm DM}h^2$).}
    \label{fig:GW_parspace}
\end{figure*}

Fig.~\ref{fig:GW_parspace} summarizes our findings in the $m_{\chi}$-$T_{\rm reh}$ parameter space for the gravitational production scenario, with both parameters expressed in GeV. As discussed in Section~\ref{sec:isocurvaturefluct}, spectator field masses below $m_{\chi}\lesssim 0.54 H_I\simeq 8.4\times 10^{12}\,{\rm GeV}$ are excluded by CMB isocurvature constraints (purple region). Our gravitational wave analysis provides additional constraints, which show that small masses combined with high reheating temperatures produce spectra conflicting with $\Delta N_{\rm eff}$ bounds (gray region), with the exclusion boundary exhibiting the expected behavior down to $T_{\rm reh} \simeq 10^{5}\,{\rm GeV}$. For reheating temperatures below this threshold, the gravitational wave signal transitions from being dominated by the late-time radiation contribution $\mathcal{J}_{\rm NR}$, which increases with decreasing reheating temperature. The teal line marks where $\chi$ produces the observed dark matter abundance. Section~\ref{sec:DMsc} analyzes the corresponding relic abundance and early-time gravitational wave contribution.

Above the $\Delta N_{\rm eff}$ exclusion region in Fig.~\ref{fig:GW_parspace}, we show constraints from the non-observation of primordial B-mode polarization by {\em Planck}~\cite{Planck:2018jri} and BICEP2/Keck Array~\cite{BICEP:2021xfz}. The resulting exclusion region (red) rules out masses $m_{\chi}\lesssim 0.64 H_I\simeq 10^{13}\,{\rm GeV}$, with the exception of very low reheating temperatures $\mathcal{O}({\rm TeV})$ where the GW signal receives comparable contributions from all terms in $\mathcal{I}(k,N)$. Spectator field masses exceeding this threshold remain viable under current constraints but could be tested by forthcoming CMB observatories such as LiteBIRD~\cite{Hazumi:2019lys} (orange region).
The GW constraints presented here are derived assuming sub-horizon observations and a fixed reheating duration. For extended inflation scenarios, the spectra would exhibit extended infrared coverage reaching down to the inflationary cutoff scale, potentially leading to problematically large integrated energy densities. These issues are naturally avoided in models producing blue-tilted spectra for scales $k<k_{\rm d}$, which occurs when  $m_{\chi}\gtrsim0.82H_I\simeq 1.28\times 10^{13}\,{\rm GeV}$, as shown in the figure.

\subsubsection{$\sigma \neq 0$}

\begin{figure*}[!t]
\centering
    \includegraphics[width=0.7 \textwidth]{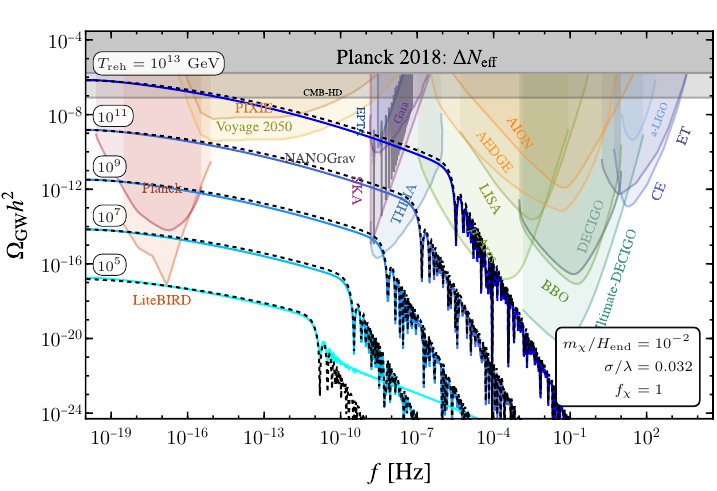}
    \caption{Gravitational wave spectrum sourced by the isocurvature fluctuations of the unstable spectator scalar field $\chi$, which decays right before matter-radiation equality, $f_{\chi}(t_{\rm d})=1$. We fix the spectator coupling and mass, and vary the reheating temperature in the range where the radiation contribution $\mathcal{J}_{\rm rad}$ is dominant. The continuous curves are the exact numerical evaluation of $\Omega_{{\rm GW},\mathcal{S}}$. The dashed curves correspond to the analytical approximation (\ref{eq:GWrad}).}
    \label{fig:GW_Ts_rad}
\end{figure*}

We now consider the scenario in which the super-horizon growth of the spectator mode functions during inflation is suppressed by the direct coupling between $\phi$ and $\chi$ with effective strength $\sigma/\lambda\neq 0$. Fig.~\ref{fig:GW_Ts_rad} shows the numerically computed $\Omega_{{\rm GW},\mathcal{S}}$ for the case of a light spectator field ($m_{\chi}/H_{\rm end}=10^{-2}$) and an inflaton coupling $\sigma/\lambda=0.032$. Here, as in the previous section, we assume that $\mathcal{J}_{\rm rad}$ dominates the GW spectrum with $f_{\chi}=1$, which for the chosen parameter values requires that $T_{\rm reh}\gtrsim (1090\,{\rm GeV})(\mathcal{J}_{\rm NR}/\mathcal{J}_{\rm reh})^{3/4}$. We use the same range of reheating temperatures as in Fig.~\ref{fig:GW_Tm_rad}. The analytical approximation (\ref{eq:GWrad}), shown as the dashed curves, closely resembles the exact result. Importantly, it accurately reproduces the spectral tilt, which for $\sigma\neq 0$ is a function of the frequency (scale), via (\ref{eq:nu}) and (\ref{eq:phik}).

For the choice of parameters in Fig.~\ref{fig:GW_Ts_rad} we note that, despite the red tilt of the GW spectrum, the amplitude does not reach the range constrained by the current observational bounds on $\Delta N_{\rm eff}$, although the case with $T_{\rm reh}=10^{13}\,{\rm GeV}$ could be ruled out by these considerations alone in the near future. However, the existence of the spectator field would be ruled out for $T_{\rm reh}\gtrsim 3\times 10^{7}\,{\rm GeV}$ by the non-detection of primordial tensors in the CMB polarization spectra. Temperatures $\gtrsim 10^{5}\,{\rm GeV}$ could be probed by future CMB observatories. Similar to the case explored in Fig.~\ref{fig:GW_Tm_rad}, the curve with the smallest reheating temperature displays a UV tail originating from the early-time contribution $\mathcal{J}_{\rm NR}$, not captured by the estimate (\ref{eq:GWrad}). 

\begin{figure*}[!t]
\centering
    \includegraphics[width=0.7 \textwidth]{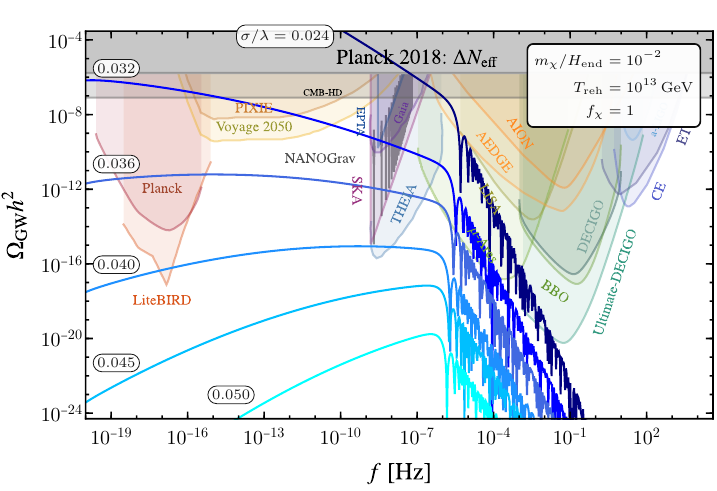}
    \caption{Gravitational wave spectrum sourced by the isocurvature fluctuations of an unstable spectator field, which decays right before matter-radiation equality. Here we fix the spectator mass and the reheating temperature, varying the effective inflaton coupling strength.}
    \label{fig:GW_sT_rad}
\end{figure*}

Fig.~\ref{fig:GW_sT_rad} shows in detail the dependence of $\Omega_{{\rm GW},\mathcal{S}}$ on the effective inflaton coupling $\sigma/\lambda$ for a fixed spectator mass and decay density ratio $f_{\chi}$. We observe that in a narrow range of couplings, $0.024\lesssim \sigma/\lambda\lesssim 0.050$, the spectrum transitions from having a steep red tilt to a blue tilt for super-horizon modes at the time of $\chi$ decay. The spectra for the two smallest couplings shown are red-tilted in the frequency range considered and are ruled out by energy density and CMB (non)observations. The couplings $\sigma/\lambda=0.036$ and $0.040$ present a non-monotonic behavior, with a transition between blue and red tilts for super-horizon scales at $t_{\rm d}$. The first coupling appears to conflict with non-detection of primordial tensor modes at CMB scales. However, the second coupling not only avoids this constraint but presents a potentially detectable spectrum for upcoming CMB observatories, PTAs, and large stellar surveys. Finally, the largest two couplings in the figure present a blue-tilted spectrum at low frequencies, with a well-defined peak at $f\simeq 10^{-6}\,{\rm Hz}$ that lies below all experimental sensitivity curves.

\begin{figure*}[!t]
\centering
    \includegraphics[width=0.7 \textwidth]{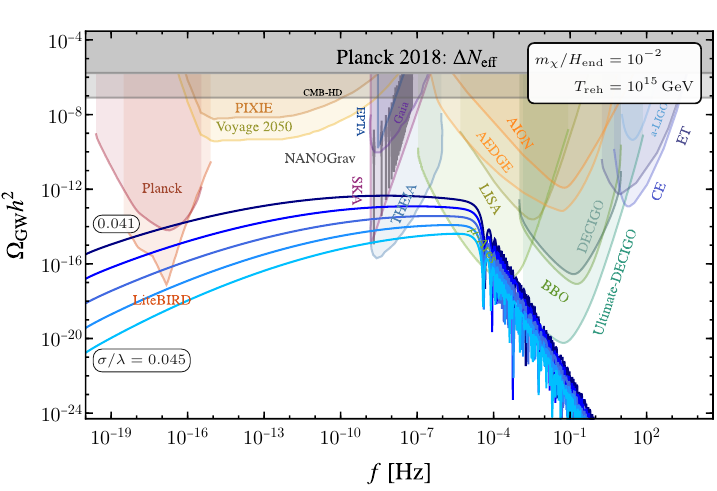}
    \caption{Gravitational wave spectrum sourced by the isocurvature fluctuations of an unstable spectator field, which decays right before matter-radiation equality, in a high reheating scenario.}
    \label{fig:GW_inst}
\end{figure*}

Fig.~\ref{fig:GW_inst} depicts an interesting parameter space region. For nearly instantaneous reheating with $T_{\rm reh}=10^{15}\,{\rm GeV}$ and $m_{\chi}/H_{\rm end}=10^{-2}$, the effective inflaton coupling range $0.041\lesssim \sigma/\lambda\lesssim 0.044$ produces GW spectra with correlated signatures detectable by CMB experiments, PTAs, stellar surveys, and space-based interferometers. 

\subsection{Dark Matter Scenarios}
\label{sec:DMsc}
In the previous section, we focused on GW spectra from spectator field fluctuations at large reheating temperatures. Due to the scalings $\mathcal{J}_{\rm NR,reh}\propto T_{\rm reh}^{-2/3}$ and $\mathcal{J}_{\rm rad}\propto T_{\rm reh}^{2/3}$, the late-time radiation contribution dominates the GW signal in that regime. We now consider the low-reheating temperature scenario, where the signal is dominated by the earliest contribution $\mathcal{J}_{\rm NR}$, which freezes simultaneously with the isocurvature power spectrum. As discussed in Section~\ref{sec:Jrad}, this contribution lacks an analytical characterization because it depends on the transition from accelerated expansion to matter domination after inflation, as well as the asymptotic oscillatory freeze-out of the isocurvature spectrum. Both effects must be characterized numerically. Nevertheless, as shown explicitly in Fig.~\ref{fig:I1}, the approximation $\mathcal{J}_{\rm NR}\propto\mathcal{J}_{\rm reh}$ remains valid. Therefore, our analytical approximation for the low $T_{\rm reh}$ scenario becomes
\begin{align} \notag
\Omega_{{\rm GW},\mathcal{S}} \; &\simeq \; g(k) \mathcal{J}_{\rm NR}^{2}(k,N)  \\  
\label{eq:GWnr}
 \; & \simeq\; \frac{ 1}{3456} \left(\frac{\mathcal{J}_{\rm NR}}{\mathcal{J}_{\rm reh}}\right)^2 \tilde{g}(\Lambda) e^{-2(N_{\rm NR}-N_{\rm end})} \left(\frac{\rho_{\rm end}}{\rho_{\rm reh}}\right)^{1/3} \left(\frac{\rho_{\chi,{\rm NR}}}{H^2_{\rm NR} M_P^2}\right)^4   \left(\frac{k_{\rm end}}{k}\right)^2 \left(\Delta_{\mathcal{S}}^2(k)\right)^2\,,
\end{align}
valid only for super-horizon modes at the end of reheating, $k\ll a_{\rm reh} H_{\rm reh}$. For sub-horizon modes, we can obtain an approximation to the spectral tilt from the integrated matter-dominated tensor Green's function in Eq.~(\ref{eq:Jrehs}). Combining these results, the low-reheating spectrum takes the form
\beq
\Omega_{{\rm GW},\mathcal{S}} \;\propto\; \begin{cases}
k^{4\Lambda-2}\ln^2(k/H_I)\,, & k/a_{\rm reh} H_{\rm reh} \ll 1 \, ,\\
k^{4\Lambda-6}\ln^2(k/H_I)\,, & k/a_{\rm reh} H_{\rm reh} \gg 1 \, .
\end{cases}
\eeq
That is, the same tilt for the curvaton-like scenario at large scales, which extends up to the reheating horizon scale, after which the suppression is stronger than in the curvaton-like scenario. 

The spectrum (\ref{eq:GWnr}) is independent of the details of the stability of $\chi$, as long as its decay occurs after reheating if $\chi$ is unstable. This contribution will therefore be the relevant one in the case of a stable $\chi$, case for which we assume it comprises the DM relic abundance of the universe. We can evaluate this density as follows,
\begin{align}
\Omega_{\rm DM}h^2 \; =\; \frac{\rho_{\chi}(a_0)}{\rho_c h^{-2}}\;=\; \frac{\rho_{\rm reh}}{\rho_c h^{-2}} \left(\frac{\rho_{\chi,{\rm reh}}}{\rho_{\rm reh}}\right)\left(\frac{a_{\rm reh}}{a_0}\right)^3\,,
\end{align}
where $\rho_c$ denotes the critical energy density at the present time. Assuming a standard thermal history after the end of reheating, this expression can be rewritten as
\begin{align}
\Omega_{\rm DM}h^2 \;=\; \left(\frac{43 \pi^2}{330}\right) \frac{T_{\rm reh} T_0^3}{\rho_c h^{-2}} \left(\frac{\rho_{\chi,{\rm reh}}}{\rho_{\rm reh}}\right) \;\simeq\; \left(\frac{43 \pi^2}{990}\right) \frac{T_{\rm reh} T_0^3}{\rho_c h^{-2}} \left(\frac{\rho_{\chi,{\rm NR}}}{H^2_{\rm NR} M_P^2}\right)\,,
\end{align}
with $T_0=2.7255\,{\rm K}$ and $\rho_c h^{-2} = 8.1\times 10^{-47}\,{\rm GeV}^4$~\cite{Fixsen:2009ug,Planck:2018vyg}. For the scenarios considered here $\rho_{\chi,{\rm NR}}\sim 10^{-12} H^2_{\rm NR} M_P^2$ (see Fig.~\ref{fig:rhos}), leading to $T_{\rm reh}=\mathcal{O}({\rm TeV})$, precisely in the range of $\mathcal{J}_{\rm NR}$ domination (see Fig.~\ref{fig:GW_parspace}). We now present the result of the evaluation of the GW signal in the case of pure gravitational production, and of inflaton interaction. 

\subsubsection{$\sigma = 0$}

\begin{figure*}[!t]
\centering
    \includegraphics[width=0.7 \textwidth]{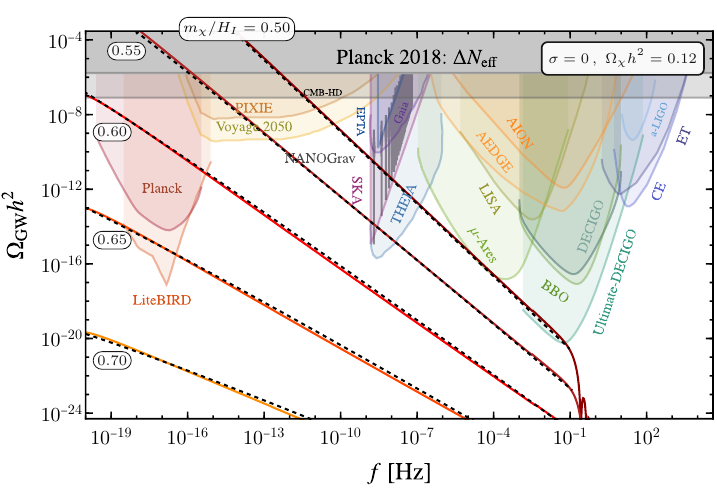}
    \caption{Gravitational wave spectrum sourced by the isocurvature fluctuations of a stable spectator scalar field $\chi$, which comprises the entirety of the DM, assuming pure gravitational production. Shown as the dashed black line is the analytical approximation (\ref{eq:GWnr}), for which $\mathcal{J}_{\rm NR}/\mathcal{J}_{\rm reh}\simeq 110-120$.}
    \label{fig:GW_DM_m}
\end{figure*}

Fig.~\ref{fig:GW_DM_m} shows the result of the numerical evaluation of $\Omega_{{\rm GW},\mathcal{S}}$ in the case of pure gravitational DM production, fixing the DM relic abundance to its observed value $\Omega_{\rm DM}h^2 = 0.1198 \pm 0.0012$~\cite{Planck:2018vyg}. For the masses $m_{\chi}/H_I=\{0.50,0.55,0.60,0.65,0.70\}$ the corresponding reheating temperatures are $T_{\rm reh}=\{700,982,1360,1867,2546\}\,{\rm GeV}$. The dashed black curves show the analytical approximation~(\ref{eq:GWnr}) for modes with $k<k_{\rm reh}$. Numerically, we find $\mathcal{J}_{\rm NR}/\mathcal{J}_{\rm reh}\simeq 110-120$, yielding good agreement with the full computation of $\Omega_{{\rm GW},\mathcal{S}}$.

Similar to the curvaton scenario, the GW spectrum exhibits a red tilt controlled by $\Lambda={\rm const.}$ for all scales with $k\ll k_{\rm reh}$, as shown in Fig.~\ref{fig:GW_DM_m}. Consequently, the strongest constraints arise at large scales. Masses below $m_{\chi}\simeq 0.59 H_I\simeq 9.3\times10^{12}\,{\rm GeV}$ violate $\Delta N_{\rm eff}$ bounds and \textit{Planck} tensor mode constraints, as shown in Fig.~\ref{fig:GW_parspace}. While larger masses could be probed by LiteBIRD and similar CMB experiments, the viable range remains narrow. For $m_{\chi}\gtrsim 0.67 H_I\simeq 10^{13}\,{\rm GeV}$, the GW signal becomes too weak for near-future detection. Notably, these GW constraints on heavy DM masses are more stringent than the isocurvature bounds alone.

\subsubsection{$\sigma \neq 0$}

\begin{figure*}[!t]
\centering
    \includegraphics[width=0.7 \textwidth]{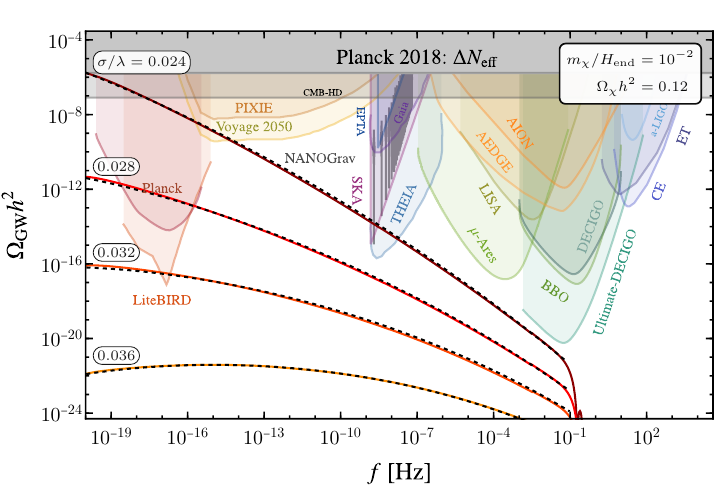}
    \caption{Gravitational wave spectrum sourced by the isocurvature fluctuations of a stable spectator scalar field $\chi$, which comprises the entirety of the DM, assuming a non-vanishing coupling to the inflaton and fixing $m_{\chi}$. Shown as the dashed black line is the analytical approximation (\ref{eq:GWnr}), for which $\mathcal{J}_{\rm NR}/\mathcal{J}_{\rm reh}\simeq 5-6$. }
    \label{fig:GW_DM_s}
\end{figure*}

We now examine the DM scenario with inflaton-spectator coupling, fixing $m_{\chi}/H_{\rm end}=10^{-2}$. Fig.~\ref{fig:GW_DM_s} shows GW spectra for couplings $\sigma/\lambda=\{0.024,0.032,0.036\}$, with corresponding reheating temperatures $T_{\rm reh}=\{435,552,664\}\,{\rm GeV}$ that yield the observed DM abundance. The dashed black curves show the analytical approximation~(\ref{eq:GWnr}) for $k<k_{\rm reh}$. Numerically, we find $\mathcal{J}_{\rm NR}/\mathcal{J}_{\rm reh}\simeq 5-6$ for all cases, yielding the good agreement shown.

The low reheating temperatures required for DM production yield smaller GW amplitudes than the high-temperature curvaton scenarios in Fig.~\ref{fig:GW_sT_rad}. Only the monotonically red-tilted spectra have sufficient amplitude to reach detector sensitivities. For the chosen mass, CMB constraints exclude $\sigma/\lambda\gtrsim 0.03$, with only $\sigma/\lambda\gtrsim 0.033$ potentially detectable by next-generation experiments. Smaller couplings produce peaked spectra far below all sensitivity curves. Similar to the $\sigma=0$ case, comparison with Fig.~\ref{fig:iso1} reveals that GW constraints exceed the isocurvature bounds.
\section{Conclusions}
\label{sec:conclusions}
We have analyzed gravitational wave signatures from spectator scalar fields during and after inflation. Heavy spectator fields with masses near the Hubble scale naturally generate blue-tilted isocurvature spectra with indices $\Lambda \simeq 0.25 - 0.5$ (where $\Lambda = 2m_{\chi,{\rm eff}}^2/3H_{\rm I}^2$), inducing secondary GWs across frequencies from $10^{-20}$ to $1$ Hz. Our approach combines analytical approximations with full numerical evolution throughout cosmic history---from inflation through reheating to radiation domination. We develop analytical expressions capturing the complete gravitational production mechanism, while numerical calculations employ normal-ordered energy density operators and regularized momentum integrals to compute both isocurvature spectra and induced GW energy densities.

Two distinct regimes emerge depending on reheating temperature and spectator field stability. In curvaton scenarios with $T_{\rm reh} \geq 10^{7} \, \rm{GeV}$, the radiation contribution $\mathcal{J}_{\rm rad}$ dominates, producing GW signals scaling as $T_{\rm reh}^{4/3}$ with amplitudes reaching $\Omega_{\rm GW}h^2 \simeq 10^{-14}$ at nHz–mHz frequencies. For dark matter scenarios with $T_{\rm reh} \lesssim 1 \, \rm{TeV}$, the early-time contribution $\mathcal{J}_{\rm NR}$ dominates, frozen at isocurvature freeze-out (approximately 5–6 $e$-folds post-inflation) with amplitudes suppressed as $T_{\rm reh}^{-2/3}$.

We have established parameter space constraints from multiple observations, presented in Fig.~\ref{fig:GW_parspace}. Isocurvature bounds from {\em Planck} observations exclude spectator masses below $m_\chi \lesssim 0.54H_I \simeq 8.4 \times 10^{12} \, \rm{GeV}$, while the non-observation of primordial tensor modes by {\em Planck} and BICEP2/Keck further constrains $m_\chi \lesssim 10^{13} \,\rm{GeV}$ except at very low reheating temperatures $T_{\rm reh} \lesssim 10^5\, \rm{GeV}$. Importantly, our GW analysis provides constraints that are generally more stringent than isocurvature bounds alone, particularly in the dark matter scenario where the constraint scales approximately as $m_\chi \gtrsim 0.7H_I$. 

The GW spectra exhibit remarkably rich phenomenology depending on the coupling between the inflaton and spectator field. For purely gravitational production ($\sigma = 0$), spectra are universally red-tilted with $\Omega_{\rm GW, \mathcal{S}} \propto f^{-0.83}$ in the infrared for the choice of $\Lambda \simeq 0.29$. Non-zero coupling introduces additional complexity: for small couplings ($\sigma/\lambda \lesssim 0.03$), spectra remain red-tilted, while intermediate values ($0.03 \lesssim \sigma/\lambda \lesssim 0.05$) can produce peaked spectra with maxima around $f \simeq 10^{-6}\, \rm{Hz}$, and larger couplings ($\sigma/\lambda \geq 0.05$) generate blue-tilted spectra in the infrared. Notably, we identified parameter regions (e.g., $m_{\chi}/H_{\rm end} = 10^{-2}, T_{\rm reh} = 10^{15}\, \rm{GeV},\ 0.041 \lesssim \sigma/\lambda \lesssim 0.044$) where correlated signatures could appear across multiple observational channels---from CMB experiments through PTAs to space-based interferometers.

Our work reveals that isocurvature perturbations from spectator fields offer a novel probe of the otherwise elusive reheating epoch. Since gravitational wave amplitudes depend strongly on the reheating temperature through competing contributions ($\mathcal{J}_{\rm NR} \propto T_{\rm reh}^{-2/3}$ and $\mathcal{J}_{\rm rad} \propto T_{\rm reh}^{2/3}$), future detection or non-detection of these signals could provide the first direct observational constraints on post-inflationary thermalization. The crossover between these regimes occurs at $T_{\rm reh} \simeq (1090 \rm{GeV})(\mathcal{J}_{NR}/\mathcal{J}_{reh})^{3/4}$, with the ratio $\mathcal{J}_{NR}/\mathcal{J}_{\rm reh}$ typically ranging from $5-120$ depending on the specific model parameters. For dark matter scenarios, we find that gravitationally produced spectator fields can precisely account for the observed relic abundance $\Omega_{\rm DM} h^2 = 0.12$ with reheating temperatures in the range $T_{\rm reh} \simeq 400-2500 \, \rm{GeV}$ (depending on $m_{\chi}/H_I$ between $0.5-0.7$), placing the mechanism squarely within the sensitivity range of proposed gravitational wave detectors such as BBO and Ultimate-DECIGO. 

The broad frequency coverage of predicted signals, spanning from 1 nHz (accessible to SKA and future PTAs) through 1-10 Hz ($\mu$-Ares, AION, AEDGE) to 1 kHz (ground-based interferometers like Einstein Telescope and Cosmic Explorer), suggests that a multi-band gravitational wave astronomy approach could be essential for fully characterizing these signatures. Our analytical results, including the complete expressions for the GW power spectrum, provide a framework for translating future observational sensitivities into fundamental constraints on early universe physics. From a theoretical perspective, our findings establish that spectator scalar fields bridge multiple frontiers in fundamental physics: they connect inflation and dark matter production, probe non-standard thermal histories, and potentially reveal physics beyond the Standard Model through coupling structures. The identification of ``sweet spots" in parameter space (e.g., $m_{\chi}/H_I \simeq 0.6-0.7$ for $\sigma = 0$, or specific coupling ranges for lighter fields) where signals are both allowed by current constraints and detectable by future experiments provides concrete targets for observational programs.

Looking forward, our analysis demonstrates that spectator scalar fields represent a theoretically well-motivated and observationally accessible source of primordial gravitational waves. As next-generation experiments come online, from CMB-HD (targeting $\Delta N_{\rm eff} = 0.014$) and LiteBIRD ($r > 0.001$) for small-scale CMB measurements to BBO ($\Omega_{\rm GW} h^2 \simeq 10^{-16}-10^{-17}$) and Ultimate-DECIGO for GWs, these signals could provide unprecedented insight into physics beyond the Standard Model operating at energy scales near $10^{13}-10^{15} \rm{GeV}$, the nature of dark matter production mechanisms, and the detailed thermal history of the universe during the poorly understood epoch between inflation and big bang nucleosynthesis. The methodology developed here, combining analytical insights with numerical precision, establishes a robust foundation for interpreting the data from this emerging multi-messenger window into the primordial universe.

\begin{acknowledgments}
\noindent
The authors would like to thank Andrew Long for a detailed discussion on the isocurvature power spectrum, and also thank Jeff Dror, David Kaiser, Rocky Kolb, Soubhik Kumar, Evan McDonough, and Vincent Vennin for helpful conversations. The work of S.V. was supported in part by DOE grant grant DE-SC0022148 at the University of Florida. MG was supported by the DGAPA-PAPIIT grant IA100525 at UNAM, the SECIHTI ``Ciencia de Frontera” grant CF-2023-I-17, and a Cátedra Marcos Moshinsky.
\end{acknowledgments}

\appendix
\newpage
\section{Regularization and Renormalization of Energy Density}
\label{app:A}
In this appendix, we address the regularization and renormalization of the energy density of the spectator field $\chi$, focusing on ultraviolet (UV) divergences arising from high-momentum modes. Using the rotated ladder operators $A_k$, $A_{-k}^\dagger$, $A_k'$, and $A_{-k'}^\dagger$, the $00$ component of the stress-energy tensor, $T_{00}^{(\chi)}$, can be written as:
\begin{equation}
    T_{00}^{(\chi)}(\eta, \mathbf{x}) \; = \; \frac{1}{2a^4} \int \frac{d^3 k d^3k'}{(2\pi)^3} \left[ \mathcal{B}_1(k, k') A_k A_{k'} + \mathcal{B}_2(k, k') A_k A_{-k'}^{\dagger} + \mathcal{B}_3(k, k') A_{-k}^{\dagger} A_{k'} + \mathcal{B}_4(k, k') A_{-k}^{\dagger} A_{-k'}^{\dagger} \right] \, .
\end{equation}
The coefficients $\mathcal{B}_i(k, k')$ encode the contributions from various mode interactions and are given as:
\begin{align}
    &\mathcal{B}_1(k,k') \; = \; \left[ \left( i \, \omega_k + \mathcal{H} \right) \left( i \, \omega_{k'} +  \mathcal{H} \right) - k k' + a^2 m_{\text{eff}}^2 - \frac{a^2 R}{6}  \right] f_k(\eta) f_{k'}(\eta) \, , \\
    &\mathcal{B}_2(k,k') \; = \; \left[ \left( -i \, \omega_k -  \mathcal{H} \right) \left( i \, \omega_{k'} -  \mathcal{H} \right) - k k' + a^2 m_{\text{eff}}^2 - \frac{a^2 R}{6} \right] f_k (\eta)f_{k'}^*(\eta)\, , \\
    &\mathcal{B}_3(k,k') \; = \;  \left[ \left( i \, \omega_k -  \mathcal{H} \right) \left( -i \, \omega_{k'} - \mathcal{H} \right) - k k' + a^2 m_{\text{eff}}^2  - \frac{a^2 R}{6} \right] f_k^*(\eta) f_{k'}(\eta)
    \, , \\
    &\mathcal{B}_4(k,k') \; = \; \left[ \left( i \, \omega_k - \mathcal{H} \right) \left( i \, \omega_{k'} -\mathcal{H} \right) - k k' + a^2 m_{\text{eff}}^2  - \frac{a^2 R}{6}  \right] f_{k'}^*(\eta) f_k^*(\eta)
    \, .
\end{align}

However, this form contains a UV divergence and must be renormalized. A detailed discussion of renormalization using the normal-ordered energy operators can be found in~\cite{Chung:2004nh, Kolb:2023ydq}. In general, renormalization requires normal-ordering with respect to the rotated basis of ladder operators, $A_k$ and $A_{-k}^\dagger$. To proceed, it is convenient to use Eqs.~(\ref{eq:rotbog1}) to express the normal-ordered operators in terms of the unrotated operators, $a_k$ and $a_{-k}^\dagger$. The resulting normal-ordered expressions are as follows:
\begin{align}
    &: a_k a_{k'} : \; = \; a_k a_{k'} + \alpha_k^* \beta_k^* \delta^{(3)}(\mathbf{k} + \mathbf{k}') \, , \\
    & : a_k a_{-k'}^{\dagger} : \; = \;  a_{-k'}^{\dagger} a_k - |\beta_k|^2 \delta^{(3)}(\mathbf{k} + \mathbf{k}') \, , \\
    & : a_{-k'}^{\dagger} a_k : \; = \; a_{-k'}^{\dagger} a_k - |\beta_k|^2 \delta^{(3)}(\mathbf{k} + \mathbf{k}') \, , \\
    & : a_{-k'}^{\dagger} a_k^{\dagger} : \; = \; a_{-k'}^{\dagger} a_k^{\dagger} + \alpha_k \beta_k \delta^{(3)}(\mathbf{k} + \mathbf{k}') \, .
\end{align}
The renormalized energy density is then expressed as:
\begin{equation}
\begin{aligned}
    \rho_{\chi}(\eta) = \langle : T_{00}^{(\chi)}(\eta, \mathbf{x}): \rangle = \frac{1}{2a^4} \int \frac{d^3k}{(2\pi)^3} \bigg[
        \mathcal{C}_1(k, -k) \alpha_k^* \beta_k^* 
        - \mathcal{C}_2(k, -k) |\beta_k|^2 - \mathcal{C}_3(k, -k) |\beta_k|^2 
        + \mathcal{C}_4(k, -k) \alpha_k \beta_k
    \bigg] \, ,
\end{aligned}
\end{equation}
where the coefficients $\mathcal{C}_i(k, k')$ are linear combinations of the $\mathcal{B}_i(k, k')$ terms, given by
\begin{equation}
\begin{aligned}
    \label{appeq:cexpressions}
    \mathcal{C}_1(k, k') &= \mathcal{B}_1(k, k') \alpha_k \alpha_{k'} + \mathcal{B}_2(k, k') \alpha_k \beta_{k'} + \mathcal{B}_3(k, k') \alpha_{k'} \beta_k + \mathcal{B}_4(k, k') \beta_k \beta_{k'} \, , \\
    \mathcal{C}_2(k, k') &= \mathcal{B}_1(k, k') \alpha_k \beta^*_{k'} + \mathcal{B}_2(k, k') \alpha_k \alpha^*_{k'} + \mathcal{B}_3(k, k') \beta_k \beta^*_{k'} + \mathcal{B}_4(k, k') \alpha^*_{k'} \beta_k \, ,\\
    \mathcal{C}_3(k, k') &= \mathcal{B}_1(k, k') \alpha_{k'} \beta^*_{k} + \mathcal{B}_2(k, k') \beta_{k'} \beta^*_{k} + \mathcal{B}_3(k, k') \alpha^*_{k} \alpha_{k'} + \mathcal{B}_4(k, k') \alpha^*_{k} \beta_{k'} \, ,\\
    \mathcal{C}_4(k, k') &= \mathcal{B}_1(k, k') \beta^*_{k} \beta^*_{k'} + \mathcal{B}_2(k, k') \alpha^*_{k'} \beta^*_{k} + \mathcal{B}_3(k, k') \alpha^*_{k} \beta^*_{k'} + \mathcal{B}_4(k, k') \alpha^*_{k} \alpha^*_{k'} \, .
\end{aligned}
\end{equation}
We can then use expressions for the Bogoliubov coefficients, 
\begin{equation}
    \alpha_k \; = \; \frac{\omega_k^* X_k + i X_k'}{2 \omega_k f_k } \, , \qquad \beta_k \; = \; \frac{\omega_k X_k -i X_k'}{2 \omega_k f_k^* } \, , 
\end{equation}
and the normal-ordered, regularized and renormalized energy density can be expressed as:
\begin{equation}
\begin{aligned}
    \rho_{\chi} (\eta) = \frac{1}{2a^2} \int \frac{d^3 k}{(2\pi)^3} \bigg[ |\omega_k X_k - i X_k'|^2 + \left(\mathcal{H}^2 - \frac{a^2 R}{6} \right) |X_k|^2 - \mathcal{H} (X_k X_k'^* + X_k' X_k^* )- \frac{1}{2\omega_k} \left(\mathcal{H}^2 - \frac{a^2 R}{6} \right)  \bigg] \, .
\end{aligned}
\end{equation}

\section{Isocurvature Computation}
\label{app:B}
In this appendix, we provide the detailed derivation of the isocurvature power spectrum given in Eq.~(\ref{eq:fullisocurvature}) in the main text. To compute the isocurvature power spectrum, we evaluate the two-point correlation function:
\begin{equation}
    \langle \delta \rho_{\chi}(\eta, \mathbf{x}) \delta \rho_{\chi}(\eta, \mathbf{y}) \rangle = \langle :\rho_{\chi}(\eta, \mathbf{x}): :\rho_{\chi}(\eta, \mathbf{y}): \rangle - \rho_{\chi}^2(\eta).
\end{equation}
First, consider the following normal-ordered operator:
\begin{equation}
\begin{aligned}
    &:\rho_{\chi}(\eta, \mathbf{x}): :\rho_{\chi}(\eta, \mathbf{y}):\; = \;  \; = \; :T_{00}^{(\chi)}(\eta, \mathbf{x}): :T_{00}^{(\chi)}(\eta, \mathbf{y}): \; = \; \\
    &\frac{1}{4a^8} \int \frac{d^3 k \, d^3 k'}{(2\pi)^3} 
    \bigg[ \mathcal{C}_1(k, k') :a_k a_{k'}: + \, \mathcal{C}_2(k, k') :a_k a_{-k'}^\dagger: + \, \mathcal{C}_3(k, k') :a_{-k}^\dagger a_{k'}: + \, \mathcal{C}_4(k, k') :a_{-k}^\dagger a_{-k'}^\dagger: \bigg] \\
    & \times \int \frac{d^3 q \, d^3 q'}{(2\pi)^3} 
    \bigg[ \mathcal{C}_1(q, q') :a_q a_{q'}: + \, \mathcal{C}_2(q, q') :a_q a_{-q'}^\dagger: + \, \mathcal{C}_3(q, q') :a_{-q}^\dagger a_{q'}: + \, \mathcal{C}_4(q, q') :a_{-q}^\dagger a_{-q'}^\dagger: \bigg] \, ,
\end{aligned}
\end{equation}
where the coefficients $C_i(k, k')$ are given by Eq.~(\ref{appeq:cexpressions}). Acting on the vacuum state, terms cancel the $\rho_{\chi}^2(\eta)$ contribution, and the remaining terms are:
\begin{equation}
    \langle \mathcal{C}_1(k, k')\mathcal{C}_4(q, q') a_k a_{k'} a_{-q}^{\dagger} a_{-q'}^{\dagger} \rangle \; = \; \mathcal{C}_1(k, k')\mathcal{C}_4(q, q') \left[\delta^{(3)}(\mathbf{k}+\mathbf{q})\delta^{(3)}(\mathbf{k}'+\mathbf{q}') + \delta^{(3)}(\mathbf{k}+\mathbf{q}')\delta^{(3)}(\mathbf{k}'+\mathbf{q}) \right] \, .
\end{equation}
Performing the integrals with respect to $q$ and $q'$, we will be left with the contribution that is proportional to $\mathcal{C}_1(k, k') \mathcal{C}_4(-k, -k') + \mathcal{C}_1(k, k') \mathcal{C}_4(-k', -k)$. Therefore, we find that the full isocurvature expression is given by\footnote{Here, we relabeled $k \rightarrow k'$ and $k' \rightarrow k''$.}
\begin{equation}
\begin{aligned}
    &\Delta_{\mathcal S}^2 (\eta, k) \; = \; \frac{1}{\rho_{\chi}^2(\eta)} \frac{1}{2a^8} \frac{k^3}{2\pi^2} \int d^3 \mathbf{r} \int \frac{d^3 k'}{(2\pi)^3}  \int \frac{d^3 k''}{(2\pi)^3} \times e^{i (\mathbf{k'} + \mathbf{k''} - \mathbf{k}) \cdot \mathbf{r}}\\
    &\times \Bigg( X_{k'} X_{k''} \left( a^2 m_{\text{eff}}^2 - k' k'' + \left(\mathcal{H}^2 -\frac{a^2 R}{6}\right) \right) + X_{k'}' X_{k''}' - \mathcal{H} \left( X_{k''} X_{k'}' + X_{k'} X_{k''}' \right) \Bigg) \\
    & \times \Bigg( X_{k'}^* X_{k''}^* \left( a^2 m_{\text{eff}}^2 - k' k'' + \left(\mathcal{H}^2 -\frac{a^2 R}{6}\right) \right) + X_{k'}^{*'} X_{k''}^{*'} - \mathcal{H}\left( X_{k''}^* X_{k'}^{*'} + X_{k'}^* X_{k''}^{*'} \right) \Bigg) \, .
\end{aligned}
\end{equation}
After performing integrals and relabeling $k' \rightarrow p$, we find
\begin{equation}
\begin{aligned}
    &\Delta_{\mathcal S}^2 (\eta, k) \; = \; \frac{1}{\rho_{\chi}^2(\eta)} \frac{1}{2a^8} \frac{k^3}{2\pi^2} \int \frac{d^3 p}{(2\pi)^3} \\
    &\times \Bigg( X_{p} X_{q} \left( a^2 m_{\text{eff}}^2 - p q + \left(\mathcal{H}^2 -\frac{a^2 R}{6} \right) \right) + X_{p}' X_{q}' - \mathcal{H} \left( X_{p}' X_{q} + X_{p} X_{q}' \right) \Bigg) \\
    & \times \Bigg( X_{p}^* X_{q}^* \left( a^2 m_{\text{eff}}^2 - p q + \left(\mathcal{H}^2-\frac{a^2 R}{6}\right) \right) + X_{p}'^{*} X_{q}'^{*} - \mathcal{H} \left( X_{p}'^{*} X_{q}^*  + X_{p}^* X_{q}'^{*} \right) \Bigg) \, ,
\end{aligned}
\end{equation}
where $q = |\mathbf{p - k}|$. This matches Eq.~(\ref{eq:fullisocurvature}) in the main text.

\section{Details of the Computation of the GW Spectrum}
\label{app:C}
In this appendix, we include the details not presented in Section~\ref{sec:gravwaves1} for the computation of the GW spectrum. We first express Eq.~(\ref{eq:s2point}) as
\begin{align} \notag
    \langle S_{ \lambda}(\eta_1, \mathbf{k}_1) & S_{\lambda}(\eta_2, \mathbf{k}_2) \rangle  \; = \; \frac{a(\eta_1)^4}{M_P^4} \frac{a(\eta_2)^4}{M_P^4} \int \frac{d^3q_1}{(2\pi)^{3/2}} \frac{d^3q_2}{(2\pi)^{3/2}} \frac{Q_{\lambda}(\mathbf{k}_1, \mathbf{q}_1) Q_{\lambda}(\mathbf{k}_2, \mathbf{q}_2)}{q_1^2 |\mathbf{k}_1 - \mathbf{q}_1|^2 q_2^2 |\mathbf{k}_2-\mathbf{q}_2|^2} \\ \notag
     &\qquad \times \mathcal{T}_{\mathcal{S}}(\eta_1) \mathcal{T}_{\mathcal{S}}(\eta_2) \langle \delta \rho_{\chi}(\mathbf{q}_1) \delta \rho_{\chi}(|\mathbf{k}_1-\mathbf{q}_1|) \delta \rho_{\chi}(\mathbf{q}_2) \delta \rho_{\chi}(|\mathbf{k}_2-\mathbf{q}_2|) \rangle \\ \notag
     &=\; \frac{a(\eta_1)^4}{M_P^4} \frac{a(\eta_2)^4}{M_P^4} \int \frac{d^3q_1}{(2\pi)^{3/2}} \frac{d^3q_2}{(2\pi)^{3/2}} \frac{Q_{\lambda}(\mathbf{k}_1, \mathbf{q}_1) Q_{\lambda}(\mathbf{k}_2, \mathbf{q}_2)}{q_1^2 |\mathbf{k}_1 - \mathbf{q}_1|^2 q_2^2 |\mathbf{k}_2-\mathbf{q}_2|^2} \\ \label{eq:SSll}
    &\qquad \times \mathcal{T}_{\mathcal{S}}(\eta_1) \mathcal{T}_{\mathcal{S}}(\eta_2) \bar{\rho}^2_{\chi}(\eta_1) \bar{\rho}^2_{\chi}(\eta_2) \langle \delta_{\chi}(\mathbf{q}_1) \delta_{\chi}(|\mathbf{k}_1-\mathbf{q}_1|) \delta_{\chi}(\mathbf{q}_2) \delta_{\chi}(|\mathbf{k}_2-\mathbf{q}_2|) \rangle 
    \, ,
\end{align}
where $\delta_{\chi} \equiv \delta \rho_{\chi}/\bar{\rho}_{\chi}$.
We then proceed by using the general expression~(\ref{eq:4pointsplit}) and split the $4$-point correlation functions in terms of a product of $2$-point correlation functions. Therefore, we have
\begin{align} \notag
    \int \frac{d^3q_1}{(2\pi)^{3/2}} &\frac{d^3q_2}{(2\pi)^{3/2}} \frac{Q_{\lambda}(\mathbf{k}_1, \mathbf{q}_1) Q_{\lambda}(\mathbf{k}_2, \mathbf{q}_2)}{q_1^2 |\mathbf{k}_1 - \mathbf{q}_1|^2 q_2^2 |\mathbf{k}_2-\mathbf{q}_2|^2}  \langle \delta_{\chi}(\mathbf{q}_1) \delta_{\chi}(|\mathbf{k}_1-\mathbf{q}_1|) \delta_{\chi}(\mathbf{q}_2) \delta_{\chi}(|\mathbf{k}_2-\mathbf{q}_2|) \rangle \\ \notag
    &=\; \int \frac{d^3q_1}{(2\pi)^{3/2}} \frac{d^3q_2}{(2\pi)^{3/2}} \frac{Q_{\lambda}(\mathbf{k}_1, \mathbf{q}_1) Q_{\lambda}(\mathbf{k}_2, \mathbf{q}_2)}{q_1^2 |\mathbf{k}_1 - \mathbf{q}_1|^2 q_2^2 |\mathbf{k}_2-\mathbf{q}_2|^2} \delta^{(3)}(\mathbf{k}_1 + \mathbf{k}_2) 
     \\ \label{eq:iiQQ}
    &\qquad\times\left[ \delta^{(3)}(\mathbf{q}_1 + \mathbf{q}_2) + \delta^{(3)}(\mathbf{k}_1 + \mathbf{q}_2 - \mathbf{q}_1) \right] \mathcal{P}_{\mathcal{S}}(\mathbf{q}_1) \mathcal{P}_{\mathcal{S}}(|\mathbf{k}_1 - \mathbf{q}_1|) \, .
\end{align}
Now the first term inside the brackets becomes
\begin{align} \notag
     \int \frac{d^3q_1}{(2\pi)^{3/2}} &\frac{d^3q_2}{(2\pi)^{3/2}} \frac{Q_{\lambda}(\mathbf{k}_1, \mathbf{q}_1) Q_{\lambda}(\mathbf{k}_2, \mathbf{q}_2)}{q_1^2 |\mathbf{k}_1 - \mathbf{q}_1|^2 q_2^2 |\mathbf{k}_2-\mathbf{q}_2|^2} \delta^{(3)}(\mathbf{k}_1+\mathbf{k}_2)  \delta^{(3)}(\mathbf{q}_1+\mathbf{q}_2)  \mathcal{P}_{\mathcal{S}}(\mathbf{q}_1) \mathcal{P}_{\mathcal{S}}(|\mathbf{k}_1 - \mathbf{q}_1|)  \\ \notag
    &= \;\int \frac{d^3q_1}{(2\pi)^{3}}  \frac{Q_{\lambda}(\mathbf{k}_1, \mathbf{q}_1) Q_{\lambda}(-\mathbf{k}_1,- \mathbf{q}_1)}{q_1^4 |\mathbf{k}_1 - \mathbf{q}_1|^4} \delta^{(3)}(\mathbf{k}_1+\mathbf{k}_2) \mathcal{P}_{\mathcal{S}}(\mathbf{q}_1) \mathcal{P}_{\mathcal{S}}(|\mathbf{k}_1 - \mathbf{q}_1|) \\
    &= \; \int \frac{d^3q_1}{(2\pi)^{3}} \frac{Q^2_{\lambda}(\mathbf{k}_1, \mathbf{q}_1)}{q_1^4 |\mathbf{k}_1 - \mathbf{q}_1|^4} \delta^{(3)}(\mathbf{k}_1+\mathbf{k}_2) \mathcal{P}_{\mathcal{S}}(\mathbf{q}_1) \mathcal{P}_{\mathcal{S}}(|\mathbf{k}_1 - \mathbf{q}_1|) \, .
\end{align}
The second term in (\ref{eq:iiQQ}) in turn yields
\begin{align} \notag
    \int \frac{d^3q_1}{(2\pi)^{3/2}} &\frac{d^3q_2}{(2\pi)^{3/2}} \frac{Q_{\lambda}(\mathbf{k}_1, \mathbf{q}_1) Q_{\lambda}(\mathbf{k}_2, \mathbf{q}_2)}{q_1^2 |\mathbf{k}_1 - \mathbf{q}_1|^2 q_2^2 |\mathbf{k}_2-\mathbf{q}_2|^2} \delta^{(3)}(\mathbf{k}_1+\mathbf{k}_2)  \delta^{(3)}(\mathbf{k}_1 + \mathbf{q}_2 - \mathbf{q}_1)  \mathcal{P}_{\mathcal{S}}(\mathbf{q}_1) \mathcal{P}_{\mathcal{S}}(|\mathbf{k}_1 - \mathbf{q}_1|) \\ \notag
    &= \; \int \frac{d^3q_1}{(2\pi)^{3}}  \frac{Q_{\lambda}(\mathbf{k}_1, \mathbf{q}_1) Q_{\lambda}(-\mathbf{k}_1, \mathbf{q}_1 - \mathbf{k}_1)}{q_1^4 |\mathbf{k}_1 - \mathbf{q}_1|^4} \delta^{(3)}(\mathbf{k}_1+\mathbf{k}_2) \mathcal{P}_{\mathcal{S}}(\mathbf{q}_1) \mathcal{P}_{\mathcal{S}}(|\mathbf{k}_1 - \mathbf{q}_1|) \\
    &= \; \int \frac{d^3q_1}{(2\pi)^{3}} \frac{Q^2_{\lambda}(\mathbf{k}_1, \mathbf{q}_1)}{q_1^4 |\mathbf{k}_1 - \mathbf{q}_1|^4} \delta^{(3)}(\mathbf{k}_1+\mathbf{k}_2) \mathcal{P}_{\mathcal{S}}(\mathbf{q}_1) \mathcal{P}_{\mathcal{S}}(|\mathbf{k}_1 - \mathbf{q}_1|) \, .
\end{align}
Combining the two pieces together and using the dimensionless isocurvature power spectrum definition~(\ref{eq:isocurvaturedimensionless}), we find that (\ref{eq:SSll}) simplifies to
\begin{align} \notag
    \langle S_{ \lambda}(\eta_1, \mathbf{k}_1) S_{\lambda}(\eta_2, \mathbf{k}_2) \rangle  \; = \; &\frac{a(\eta_1)^4}{M_P^4} \frac{a(\eta_2)^4}{M_P^4} \mathcal{T}_{\mathcal{S}}(\eta_1) \mathcal{T}_{\mathcal{S}}(\eta_2) \bar{\rho}^2_{\chi}(\eta_1) \bar{\rho}^2_{\chi}(\eta_2)\\
     &\times \pi \int d^3q_1 \frac{Q^2_{\lambda}(\mathbf{k}_1, \mathbf{q}_1)}{q_1^7 |\mathbf{k}_1 - \mathbf{q}_1|^7} \delta^{(3)}(\mathbf{k}_1+\mathbf{k}_2) \Delta^2_{\mathcal{S}}(\mathbf{q}_1) \Delta^2_{\mathcal{S}}(|\mathbf{k}_1 - \mathbf{q}_1|)\,.
\end{align}
Therefore, the full contribution to the tensor 2-point function (\ref{eq:gwspectrgenexp}) sourced by isocurvature can be expressed as
\begin{align} \notag
   \langle h_\lambda(\eta, \mathbf{k}_1) h_\lambda(\eta, \mathbf{k}_2) \rangle \; = \; &\frac{\pi}{M_P^8} \int_0^{\eta}  d\eta_1 \mathcal{G}_{\mathbf k}(\eta, \eta_1) \mathcal{T}_{\mathcal{S}}(\eta_1) \bar{\rho}_{\chi}^2(\eta_1)  a(\eta_1)^4 \int_0^{\eta} d\eta_2 \,  \mathcal{G}_{\mathbf k}(\eta, \eta_2) \mathcal{T}_{\mathcal{S}}(\eta_2) \bar{\rho}_{\chi}^2(\eta_2)  a(\eta_2)^4 \\
   & \times \int d^3q_1 \frac{Q^2_{\lambda}(\mathbf{k}_1, \mathbf{q}_1)}{q_1^7 |\mathbf{k}_1 - \mathbf{q}_1|^7} \delta^{(3)}(\mathbf{k}_1+\mathbf{k}_2) \Delta^2_{\mathcal{S}}(\mathbf{q}_1) \Delta^2_{\mathcal{S}}(|\mathbf{k}_1 - \mathbf{q}_1|) \, .
\end{align}
In practice, it is often more convenient to reformulate this calculation in terms of $e$-folds rather than conformal time, i.e.~$\langle h_\lambda(N, \mathbf{k}) h_\lambda(N, \mathbf{k}') \rangle$. Below, we detail the full procedure for obtaining both the time-dependent and wavenumber-dependent components from the isocurvature contribution to the power spectrum. 

In cosmic time, the equation governing the secondary gravitational waves corresponds to
\begin{equation}
    \ddot{h}_{ij} + 3H \dot{h}_{ij} - \frac{\nabla^2 h_{ij}}{a^2} h_{ij} \; = \; g(t) \, ,
\end{equation}
where $g(t)$ is a general source term, analogous to Eq.~(\ref{eq:hijincosmictime}). We now switch to the $e$-fold variable,
\begin{equation}
    N \; = \; \log a \, , \qquad dN = \frac{\dot{a}}{a} dt \; = \; H dt \, .
\end{equation}
so that derivatives are expressed as,
\begin{equation}
\dot{h} \; = \; \frac{d}{dt} h \; = \; \frac{dN}{dt} \frac{dh}{dN} = H \frac{dh}{dN} \, ,\qquad \ddot{h} = \frac{d}{dt} \left( H \frac{dh}{dN} \right) = \dot{H} \frac{dh}{dN} + H^2 \frac{d^2 h}{dN^2} \, .
\end{equation}
Substituting these expressions, we obtain
\begin{equation}
\frac{d^2 h}{dN^2} + (3 - \varepsilon_H) \frac{dh}{dN} + \left(\frac{k}{aH}\right)^2 h \; = \; \frac{g(t)}{H^2} \, ,
\end{equation}
where the slow-roll parameter is defined as $\varepsilon_H = - \frac{\dot{H}}{H^2}$.

In the case of conformal time, we have in turn
\begin{equation}
h' \; = \; \frac{d}{dt} h \; = \; \frac{dt}{d\eta} \frac{dh}{dt} \; = \; \frac{1}{a} \frac{dh}{d\eta}\, \qquad h'' \; = \; \frac{1}{a} \frac{d}{d\eta} \left( \frac{1}{a} \frac{dh}{d\eta} \right) 
\; = \; -\frac{a'}{a^3} \frac{dh}{d\eta} + \frac{1}{a^2} \frac{d^2 h}{d\eta^2} \, .
\end{equation}
Thus, the gravitational wave equation of motion in conformal time is:
\begin{equation}
\frac{1}{a^2} h'' - \frac{\mathcal{H}}{a^2} h' + \frac{3 H}{a} h' + \frac{k^2}{a^2} h \; = \; g(t) \, ,
\end{equation}
or equivalently:
\begin{equation}
h'' + 2\mathcal{H} h' + k^2 h \; = \; a^2 g(t) \, .
\end{equation}

To translate between Green's functions in cosmic time, conformal time, and $e$-folds, we use:
\begin{equation}
    \int dt' g(t') \mathcal{G}_{\mathbf{k}}(t, t')  \; = \; \int d\eta' a^2 g(\eta') \mathcal{G}_{\mathbf{k}}(\eta, \eta')  \; = \; \int dN'  g(N') \frac{\mathcal{G}_{\mathbf{k}}(N, N')}{H^2} \, ,
\end{equation}
which can be rewritten as
\begin{equation}
    \int dN' a^2 g(\eta') \mathcal{G}_{\mathbf{k}}(\eta, \eta') \bigg|\frac{d \eta'}{dN'} \bigg| \; = \; \int dN' \frac{a}{H} g(N')\mathcal{G}_{\mathbf{k}}(N, N') dN' \, ,
\end{equation}
where 
\begin{equation}
    \frac{d \eta}{d N } \; = \; \frac{d \eta}{dt} \frac{dt}{dN} \; = \;  \frac{1}{aH} \, .
\end{equation}
Thus, the full contribution in the $e$-fold variable can be expressed as $\langle h_\lambda(N, \mathbf{k}) h_\lambda(N, \mathbf{k}') \rangle = \delta^{(3)}(\mathbf{k} + \mathbf{k}') \mathcal{P}_{\lambda}(N,k)$, where
\begin{align} \notag
   \mathcal{P}_{\lambda}(N, k) \;=\;  &\frac{\pi}{M_P^8} \left[ \int_0^{N} dN' \mathcal{G}_{\mathbf{k}}(N,N') \mathcal{T}_{\mathcal{S}}(N') \frac{a^4(N') \bar{\rho}_{\chi}^2(N')}{(a(N') H(N'))^2} \right]^2 \\ 
   &\times  \int d^3p \frac{Q^2_{\lambda}(\mathbf{k}, \mathbf{p})}{p^7 |\mathbf{k} - \mathbf{p}|^7} \Delta^2_{\mathcal{S}}(\mathbf{p}) \Delta^2_{\mathcal{S}}(|\mathbf{k} - \mathbf{p}|)  \, .
\end{align}

We now aim to simplify the momentum integral. Using Eqs.~(\ref{eq:qexpressions}) in the main text we find that
\begin{equation}
    \sum_{\lambda} Q^2_{\lambda}(\mathbf{k}, \mathbf{p}) \; = \; \frac{p^4}{2} \sin^4 \theta \cos^2(2 \phi) + \frac{p^4}{2} \sin^4 \theta \sin^2(2 \phi) \; = \; \frac{p^4}{2} \sin^4 \theta \, .
\end{equation}
Next, we express the integral in spherical coordinates:
\begin{equation}
    \int d^3p \frac{p^4}{2} \sin^4 \theta \, f(p,|\mathbf{k - p}|)\; = \; \pi \int dp \, p^6  d \cos \theta  \sin^4 \theta\, f(p,|\mathbf{k - p}|) \, .
\end{equation}
To simplify the integration, we switch to new variables $(p, q)$, where the relationship between $p$, $q$, and $\cos\theta$ is given by
\begin{equation}
q^2 \; = \;  p^2 + k^2 - 2pk \cos\theta \, ,
\end{equation}
derived from the triangle relation. Solving for $\cos\theta$, we find
\begin{equation}
\cos \theta \; = \;  \frac{k^2 + p^2 - q^2}{2kp} \, ,
\end{equation}
where $f(p, q)$ is some arbitrary function. The sine of the angle $\theta$ can then be written as
\begin{equation}
\sin^4\theta = (1 - \cos^2\theta)^2 = \frac{(k^4 - 2k^2(p^2 + q^2) + (p^2 - q^2)^2)^2}{16k^4p^4} \, .
\end{equation}
With the transformed volume element given by
\begin{equation}
dp \, d\cos\theta = \frac{q}{kp} \, dp \, dq\,,
\end{equation}
and the triangle inequality $|k - p| \leq q \leq k + p$, we can then simplify the momentum integral as
\begin{equation}
  \int dp \, \frac{p^{6}}{2} d \cos \theta  \sin^4 \theta\, f(p,|\mathbf{k - p}|) \; = \; \frac{1}{16 k^5} \int_0^\infty dp \, p \int_{|k-p|}^{k+p} dq \, q \, (k^4 - 2k^2(p^2 + q^2) + (p^2 - q^2)^2)^2\, f(p, q) \, . 
\end{equation}
Finally, defining $\Delta_h^2 = \sum_{\lambda} \Delta^2_{\lambda}$, the full gravitational wave power spectrum is:
\begin{align}
\Delta_h^2(k) \;=\; &2 \left[\frac{1}{8}\int^NdN'\, \mathcal{G}_{\mathbf{k}}(N,N')\,\mathcal{T}_{\mathcal{S}}(N')\left(\frac{a(N')H(N')}{k}\right)^2\left(\frac{\bar{\rho}_{\chi}(N')}{H^2(N')M_P^2}\right)^2\right]^2 g(k)
\, ,
\end{align}
where 
\beq
\label{eq:f1gen}
g(k) \;=\; k^2\int_0^{\infty} dp\, p\int_{|k-p|}^{k+p}dq\, q\,\frac{(k^4-2k^2(p^2+q^2)+(p^2-q^2)^2)^2}{p^7q^7}\,\Delta_{\mathcal{S}}^2(p)\Delta_{\mathcal{S}}^2(q)\,.
\eeq

\addcontentsline{toc}{section}{References}
\bibliographystyle{utphys}
\bibliography{references}

\end{document}